\documentclass[12 pt,fleqn,thmsa]{article}
\usepackage{amsfonts}
\usepackage{amssymb,hyperref,backref}
\usepackage{amsmath,harvard,tikz,verbatim,epsfig}
\usepackage{pdfpages}
\usepackage{amsthm}
\usepackage{graphicx}
\usepackage{float}
\usepackage{comment}
\usepackage{adjustbox}
\usepackage{amsmath,amsfonts,esint}
\usepackage{amsmath,amstext,rotating}
\usepackage{amsfonts,amssymb,graphics,xspace,endnotes}
\usepackage{fancyhdr}
\usepackage{epsfig}
\usepackage{srcltx}
\usepackage{amsthm}
\usepackage{geometry}
\usepackage{subfig,graphicx}
\usepackage{tikz} 
\usetikzlibrary{patterns}
\usetikzlibrary{backgrounds}

\newcommand{\sgn}{sgn}
\newcommand{\be}{\begin{eqnarray*}}
\newcommand{\ee}{\end{eqnarray*}}

\newcommand{\avg}{\frac{1}{n}\sum_{i=1}^n}

\newcommand{\usum}{ \sum_{i \neq j} }

\renewcommand{\theequation}{\thesection.\arabic{equation}}
\newtheorem{assumption}{Assumption}
\newtheorem{definition}{Definition}

\newtheorem{theorem}{Theorem}[section]

\newtheorem{remark}{Remark}

\newcommand{\mcA}{{\mathcal A}}
\newcommand{\mcF}{{\mathcal F}}

\newcommand{\mcC}{{\mathcal C}}

\newtheorem{assumpB}{Assumption}

\DeclareMathOperator*{\argmax}{arg\,max}
\DeclareMathOperator*{\sign}{sign}
\newcommand{\real}{{\mathbb R}}
\newcommand{\PP}{{\mathbb P}}
\newcommand{\EE}{{\mathbb E}}
\input{tcilatex}

\begin{document}

\bibliographystyle{econometrica}


\begin{center}

\noindent 
 {\bf  Sharp and Robust  Estimation of \\ Partially Identified Discrete Response  Models}

\noindent
 { Shakeeb Khan (Boston College)\\ Tatiana Komarova (University of Manchester)\\ Denis Nekipelov (UVA)}

 \
 
 First Version: May 2022

This Version: May 20, 2024\footnote{This paper previously circulated under the title
``On Optimal Set Estimation for Partially identified Binary Choice Models". We are grateful for helpful comments from Ilya Molchanov, conference participants at the 2019 Asian ES meetings in Xiamen, the 2019 Chinese ES meetings in Guangzhou,  the 2019 Midwest Econometric Study Group in Columbus, Ohio, the 2022 Montreal Econometrics summer conference, the 2022 Advances in Econometrics conference at Toulouse School of Economics, and seminar participants at Georgetown, UC Berkeley, UC Louvain, University of Bristol, University of Warwick, UC Riverside, Yale, University of Iowa, Boston University, University of Arizona, University of Miami. }
 

\

\noindent
{\bf Abstract:}

\end{center}

\noindent
{\small Semiparametric discrete choice models are widely used in a variety of practical applications. While these models are point identified in the presence of continuous covariates, they can become partially identified when covariates are discrete. In this paper we find that classical estimators, including the maximum score estimator, (\citeasnoun{manski1975}), 
loose their attractive statistical properties without point identification. First of all, they are not {\it sharp} with the estimator converging to an {\em outer region} of the identified set, (\citeasnoun{komarova2013}), 
and in many discrete designs it weakly converges to a {\em random set}. Second, they are not {\it robust,} with their distribution limit discontinuously changing with respect to the parameters of the model.
We propose a novel class of estimators based on the concept of a quantile of a random set, 
which we show to be both sharp and robust.
We demonstrate that our approach  extends from cross-sectional settings to classical  static and dynamic discrete panel data models.
}



\pagebreak

\section{Introduction}
Early work on modeling discrete choices of economic agents with random utility over those choices led to the creation of an important class of semiparametric econometric models where discrete outcome variable is a parametric function of observed regressors, but also depends on the random noise whose distribution is unknown and is not specified parametrically. Pioneering this line of work, \citeasnoun{manski1975} advanced a series of foundational papers, elucidating conditions under which the parameters of such models attain point identification. Specifically, by stipulating the conditional median of the random noise to be zero, coupled with, most notably, a continuity assumption regarding the distribution of at least one regressor, he established the precise conditions sufficient for parameter point identification. Discrete-only data, however, is commonplace in a range of policy-relevant applications and in this paper, we focus on the settings of the discrete choice model where regressors have only a discrete support. This means that the continuity assumption used in the prior literature is violated which may lead to the failure of point identification, the case analyzed in \citeasnoun{komarova2013}.  We demonstrate that depending on the data generating process parameters, our model can be either point- or partially-identified, resulting in the identified set being either a singleton or a non-singleton.

Within the discrete-only regressors framework, we propose a new class of estimators for discrete choice models that can be applied in either partially or point identified scenarios. These estimators are based on the concept of a quantile of a random set from the random set theory (e.g., see \citeasnoun{molchanov2006book}). We show that they successfully address the changing structure of the identified set over the parameter space from singleton to non-singleton sets. We compare our new estimator with 
the maximum score estimator and  the closed-form estimator based on ideas of \citeasnoun{ichimura1994}, which were both  originally intended for point identification scenarios. Additionally, we contrast our new estimator with two novel variants of the maximum score estimator and the closed-form estimator, designed to combine the strengths of both approaches. 

We use two criteria for evaluation and comparison of estimation methodologies. First, {\em sharpness} is the property of an estimator to approximate the identified set with high probability for a given value of parameter of the data generating process. The rationale behind employing this criterion is clear-cut; our objective is to ascertain the truth in the probability limit. Second, {\em robustness} is the property that the distribution of an estimator varies continuously in the small neighborhood of a particular parameter value of the data generating process.  This property is important as the lack of the local continuity in distribution required for robustness can also result in failure of bootstrap and other resampling-based methods for inference. This resembles the case of the parameter on the boundary of the parameter space leading to discontinuity in the distribution limit and the failure of bootstrap, e.g., characterized by \citeasnoun{andrews-boundary}.

We find that the maximum score and closed-form estimators are not sharp over the whole parameter space. E.g., the maximum score estimator can converge in probability to a singleton or a set, but it can also weakly converge to a random set. However, the maximum score estimator is robust whereas the closed-form estimator is not. Our two novel variants of the maximum score estimator and the closed-form estimator, as mentioned earlier, aim to amalgamate the strengths of both approaches, potentially yielding improved properties. One variant combines the objective functions of the the maximum score the closed-form estimator. The other variant combines the set estimates produced by those estimators. The former combination idea leads to sharp estimation, whereas the second one does not. At the same time, both of these combination estimators fail to be robust. 

In contrast, our new estimator rooted in the concept of a quantile of a random set proves to be \textit{both} sharp and robust, establishing its superiority over all previously discussed estimation methodologies in our discrete-only setting. To elaborate,  this estimator incorporates the classic maximum score estimator and defines a new estimator based on the quantile of the random set produced by the maximum score method. It can be implemented in practice using a simple re-sampling procedure. We posit that the  estimators of this kind may be appealing in other partially identified models with discrete variables.

Our paper draws on rich prior literature on identification and inferences for point and partially
identified semiparametric discrete choice models. This includes classic work on the maximum score estimator in \citeasnoun{manski1975} and \citeasnoun{manski1985} with the distribution theory developed in \citeasnoun{kimpollard} and
the smoothed version of the estimator \citeasnoun{horowitz-sms} yielding asymptotically normally distributed parameters.  This model was further studied in setting with heteroskedastic errors in \citeasnoun{khan:13}, and in setting with discrete regressors in \citeasnoun{komarova2013}. \citeasnoun{ichimura1994} and \citeasnoun{ahnetal2018} consider an alternative set of 
procedures focusing on the implication for the choice
probability to be close to 0.5 for particular values of covariates. On the side of inference, important relevant
results are discussed in \citeasnoun{abrevayahuang2005} showing the failure of the standard bootstrap for the classic maximum score estimator, and later \citeasnoun{cattaneojansson2020} developed a valid bootstrap
procedure for the maximum score.  \citeasnoun{rosenura2022} considered finite sample properties of a related estimator that is based on moment inequalities.

The rest of the paper is organized as follows. In  Section \ref{sec:identification} we present the general model, underlying technical assumptions and its simple
version which we showcase the main technical points throughout the paper. We demonstrate that, depending
on the values of parameter of the data generating
process, the model can either be point- or partially
identified. We create a toolkit for analysis of 
estimators which can capture this behavior of the identified set and discuss in detail two main criteria which we use for evaluation in the paper --  {\it sharpness} and {\it robustness} introduced earlier. 

In Section \ref{sec:classic} we consider 
the maximum score estimator \citeasnoun{manski1975} and the closed-form estimator which stems from the ideas in  \citeasnoun{ichimura1994} and establish their properties from the perspectives of our sharpness and robustness criteria.  As mentioned earlier, neither of these is  sharp over the whole parameter space but  the patterns of their non-sharpness are very different. To elaborate, the maximum score estimator is sharp on the set of full Lebesgue measure in the parameter space whereas the closed-form estimator is sharp on the set of a zero Lebesgue measure in the parameter space. On the second criterion, the maximum score estimator is robust over the whole parameter space whereas the closed-form estimator is not robust everywhere except for the points  in that set of Lebesgue measure zero  where it is sharp. This section also introduces two novel combination estimation approaches designed to build on the respective strengths of the maximum score and closed-form estimators and establishes that one of them sharp over the whole parameter space, the other one is not, and neither of them is robust.


In Section \ref{sec:quantile} we develop our main novel estimation methodology based on the  concept of a quantile of a random set to construct a new class of estimators which are shown to be both
sharp and robust everywhere on the parameter space.


In Section \ref{application} we  (a) provide  an empirical
illustration highlighting results from all the estimation approaches discussed in the paper; and (b) propose an idea for a feasible implementation of a random set quantile and  demonstrate the performance of our random set quantile estimator in a series of Monte 
Carlo simulations.

Section \ref{sec:panel data} demonstrates how ideas from our simple cross-sectional model extend to  single-index models as well as static and dynamic panel data models enabling inference when those
models are partially identified. 

Finally, Section \ref{sec:conclusion} concludes by summarizing results and discussing areas for future research, and the appendix collects all proofs of the main theorems.

\section{Identification, Sharpness and Robustness}
\label{sec:identification} 


In this section, we illustrate our central ideas by reviewing the fundamental concept of identification and introducing set estimation relevant to the binary choice models under consideration. This will serve as a foundation for formally defining the concepts of sharpness and robustness in the context of our estimators.

To introduce the formal concepts we turn to the simplest cross-sectional 
form of the semiparametric discrete choice model
\begin{equation}\label{discrete:choice}
Y={\bf 1}\left(\widetilde{X}'\widetilde{\alpha}_0-\epsilon \geq 0\right),
\end{equation}
where the outcome random variable $Y$ with range $\{0,\,1\}$ is generated from the
vector of covariates $\widetilde{X}$ and an unobserved disturbance $\epsilon.$ To avoid confusion, we use capital letters throughout the paper for random variables and small letters to denote their specific 
realizations.
We impose the following assumption on the structure of the model and the data generating process:
\begin{assumption}\label{assume:MS}
\begin{itemize}
 \item[(i)] $(y_i,\widetilde{x}_i)^{n}_{i=1}$ is an i.i.d. random sample from the joint distribution $(Y,\widetilde{X})$ induced by (\ref{discrete:choice}) for some $\widetilde{\alpha}_0 \in \mcA
  \subset \real^K.$
  \item[(ii)] Parameter space $\widetilde{\mcA}$ is a compact subset of $\real^{K}$ such that for
  some dimension $k \in \{1,2,\ldots,K\},$ $\alpha^k \equiv 1$ for all $\alpha \in \widetilde{\mcA}.$
  \item[(iii)] Distribution of regressors $\widetilde{X}$ is discrete with the support $\mathcal{X}$. The distribution density $f_{\epsilon|\widetilde{X}}(\cdot\,|\,\widetilde{X}=\widetilde{x})$ is above zero
  and in some fixed neighborhood of zero $f_{\epsilon|\widetilde{X}}(\cdot\,|\,\widetilde{X}=\widetilde{x})>L>0.$
  \item[(iv)] $\mbox{Median}\left( \epsilon\,\big|\,\widetilde{X}=\widetilde{x}\right)=0$ for each 
  $\widetilde{x}$ in the range of $\widetilde{X}.$
\end{itemize}
\end{assumption}
Assumptions \ref{assume:MS} (i), (ii) and (iv) are classic assumptions of \cite{manski1975} and following literature while Assumption \ref{assume:MS} (iii) extends the original framework to the case where regressors $\widetilde{X}$ are discrete under 
additional smoothness assumption on the distribution of the unobserved 
shock $\epsilon$. This, in particular, guarantees that the conditional c.d.f. $F_{\epsilon|\widetilde{X}}(\cdot\,|\,\widetilde{X}=\widetilde{x})$ is strictly increasing around 0, which has implication on  the characterization of the identified set. 

Under Assumption \ref{assume:MS}, parameter $\widetilde{\alpha}_0$ characterizing the 
data generating process can, in general, be no longer identified. In fact, as can be deduced from \citeasnoun{manski1975} and \citeasnoun{manski1985}, the 
identified set for $\widetilde{\alpha}_0$ -- which we will denote as $\mcA_0 $ -- is characterized as set of the values of $\widetilde{\alpha}$ such that 
\begin{equation} \label{threshold:crossing}
P(Y=1|\,|\,\widetilde{X}=\widetilde{x}) \lessgtr 0.5 \; \iff \widetilde{x}'\widetilde{\alpha} \lessgtr 0, \;\mbox{and}\; 
P(Y=1|\,|\,\widetilde{X}=\widetilde{x}) = 0.5 \; \iff \widetilde{x}'\widetilde{\alpha}=0,
\end{equation} 
and that satisfy the normalization restriction in  Assumption \ref{assume:MS}(i). Let $\mcA$ and $\mcA_{0}$ denote the projections of the parameter space $\widetilde{\mcA}$ and the identified set $\widetilde{\mcA}_0$, respectively, onto the $\real^{K-1}$ excluding the normalized $k$-th component in the original sets.

To explain our arguments, we will use two illustrative designs. 
\begin{definition}[Illustrative Design 1]
Take $K=2$ and let the vector of regressors have a fixed first component with $\widetilde{X}=(1,X)$, where $X \in \{0,1\}$.  Suppose $P(X=1)=q
 \in (0,1).$ We take $\widetilde{\alpha}_0 \equiv (\alpha_0,1)$. Let $\mbox{Med}(\epsilon\,|\,X)=0.$
\end{definition}

\begin{definition}[Illustrative Design 2]
Take $K=3$ and let the vector of regressors have a fixed first component with $\widetilde{X}=(1,X)$, where $X=(X_1,X_2) \in \{(0,1),(1,0)\}^2$.  Suppose $P(X=(0,1)')=q>\frac12$ (without loss of generality). We take $\widetilde{\alpha}_0 \equiv (\alpha_{0,1},1,\alpha_{0,2})$ (thus, $\alpha_0=(\alpha_{0,1},\alpha_{0,2})$. 
 Let $\mbox{Med}(\epsilon\,|\,X)=0.$
\end{definition}
Normality of the error distribution on these designs is not important and is simply used for convenience to fully characterize the heteroskedastic distribution of the unobserved term. 

It is easy to see from our characterization of the identified set\footnote{See Appendix \ref{appendix:proofs} for technical characterization of the identified set.} above that in our Illustrative Design 1 the following holds (suppose the parameter space $\mcA$ is large enough to contain values $0$ and $-1$): 
\begin{enumerate}
\item[(1A)] If $\alpha_0=0$ (or $\alpha_0=-1$), then $\mcA_0=\{0\}$ (or $\{-1\},$
    respectively). 
\item[(1B)] If $\alpha_0 \not\in \{0,\,-1\}$, and, without loss of generality,
    $\alpha_0>0$ then $\mcA_0=(0,\,+\infty) \cap \mcA.$
\end{enumerate} 

For simplicity, the case of $\alpha_0>0$ will be our main case for partial identification in the context of this design (case $\alpha_0<-1$ results in the identified set $\mcA_0=(-\infty,-1) \cap \mcA$, and the case $\alpha_0 \in (-1,0)$ gives the identified set $\mcA_0=(-1,0)$. 

In Illustrative Design 2 the vector of regressors has only 2 support points and the linear index
$\widetilde{X}'\widetilde{\alpha}$ forms two hyperplanes for those support points: $\alpha_1+\alpha_2=0$ and $\alpha_1+1=0.$ If the parameter vector takes a value on a given hyperplane, then the corresponding probability 
$\PP(Y=1\,|\,\widetilde{X})=\frac12.$ 
If the parameter vector is not on particular hyperplane, we can only identify which side of the 
hyperplane it is on, but not its value.
Using the 
convention for the sign function that 
$\sign(0)=0,$ we can express the identified set as
$$
\mcA_0=\left\{
(\alpha_1,\alpha_2)\,:\,
\sign(\alpha_1+\alpha_2)=
\sign(\alpha_{1,0}+\alpha_{2,0}),\,
\sign(\alpha_1+1)=\sign(\alpha_{1,0}+1)
\right\} \cap \mcA.
$$
Suppose the parameter space $\mcA$ is large enough to contain $(-1,1)$, then there are 4 following cases for the identified set
$\mcA_0:$
\begin{enumerate}
\item[(2A)] If $\alpha_0=(-1,1)$, then $\mcA_0=\{(-1,1)\}$ (\textit{the only point identification case}). 
\item[(2B)] If $\alpha_{01}=-1$ and $\alpha_{02}\neq 1$, then 
$$\mcA_0= \left\{(-1, \alpha_{2}): \sign(-1+\alpha_{2})=\sign(-1+\alpha_{2,0})\}\right\} \cap \mcA.$$ $\mcA_0$ is  fully contained in one-dimensional hyperplane $\alpha_{01}=-1$ in $\real^{2}$ (\textit{partial  identification case; identified set has empty interior in $\real^{2}$}).
\item[(2C)] If $\alpha_{01}\neq -1$ and $\alpha_{01}+\alpha_{02}=0$, then 
$$\mcA_0= \left\{(\alpha_{1}, -\alpha_{1}): \sign(\alpha_{1}+1)=\sign(\alpha_{0,1}+1)\}\right\} \cap \mcA.$$ $\mcA_0$ is  fully contained in one-dimensional hyperplane $\alpha_{01}+\alpha_{02}=0$ in $\real^{2}$ (\textit{partial  identification case; identified set has empty interior in $\real^{2}$}).
\item[(2D)] If $\alpha_{01}\neq -1$ and $\alpha_{01}+\alpha_{02}\neq 0$, then only the general expression provided above applies
(\textit{partial  identification case; identified set has a non-empty interior in $\real^{2}$}).
\end{enumerate}
As evident, Illustrative Design 2 exhibits greater complexity, wherein even in partially identified cases, certain conditional probabilities of choice may remain equal to $1/2$. Consequently, this leads to identified sets characterized by an empty interior in $\mathbb{R}^{K-1}$.

For a substantial portion of our findings, drawing upon the insights derived from Illustrative Design 1 will prove sufficient. Thus, this design will serve as our primary illustrative tool throughout the paper. However, a select few results will be more effectively explained by employing Illustrative Design 2.

One  a more general note, note that (\ref{threshold:crossing}) defines a convex polyhedron.
Its dimension $K-1-r_0$ depends on the rank $r_0$ of the system of equations $\widetilde{x}'\widetilde{\alpha}=0$ in ((\ref{threshold:crossing})) (here we use the normalization of the parameter $\widetilde{\alpha}$ to obtain $K-1$ as the largest dimension). When $K-1-r_0>0$, the boundary of this polyhedron when considered in $\real^{K-1-r_0}$ is excluded due to strict inequalities. The identified set $\mcA_0$ is obtained as the intersection of this polyhedron with $\mcA$.

Having demonstrated that in general our model is  partially
identified, we now turn to the characterization of estimators for the 
identified set $\mcA_0.$ Let $\widehat{\mcA}$ denote an estimator for the 
identified set obtained from sample $(y_i,\,\widetilde{x}_i)^n_{i=1}$. 

\begin{definition}\label{probability:limit}
A set estimator $\widehat{\mcA} \subset {\bf R}^{K-1}$  converges in probability to $\mcA^* \subset {\bf R}^{K-1} $  if $d_H(\widehat{\mcA},\mcA^*)\stackrel{p}{\rightarrow} 0,$ where
$d_H(\cdot,\cdot)$ is the Hausdorff distance.\footnote{Recall that for two sets $S_1, S_2 \subset \bf R^K $ the Hausdorff distance is 
$ d_H(S_1,S_2)\equiv \max\{\sup _{{x\in S_1}}\inf _{{y\in S_2}}d_E(x,y),\,\sup _{{y\in S_2}}\inf _{{x\in S_1}}d_E(x,y)\},$
where $d_E(x,y)$ denotes the Euclidean distance between vectors $x$ and $y$.}
\end{definition}
In this definition $\mcA^*$ is a probability limit of estimator $\widehat{\mcA}$ 
treated as a random set and coincides with the definition of convergence in probability from the random set theory (see \cite{molchanov2006book}, Definition 6.19). This definition allows us to introduce the concept of {\it sharpness}
of the set estimator $\widehat{\mcA}.$

\begin{definition}[Sharpness]
\label{def:sharpness}
A set estimator $\widehat{\mcA}$ is sharp for parameter $\alpha_0$ of the data generating process if its probability limit $\mcA^*$ exists and satisfies 
$d_H(\mcA^*,\,\mcA_0)=0.$
\end{definition}

Definition \ref{def:sharpness} characterizes sharpness by two essential properties of a set estimator. First, a sharp estimator needs to converge in probability to a deterministic  set.
Second, the limit set has to be within Hausdorff distance 0 from the identified set.
As stated in the definition, sharpness is a pointwise property and the same set 
estimator may not necessarily remain sharp for different values of  
parameter $\alpha_0$ of the DGP. 

The next step in characterizing behavior of set estimators $\widehat{\mcA}$ involves examining scenarios where a particular estimator lacks sharpness. Since sharpness is linked to both the convergence of an estimator in probability and the probability limit being the identified set, the absence of sharpness typically implies a lack of convergence in probability. In the absence of sharpness, we  maintain the assumption that for  parameter $\alpha_0$ of the DGP  the estimator 
$\widehat{\mcA}$ still converges weakly to a random set ${\bf A}(\alpha_0).$ Weak convergence of random sets as discussed in \citeasnoun{molchanov2006book} can be
characterized as the convergence of the sequence of  Choquet capacities $T^{\widehat{\mcA}}_{\alpha_0}(\mathcal{K})$ of
random sets $\widehat{\mcA}$ to the capacity of the limit random set $T^{{\bf A}(\alpha_0)}_{\alpha_0}(\mathcal{K})$
for the parameter $\alpha_0 \in \mcA$  of the DGP 
for all sets $\mathcal{K}$ in the Fell topology on $\mcA.$\footnote{Recall
that Choquet capacity of the random set ${\mathcal B}$ is defined as
$T^{\mathcal B}(\mathcal{K})=\PP ({\mathcal B} \cap \mathcal{K}).$}

To motivate our second criterion, recall that  the original Hodges estimator (e.g., see \cite{hodges}) \footnote{For important work on properties of the original Hodges estimator, see for example \citeasnoun{LeebPots2005},\citeasnoun{LeebPots2006},\citeasnoun{LeebPots2008}.} illustrated  the role of continuity of the estimator's limit  when comparing its properties to MLE. We aim to use a similar principle here, though now in the partially identified setting, to evaluate another aspect of the behavior of the set estimators in our settings. Namely,  we are concerned
with potential dependence of weak limit of set estimators   $\widehat{\mcA}$ on the underlying parameter $\alpha_0$ of the DGP as  the limit may change discontinuously analogously to the behavior of the identified set in our illustrative designs above.

Drawing from the analysis for the point identified case in \citeasnoun{ibragimov}, we consider weak convergence of set estimators $\widehat{\mcA}$ under locally varying parameters of the DGP  (\ref{discrete:choice}). Let $\alpha(n,t;\alpha_0) \in \mcA$ be a sequence of parameter values
indexed by constant $t \in \real$ and sample size $n$ converging to $\alpha_0$ as $n \rightarrow \infty.$ Parameter $\alpha(n,t;\alpha_0)$ for each fixed
$t \in \real$ and $n$ determines the probability distribution $\PP_{\alpha(n,t;\alpha_0)}$ for the 
random vector of 
observable variables $(Y,\widetilde{X}).$ We then construct the Choquet capacity of the set 
estimator
$\widehat{\mcA}$ induced by this probability distribution as $T^{\widehat{\mcA}}_{\alpha(n,t;\alpha_0)}(\mathcal{K})=\PP_{\alpha(n,t;\alpha_0)}\left(\widehat{\mcA} \cap \mathcal{K} \right).$

\begin{definition}[Local robustness]
\label{def:local:robustness}
For a sequence of parameters $\alpha(n,t;\alpha_0) \rightarrow \alpha_0$ for all $t >0,$ the set estimator $\widehat{\mcA}$ is locally robust at $\alpha_0$ with respect to sequence $\alpha(n,t;\alpha_0)$ if the sequence of capacities $T^{\widehat{\mcA}}_{\alpha(n,t;\alpha_0)}(\mathcal{K})$ converges to capacity $T^{{\bf A}_t}_{t}(\mathcal{K})$ of the random set ${\bf A}_t$ 
such that 
$$\sup\limits_{t>0}\lim\limits_{\alpha \rightarrow \alpha_0}\PP\left(
d_H({\bf A}_t,\,{\bf A}(\alpha))=0\right)=1,
$$
where ${\bf A}(\alpha)$ is the limit random set of
$\widehat{\mcA}$ for a fixed parameter value $\alpha$ of the DGP.
\end{definition}

Thus, robustness is the property of the continuity of the distribution of the  estimator's limit 
at a given point $\alpha_0$ with respect to some sequences of parameters of the DGP converging to that point.

The combined concepts of sharpness
and robustness allow us to analyze properties of set estimators.
In particular, if an estimator is
sharp at a given $\alpha_0 \in \mcA,$
then it is not necessarily robust with respect
to a certain range of 
sequences $\alpha(n,t;\alpha_0),$  as we show in the next section for specific instances of set estimators.
At the same time, an estimator which is robust at a given point
of the parameter space is not necessarily sharp because robusness is the property of continuity
of the limiting distribution. In case where an estimator only converges in distribution and not in 
probability, it cannot converge to a fixed (identified) set.


An important question in our analysis
will be the selection of the drifting
rates for the sequences $\alpha(n,t;\alpha_0)$. Since the the shape of the identified
set is determined  by conditional choice probabilities $p(\widetilde{x})=P(Y=1\,|\,\widetilde{X}=\widetilde{x})$
(through the threshold
crossing condition (\ref{threshold:crossing})),  the question of robustness of the estimators can be approached by analyzing the limits of estimators 
along the sequences 
$\alpha(n,t;\alpha_0)$ that induce the sequences of conditional probabilities $p^{(n)}(\widetilde{x})$ varying with the sample
size $n$ and approach $p(\widetilde{x})$ in the limit, with cases when some $p(\widetilde{x})$ are equal to $\frac12$ being particularly interesting as they result either in the case of point identification or that of partial identification but with the identified set having an empty interior in the parameter space.  

Due to the discrete nature of our regressors, the conditional
probability $P(Y=1\,|\,\widetilde{X}=\widetilde{x})$ can be estimated as the ratio of two sample means, which has $\sqrt{n}$-rate of convergence. This can be translated into the parameter sequences with the 
behaviour $\|\alpha(n,t;\alpha_0)-
\alpha_0\|=O(1/\sqrt{n})$. Such sequences  will
play the central role in
our subsequent robustness analysis.

\section{Classic estimators and their extensions}\label{sec:classic}
In this section we analyze two classic estimators --  maximum score and the closed-form estimator -- for the semiparametric discrete choice models originally 
developed for the point identified
models. We study their sharpness
and robutsness properties. 
We also propose and analyze two novel estimators that build on the bases of those two classic estimators and are designed to combine their respective strengths. 

In order to emphasize the key technical aspects, we will provide a detailed proof of the results in this section specifically for one of our illustrative designs. Results for generic discrete settings will also be presented, accompanied by an overview of how their proofs align with the cases in illustrative designs and the additional algebraic considerations they entail.



\subsection{Maximum Score Estimator}\label{MS:section}
\subsubsection{Review}
Maximum score estimator was among the first and most influential estimation methods proposed in \citeasnoun{manski1975} for semiparametric discrete choice models.
The key technical assumption 
behind the maximum score estimator
is the median condition (\ref{assume:MS}) (iv) leading to the 
point identification of the model coefficients in case where a regressor 
has continuous distribution and has a non-trivial impact on the index.  We show that in the discrete regressors setting (thus, with point identification generally failing), the maximum score
estimator is robust everywhere on the parameter space, but not uniformly sharp.

The  objective function for the maximum score estimator constructed from the sample $(y_i,\widetilde{x}_i)^n_{i=1}$ takes the form
\begin{equation}
\label{MS:objective}
MS_n({\alpha})=\avg y_i {\bf 1}( \widetilde{x}_i' \widetilde{\alpha} \geq 0)+(1-y_i) {\bf 1}( \widetilde{x}_i' \widetilde{\alpha}  < 0). 
\end{equation}
The estimator yields the maximum
to this objective function: 
$
\widehat{\cal A}_{ms}= \arg \max_{\alpha \in {\cal A}} MS_n(\alpha).$
An alternative way to define the objective function proposed in \citeasnoun{komarova2013} is:
\begin{equation*}  
MS_n(\alpha)= \sum_{\widetilde{x} \in \mathcal{X}}  (\hat p(\widetilde{x})-0.5)\cdot \sign(\widetilde{x}_i' \widetilde{\alpha})  \widehat{P}(\widetilde{X}=\widetilde{x})
\end{equation*}
where $\hat  p(\widetilde{x})$ and $ \widehat{P}(\widetilde{X}=\widetilde{x})$  denote, respectively, a conditional choice probability estimator (estimated as a group average) and the sample frequency of $\widetilde{X}$ at a given support point $\widetilde{x}$. 



In general discrete regressors settings, $\widehat{\cal A}_{ms}$ will be a non-singleton set in  the parameter space $\mcA$.
The question we are interested in what this set converges to as the sample size, denote by $n$, gets arbitrarily large.

\subsubsection{Analysis of the sharpness of the maximum score estimator}\label{MS:sharpness}


We start our analysis by considering the  infeasible version of the maximum score
estimator as the 
maximizer of 
$
MS_{n,INF}(\alpha) =\sum_{\widetilde{x} \in \mathcal{X}}  \left(p(\widetilde{x})-1/2 \right) \sign(\widetilde{x}' \widetilde{\alpha})  \widehat{P}(\widetilde{X}=\widetilde{x}), $
which takes conditional choice probabilities $p(\widetilde{x})$ as known. This infeasible maximum score
estimator is denoted as $\widehat{\cal A}_{ms, INF}$. 

By arguments completely analogous to  \citeasnoun{komarova2013}, we can show that $\widehat{\cal A}_{ms, INF}$ can be a strict superset of $\mcA_{0}$ -- namely, this may happen whenever there  are $\widetilde{x}$ in the discrete support for which  $p(\widetilde{x})=1/2$ as these cases are simply ignored by $MS_{n,INF}(\alpha)$, as evident from the representation above. As shown in the Appendix, for all values of parameter
$\widetilde{\alpha}_0$ of the DGP, the infeasible estimator $\widehat{\cal A}_{ms, INF}$   converges in probability to the
maximizer ${\cal A}_{ms}$ of the population objective function
\begin{equation}
\label{MS:population} \textstyle
MS(\alpha) = \sum_{\widetilde{x} \in \mathcal{X}}  \left(p(\widetilde{x})-1/2 \right) \sign(\widetilde{x}' \widetilde{\alpha}) {P}(\widetilde{X}=\widetilde{x}).
\end{equation}

To give more details, $\mcA_{ms}$ is defined as the set of $\widetilde{\alpha} \in \mcA$ that satisfies the following: 
\begin{equation} \label{threshold:crossing2}
p(\widetilde{x}) < 0.5 \;  \iff \widetilde{x}'\widetilde{\alpha} <0, \qquad 
p(\widetilde{x}) > 0.5 \;  \iff \widetilde{x}'\widetilde{\alpha} > 0, 
\end{equation} 
and, thus, ignores the cases $p(\widetilde{x})=1/2$ at the decision making boundary. $\mcA_{ms}$ always has a non-empty interior in $\real^{K-1}$ (recall that $\mcA_0$ can have smaller dimension $K-1-r_0$ which depends on equality constraints). Description (\ref{threshold:crossing2}) by itself presents a convex polyhedron with its boundary excluded. This polyhedron is then intersected with $\mcA$. This description of $\mcA_{ms}$ is sufficient to conclude that ${\mcA}_{ms}$ will coincide with ${\mcA}_{0}$ when $p(\widetilde{x}) \neq 1/2$ for all $\widetilde{x} \in \mathcal{X}$, and will be a superset of ${\mcA}_{0}$ otherwise. 





In Illustrative Design 1, as we discussed in Section \ref{sec:identification}, we distinguish between the case of point identification (whenever
$\alpha_0 \in \{-1,0\}$) and the case of partial identification of the parameter of interest.
While we defer formal derivations to the appendix,
we can characterize the maximizer of the population maximum score objective as follows.
When $\alpha_0=0$, then $\mcA_{ms}=[-1,+\infty) \cap \mcA$. If $\alpha_0=-1$, then $\mcA_{ms}=[-\infty,1)$.
If $\alpha_0 \not\in \{0,\,-1\}$ and, without loss of generality,
    $\alpha_0>0$ then $\mcA_{ms}=\overline{\mcA}_{0} =[0,\,+\infty) \cap \mcA.$
Note that cases $\alpha_0
 \in \{0,\,-1\}$ in Illustrative Design 1, of course, correspond to situations when there is  $x$ in the support such that $p({x})=\frac12$. Such $x$ drives point identification of $\alpha_0$ in the the formal identification analysis but at the same time is fully disregarded by the population maximum score objective function $MS(\alpha)$ (as well as by $MS_{n,INF}(\alpha)$).  Thus, in these cases the infeasible maximum score estimator is not sharp. In cases $\alpha_0
 \not\in \{0,\,-1\}$, the maximizer
 of the  infeasible maximum score  objective function is
 the closure of the identified set with probability approaching 1.
Consequently, the unfeasible maximum score
estimator can be sharp for those values
of parameters of the DGP.

Now, let's shift our focus to the original (feasible) maximum score estimator, which maximizes the (feasible) sample objective function $MS_{n}(\alpha)$. In this context, situations where $p(\widetilde{x})=1/2$ once again present challenges, albeit in a distinct manner compared to the issues encountered in $MS(\alpha)$. Here, these cases are not disregarded in the objective function but lack consistent consideration. This is due to the fact that, for arbitrarily large samples, the estimator $\hat p(\widetilde{x})$ fluctuates on different sides of $1/2$, with probabilities approaching $1/2$. As turns out, this will translate in the fluctuating behavior of the maximum score estimator $\widehat{\mcA}_{ms}$ itself. We first describe this fluctuating behavior occurs in the context of Illustrative Design 1.
While the technical details of this derivation
are deferred to the appendix, when can can show
that while $\alpha_0 \not 
\in \{0,-1\},$ then the probability limit
of $\widehat{\mcA}_{ms}$ is the set $[0,\,+\infty),$
which is the closure of the identified set.
However, whenever $\alpha_0=0$ (or $\alpha_0=-1$) , then 
%
%
whenever 
$\alpha_0 \in \{0,\,-1\},$ then
$\widehat{\mcA}_{ms}$ neither converges to $\mcA_{ms}$ nor to
$\mcA_0.$ Using the terms of the 
random set theory in \cite{molchanov2006book}, we can
characterize the limit as a random set
\begin{equation} 
\label{MSlimitID1} \widehat{\mcA}_{ms} \stackrel{d}{\to} {\bf A} \equiv B \cdot [0,+\infty) \cap \mathcal{A} + (1-B) \cdot [-1,0) \cap \mathcal{A},
\end{equation}
where $B$ is a Bernoulli random variable with parameter $\frac12$.


This means that the maximum score estimator is {\it not sharp} at points $\alpha_0 \in \{0,\,-1\}
$ and is sharp elsewhere in the 
parameter space.

To give a deeper statistical interpretation of the the limit of the maximum score estimator, we rely on our discussion from Appendix \ref{appendix:uniform}, analyzing  uniform convergence of $MS_n(\alpha)$ to $MS(\alpha)$. One useful observation from there is that 
for the normalized empirical process
$\sqrt{n}\left(
MS_n(\alpha)-MS(\alpha)
\right)$
we can establish the following
weak convergence in $\ell_{\infty}(A)$
(the space of bounded functions
in $\infty$-norm on $A$) for all
$A \subset \mcA:$
\begin{equation}\label{stoch:convergence}
\sqrt{n}\left(
MS_n(\alpha)-MS(\alpha)
\right)\rightsquigarrow
\mbox{sign}(\alpha)\,Z^0+
\mbox{sign}(1+\alpha)\,Z^1,
\end{equation}
where $Z^0 $
and $Z^1 $ and are independent mean
zero Gaussian random variables. 

The convergence result (\ref{stoch:convergence})
sheds the light on the mechanics
of the formal characterization of the 
maximum score estimator for Illustrative Design 1.
The right-hand side of (\ref{stoch:convergence}) is often
referred to as ``stochastic residual."
In fact, whenever $\alpha_0
\not\in \{0,\,-1\},$ then we can simply rely on the finite variance
of the Gaussian variables and boundedness of the sign function 
to conclude that 
$\sup\limits_{\alpha \in \mcA}\|MS_n(\alpha)-MS(\alpha)\|=o_p(1)$
leading to the conclusion that the 
limit of the maximizer of the sample
maximum score objective function 
is the same as the maximizer of the 
population objective function.

However, whenever $\alpha_0=0$
(or $\alpha_0=-1$) the right-hand side
no longer plays the role of the 
``residual." The population objective
function $MS(\alpha)$ in that case is equal to 
zero whenever $\alpha<-1$
and is equal to the strictly positive constant for $\alpha \geq -1,$ thus,
attaining its maximum on the set
$[-1,+\infty).$
Due to the presence of Gaussian variables $Z^0$ and $Z^1$ symmetrically distributed around zero, the term $\mbox{sign}(\alpha)\,Z^0+
\mbox{sign}(1+\alpha)\,Z^1$
adds a positive weight on either
set $\alpha<0$ or $\alpha \geq 0$
with equal probabilities each.
This means that, even though, the ``stochastic residual" term is
infinitesimal, it randomly selects one of two subsets in the partitioning of the set $\mcA_{ms}$ (which, as a reminder, maximizes the population $MS(\alpha)$), designating that subset as the maximizer of the sample objective function $MS_n(\alpha)$. It is important to note that the partitioning of $\mcA_{ms}$ occurs at the parameter value $\alpha_0$ in the DGP, carrying significant implications for our search for a superior estimator compared to the maximum score at a later stage.

Our  findings here extend to general discrete-only scenarios. For instance, in Illustrative Design 2 we can partition the parameter space into the following sets:
$\mcC_1=\{\alpha_1+\alpha_2<0,\,\alpha_1+1<0\},$
$\mcC_2=\{\alpha_1+\alpha_2<0,\,\alpha_1+1>0\},$
$\mcC_3=\{\alpha_1+\alpha_2>0,\,\alpha_1+1>0\},$
$\mcC_4=\{\alpha_1+\alpha_2>0,\,\alpha_1+1<0\},$
as well as hyperplanes $\alpha_1+\alpha_2=0,$
$\alpha_1+1=0$ with and without the point $(-1,1).$

In Case 2A the maximizer of the population objective is the entire parameter
space $\mcA$ and the maximum score estimator
has the asymptotic distribution outputting sets $\overline{\mcC}_1$ -- $\overline{\mcC}_4$ with probabilities
$\frac14$ each.

In case 2B, the maximizer of the population objective is a half-space 
$\{(\alpha_1,\alpha_2)\,:\,\sign(\alpha_1+\alpha_2)=
\sign(\alpha_{0,1}+\alpha_{0,2})\}$
and the maximum score estimator has 
asymptotic distribution outputting
sets $\overline{\mcC}_3$ and $\overline{\mcC}_4$ (if $\alpha_{0,1}+\alpha_{0,2}>0$) or $\overline{\mcC}_1$
and $\overline{\mcC}_3$ (if $\alpha_{0,1}+\alpha_{0,2}<0$)
with probabilities $\frac12$ each.

In case 2C, the maximizer of the population
objective function is a half-space 
$\{(\alpha_1,\alpha_2)\,:\,\sign(\alpha_1+1)=
\sign(\alpha_{0,1}+1)\}$ and the maximum score estimator has 
asymptotic distribution outputting
sets $\overline{\mcC}_2$ and $\overline{\mcC}_3$ 
(if $\alpha_{0,1}+1>0$) or $\overline{\mcC}_1$
and $\overline{\mcC}_4$ (if $\alpha_{0,1}+1<0$)
with probabilities $\frac12$ each.

In case 2D the maximum score estimator converges
in probability 
to one of the 
sets $\overline{\mcC}_1$ -- $\overline{\mcC}_4,$ coinciding with the closure
of the identified set.

Theorem \ref{th:MSgeneral} formulates a general result. 

\begin{theorem}\label{th:MSgeneral}
Suppose Assumption \ref{assume:MS} holds. The maximum score estimator is sharp for any parameter value  $\widetilde{\alpha}_0$ in the DGP that results in $p(\widetilde{x})\neq 1/2$, for all $\widetilde{x} \in \mathcal{X}$. For other parameter values in the DGP, the maximum score estimator does not have have a probability limit and converges weakly to a random set. 
\end{theorem}

Or next theorem gives a more elaborate result regarding the asymptotic behaviour of the maximum score estimator when $p(\widetilde{x})= 1/2$ for some  $\widetilde{x} \in \mathcal{X}$, and gives a form of its distribution limit. 

\begin{theorem}
\label{th:MSgeneral2} 
Suppose that Assumption \ref{assume:MS} holds. 
If for a parameter value $\alpha_0$ in the DGP there are choice probabilities $p(\widetilde{x})=1/2$ for some $\widetilde{x}$ in the support, then 
\[ \widehat{\mcA}_{ms}
\stackrel{d}{\rightarrow} B_1 \mcC_1 + \ldots +B_L \mcC_L\]
for some  non-empty deterministic sets $\mcC_1$, $\ldots$, $\mcC_L$, $L\geq 2$, that partition the maximizer of the population maximum score objective function $\mcA_{ms}$ (as proven earlier,  $\mcA_{ms}$ is generally a superset of the identified set $\mcA_0$). That is, 
\begin{itemize}
\item $\mcC_1$, $\ldots$, $\mcC_L$ are pairwise disjoint; 
\item $\cup_{\ell=1}^L \mcC_{\ell} = \mcA_{ms}$. 
\end{itemize}

In addition, 
\begin{itemize}
\item $B_1$, \ldots, $B_L$ are dummy variables such that 
$B_{\ell} B_m =0$ for  $\ell \neq m$ (mutually exclusive), and $B_1+\ldots+B_L=1$ (collectively exhaustive),  
and $p(B_{\ell}=1)\in (0,1)$ for each $\ell$. 

\item The boundary of each $\mcC_{\ell}$ includes the identified set $\mcA_0$ (hence, each $\overline{\mcC}_{\ell}$ includes $\mcA_0$). 

\item The intersection of any $[L/2]+1$\footnote{ Here $[a]$ stands for the integer part of $a$.}  closed sets $\overline{\mcC}_{\ell}$     coincides with the closure $\overline{\mcA}_{o}$ of the identified set $\mcA_0$. 

\end{itemize} 
\end{theorem}

Our final discussion here relates to the nature of the sets $\mcC_{\ell}$.  Without a loss of generality, let the first $M$, $M\geq 1$, points in $\cal{X}$ be those at the decision making boundary  $p(\widetilde{x})= 1/2$ while the rest are not. Let us denote the collection of these first $M$ support points as $\mathcal{X}_{db}$. As explained earlier, the maximizer $\mathcal{A}_{ms}$ of the population maximum score objective function is purely defined by $\widetilde{x} \notin \mathcal{X}_{db}$ through inequalities 
(\ref{threshold:crossing2})  for all $\widetilde{x} \notin \mathcal{X}_{db}$.

Consider any combination of inequalities $\widetilde{x}'\widetilde{\alpha} \geq 0$ or  $\widetilde{x}'\widetilde{\alpha}<0$ for $\widetilde{x} \in \mathcal{X}_{db}$ -- a combination  has to include an inequality for each $\widetilde{x} \in \mathcal{X}_{db}$. 
Each this combination results in a maximum score estimate $\mcC_{\ell}$ which is a subset of $\mcA_{ms}$ (since (\ref{threshold:crossing2})  hold asymptotically for all $\widetilde{x} \notin \mathcal{X}_{db}$ and, thus, are taken as given). Different combinations of signs of $\widetilde{x}'\widetilde{\alpha}$ for $\widetilde{x} \in \mathcal{X}_{db}$ result either in disjoint or identical maximum score estimates $\mcC_{\ell}$. We denote the collection of all these unique maximum score estimates as $\mcC_{1}$, \ldots, $\mcC_{L}$. For more details see the proof of Theorem \ref{th:MSgeneral2}.


\subsubsection{Robustness of the 
maximum score estimator}
In the previous subsection, we established that the maximum estimator
is not sharp whenever there exist points $\widetilde{x}$ in the support of covariate $\widetilde{X}$ where $p(\widetilde{x})= \frac12.$ If such points do not exist, it is sharp and, moreover, it converges
in probability to the identified set $\mcA_0.$ However, at those values the maximum score
estimator only converges weakly to a random set. This means, that the weak limit of the 
maximum score estimator in our partially identified is {\it discontinuously} changes with respect
to the parameter of the underlying data generating process. 

In this subsection we study {\it robustness} of the maximum score estimator (\ref{MS:objective}) for our
simple discrete design at $\alpha_0 \in \{0,-1\}$ as the property of continuity
of the weak limit of the estimator with respect to specific class of sequences of
parameters of the data generating process. As we discussed in Section \ref{sec:identification}, parameter sequences of particular interest to us take
the form $\alpha(n,t;\alpha_0)=\alpha_0+t/\sqrt{n}.$

The choice of local data generating processes indexed by these sequences reflects
the foundational property of the maximum score estimator allowing us to interpret it
as an ensemble of
weak learners (e.g., see section
10.1 in \cite{shalev:14}). To see this, we take
our Illustrative Design 1.
Objective 
function (\ref{MS:objective}) can be viwed as
the aggregator of ``votes" where each observation casts a vote for 
one of the sets $(-\infty,-1),$
$[-1,0)$ and $[0,+\infty).$
To see this, consider a single element of the 
maximum score objective function
$
y_i\,I[\alpha+x_i \geq 0]+(1-y_i)\,I[
\alpha+x_i<0
].
$
For each possible combination
$(y_i,x_i)$ the element of the 
objective function selects the 
intervals $[-1,+\infty),$ $[0,+\infty),$  $(-\infty,-1)$
and $(-\infty,0).$

This element is
a classifier\footnote{This classifier is referred to as a ``weak learner" in the terminology of the Machine learning literature. The term ``weak learner" is used because the probability that it classifies the object of interest correctly is 
not diminishing
in larger samples.} which can select one of the sets $(-\infty,-1),$
$[-1,0)$ and $[0,+\infty)$
or their pairwise unions. For the
pairwise union, the classifier selects each set 
in the union.

The classification outcome for observation $i$ is 
$(-\infty,-1)$ with probability
$(1-p(1))q,$ $(-\infty,-1) \cup [-1,0)$
with probability $(1-p(0))(1-q),$ 
$[0,+\infty)$ with probability $p(0)(1-q)$
and $[-1,0)\cup (0,+\infty)$ with probability $p(1)q$ (where
$p(x)=P(Y=1\,|\,X=x)$ and $q=P(X=1)$).  We consider the concept of the weak learner in context of partial identification, novel
to the machine learning literature.

Let $v_i$ be a three-dimensional vector with elements $1/0$ depending
on whether the corresponding set $(-\infty,-1),$
$[-1,0)$ or $[0,+\infty)$ (for each of the three dimensions) 
was selected by observation $i.$ Then $\bar{v}=\frac1n \sum^n_{i=1}v_i$
produces a vector of collective ``votes" of $n$ classifiers 
for each observation $i$ and
the maximum score estimator can be
written as 
$$
\widehat{\mcA}_{ms}=
\mbox{(}-\infty,-1\mbox{)}\cdot {\bf 1}\{\argmax \bar{v}=1\}+
\mbox{[}-1,0 \mbox{)}\cdot {\bf 1}\{\argmax \bar{v}=2\}+
\mbox{[}0,+\infty\mbox{)}\cdot {\bf 1}\{\argmax \bar{v}=3\}.
$$

The estimator is a function of the sample mean $\bar{v}$ converging
at the standard parametric rate
$\sqrt{n}$ to a normal random variable. For the analysis of continuity of distribution limit of such parameters the literature
(e.g., \cite{ibragimov}, \cite{Lecam1953}) suggest considering
sequences of the population distributions of the underlying random variable with its expectation 
drifting at the same parametric rate.
Since the expectation of the random vector $v_i$ is linear in the probability $p(0),$ sequences 
of probabilities $p^{(n)}(0)$ drifting at $1/\sqrt{n}$ rate will result in an equivalent drifting of the expectation
of $v_i$ at that rate.

We now establish the general result for the maximizer of (\ref{MS:objective}).

\begin{theorem}
\label{th:MSrobustness} 
Suppose that Assumption \ref{assume:MS} holds. 
Consider sequence of the data generating processes
$\alpha(n,t;\alpha_0)=\alpha_0+t /\sqrt{n}$ such that for a parameter value $\alpha_0$ in the DGP there are choice probabilities $p(\widetilde{x})=1/2$ for some $\widetilde{x}$ in the support, and $\alpha(n,t;\alpha_0) \in \mcC_k$ for $\mcC_k$ from the set system specified in Theorem \ref{th:MSgeneral2}. Then 
$$\widehat{\mcA}_{ms}
\stackrel{d}{\rightarrow} B_1(t)\, \mcC_1 + \ldots +B_L(t)\, \mcC_L,$$
where $\mcC_1,\ldots,\mcC_L$ are the sets from the 
set system specified in Theorem \ref{th:MSgeneral2}
and $B_1(t),\ldots,B_L(t)$ are dummy variables
such that $B_j(t)B_j(t)=0$ for $i \neq j,$ 
$\sum^L_{\ell=1}B_\ell(t)=1$ and
$$
\lim\limits_{\|t\| \rightarrow +\infty}\PP(B_{k}(t)=1) =1\;\;\mbox{and}\;\;\lim\limits_{\|t\| \rightarrow 0}\PP(B_{j}(t)=1) =P(B_j),\;\;j=1,\ldots,L,
$$
where $B_j$ are dummy variables specified in
Theorem \ref{th:MSgeneral2}.
\end{theorem}



Theorem \ref{th:MSrobustness} has the following two major implications for Illustrative Design 1. First, when the 
sequence of parameters of the data generating process converge to parameter values
where the maximum score estimator is sharp for the Illustrative Design 1, drifting does not impact its limit and
it converges to the identified set $\mcA_0.$ Second, when the sequence of parameters
of the data generating process converges to the value where the maximum score estimator
is not sharp, the maximum score estimator converges to a random set whose distribution
depends on the constant indexing a particular parameter sequence. In particular, it has the property that 
$\sup\limits_{t>0}\lim\limits_{n \rightarrow \infty}\PP\left( d_H(\widehat{\mcA}_{ms},\,[0,+\infty)\cap \mcA)=0\right) =1,$ i.e., it converges to limit of the estimator at parameter values of the data generating process not equal to zero (but, possibly, arbitrarily close to zero). This means that the maximum score estimator is {\it locally robust} respect to 
parameter sequences $\alpha(t,n;\alpha_0)=\alpha_0+t/\sqrt{n}$ at points where it is not sharp, according to Definition \ref{def:local:robustness}.

Parameter drifting to the parameter values where the maximum score estimator is not sharp ranges between two regimes. The first regime corresponds to the distribution limit of the maximum score estimator 
which is a random set taking sets $[-1,0)$ and $[0,+\infty)$ with equal probabilities (corresponds to $t=0$). In the
second regime is where the maximum score estimator puts a point mass of 1 on the set $[0,+\infty),$  which is the maximizer of the population maximum score objective function and the closure of the identified set whenever parameter of the data generating process is fixed at values outside of $-1$ and $0$ (corresponding to $t=+\infty$).

To summarize, for parameter of the data generating process drfiting towards $\alpha_0=0$ as $t$ varies from $0$ 
to $+\infty,$ the distribution of the limit random set varies from equal randomization between $[-1,0)$
and $[0,+\infty)$ to selecting
a fixed set $[0,+\infty).$ Thus, this
choice of the drifting sequence bridges the two cases. As a result, even though the maximum score estimator is not sharp at that point, it is locally robust.

For Illustrative Design 2, Theorem \ref{th:MSrobustness} leads to a similar
behavior of the limit as for Illustrative Design 
 1, albeit, with more complex structure of limit. When $\alpha_0$ corresponds to the value where $p(\bar{x})=\frac12$ for one or more points in the support of $\widetilde{X},$ the distribution limit of the maximizer of (\ref{MS:objective}) takes value on two or more sets
$\mcC_1, \mcC_2,\mcC_3,\mcC_4$  formed by intersections of half-spaces with boundary hyperplanes $\alpha_1+\alpha_2=0$ and $\alpha_1+1=0.$

%
%
%

Whenever $\alpha(t,n;\alpha_0)=
\alpha_0+t/\sqrt{n} \in \mcC_k$ and $\|t\|$ increases, the probability that random set has a realization $\mcC_k$ increases. Consequently, $\sup\limits_{t,\,\alpha(t,n;\alpha_0) \in \mcC_k}\lim\limits_{n \rightarrow \infty}\PP\left( d_H(\widehat{\mcA}_{ms},\,\mcC_k\cap \mcA)=0\right) =1.$
Thus, the maximum score estimator is robust.

\subsection{Two-Step Closed Form Estimator}\label{ichimura}
We next consider the second classic estimator in our semiparametric binary choice model with discrete regressors.
 This estimator relates directly to procedures introduced in \citeasnoun{ichimura1994} and \citeasnoun{ahnetal2018}.
They are based on the following implication of Assumption \ref{assume:MS} 
(including strict monotonicity of $F_{\epsilon|X}$ at 0): 
\begin{equation*} p(\widetilde{x})=1/2 \iff \widetilde{x}'\widetilde{\alpha}_0=0. \end{equation*}
Based on the above implication, \citeasnoun{ichimura1994} proposed the estimator to be minimizer of the  $\avg \hat w_i(\widetilde{x}_i'\,\widetilde{\alpha})^2,$ subject to 
normalization on $\widetilde{\alpha}$ imposed by Assumption \ref{assume:MS} (ii). In this 
setup $\hat w_i$ is a smoothing function which uses a nonparametrically estimated conditional
probability $p(\widetilde{x})$ for each observation and puts higher weight on the observations for which this predicted probability is close to $1/2$. 

This estimator is easy to implement, as it is of closed form in each stage.
Under stated conditions ensuring point identification, this estimator was shown to have desirable asymptotic properties,
specifically being asymptotically equivalent to maximum score or smoothed maximum score (\citeasnoun{horowitz-sms}).
Its disadvantage compared to maximum score is that because it involves nonparametric procedures, it requires one to make the choice of kernel function and bandwidth sequence.

In our discrete regressors setup, we can e.g. take $w_i={\bf 1}\left\{|\widehat{p}(\widetilde{x}_i)-\frac12|<h_n\right\}$ where sequence $h_n \rightarrow 0$ is the tuning parameter of the estimator. 

With $w_i$ chosen in this way, we denote the minimizer of $\avg  w_i(\widetilde{x}_i'\,\widetilde{\alpha})^2,$ subject to the required normalization as $\widehat{\mcA}_{CF}.$ If 
$|\widehat{p}(\widetilde{x}_i)-\frac12|<h_n$ is not satisfied for any observations we set
$\widehat{\mcA}_{CF}\equiv \mcA$ to ensure the estimation procedure does not halt.

\subsubsection{Analysis of sharpness of the closed form estimator} 
Note that the closed form estimator can be considered as maximizing the sample objective function 
\begin{equation}\label{Ich:objective} 
I_n({\alpha})=-\avg  w_i\,(\widetilde{x}_i' \widetilde{\alpha})^2 
\end{equation}
with $w_i$ chosen as above. The respective population objective function is then 
\begin{equation}\label{Ich:population}
I(\alpha)=-{\mathbb E}\left[{\bf 1}\{p(\widetilde{X})=1/2\}(\widetilde{X}'\widetilde{\alpha})^2 \right]
\end{equation}
whose maximizer in $\real^{K-1}$ we denote as   $\mcA_{CF}$.  We take $\mcA_{CF}$ to 
the entire parameter space $\mcA$ where ${\mathbb P}(p(\widetilde{X})=1/2)=0.$ 

Let us turn to Illustrative designs to exemplify closed form estimation. 

In Illustrative Design 1, 
$$
\mcA_{CF}=\left\{
\begin{array}{l}
\alpha_0,\;\;\mbox{if}\;\alpha_0\in \{0,1\}\\
\mcA,\,\;\;\mbox{if}\;\alpha_0 \not\in \{0,\,-1\}.
\end{array}
\right.
$$

This description indicates that the population objective function attains its optimum, equivalent to the identified set, solely when the parameter $\alpha_0$ in the DGP takes on values from the set ${0, -1}$ and, thus, when we have the case of point identification. When $\alpha_0$ assumes any other value, thus, we are in the situation of partial identification, then the population objective function produces the default of the entire parameter space, which is, of course, a superset of the identified set.

In Illustrative Design 2, 
$$
\mcA_{CF}=\left\{
\begin{array}{l}
\alpha_0,\;\;\mbox{if}\;\alpha_0=(-1,1) \; \; \mbox{(Case 2A)}\\
(-1,\real) \cap \mathcal{A},\;\;\mbox{if}\;\alpha_{01}=-1, \alpha_{02}\neq 1 \; \; \mbox{(Case 2B)}\\
\left\{(\alpha_{1}, -\alpha_{1}): \alpha_{1} \in \real\}\right\} \cap \mcA ,\;\;\mbox{if}\;\alpha_{01}\neq -1, \alpha_{01}+\alpha_{02}=0 \; \; \mbox{(Case 2C)} \\
\mcA,\,\;\;\mbox{if}\; \alpha_{01}\neq -1, \alpha_{01}+\alpha_{02}\neq 0. \; \; \mbox{(Case 2D)}
\end{array}
\right.
$$
In line with our conclusions for Illustrative Design 1, $\mcA_{CF}$ coincides with the identified set $\mcA_0$ only in the point identification case (Case 2A). In situations of partial identification (Cases 2B-2D), $\mcA_{CF}$ is a strict superset of $\mcA_0$. 

These conclusions are 
intuitive. Maximizer of (\ref{Ich:objective}) disregards values $\widetilde{X}$ that do not lie on the decision-making boundary characterized by $p(\widetilde{X})=1/2$. In scenarios involving point identification (like 2A in Illustrative Design 2 or 1A in Illustrative Design 1), such point identification is obtained because there are enough of $\widetilde{X}$ instances at the decision boundary. The closed-form method then efficiently utilizes all such $\widetilde{X}$ instances. However, in cases where there are insufficient $\widetilde{X}$ values at the decision boundary to achieve point identification (as observed in Cases 2B and 2C in Illustrative Design 2), or none at all (as seen in Case 2D in Illustrative Design 2 or Case 1B in Illustrative Design 1), the closed-form method relies solely on limited information (or none at all in Cases 2D and 1B). It overlooks information from all other $\widetilde{X}$ values not on the decision boundary, even though they may contribute to the structure and shape of the identified set.

A result for a generic discrete regressors setting is given in Theorem \ref{th:closedformPOPgeneral}.

\begin{theorem} 
\label{th:closedformPOPgeneral}
Suppose Assumption \ref{assume:MS} holds. 

If the parameter value $\alpha_0$ in the DGP is such that the model is point identified, then $\mcA_{CF}=\mcA_0$. 

If the parameter value $\alpha_0$ in the DGP is such that the model is not point identified, then $\mcA_{CF}$ is a superset of $\mcA_0$. 
\end{theorem}


Our next focus is on the sampling behavior of the closed-form estimator formally established in Theorem 
\begin{theorem} 
\label{th:closedformtoy1}
Suppose Assumption \ref{assume:MS} holds 
and the tuning parameter is set such that $h_n \rightarrow 0$ and $h_n\,\sqrt{n}
\rightarrow \infty$. 
Then $d_H(\widehat{\mcA}_{CF},\,
{\mcA}_{CF} ) \stackrel{p}
{\longrightarrow} 0$. 
\end{theorem}

Results of Theorems \ref{th:closedformPOPgeneral} and \ref{th:closedformtoy1} imply that the closed-form estimator is not sharp when $\alpha_0$ in the DGP does not result in point identification. 

Some further insights can be obtained from considering illustrative designs. The maximum score estimator in Illustrative Design 1 was sharp only when $\alpha_0 \notin \{0,-1\}$ (1B) in the DGP and was not sharp otherwise (1A). Conversely, the closed-form estimator in the same design exhibits contrasting behavior -- it is sharp solely when $\alpha_0 \in {0,-1}$ (1A) within the DGP, and loses sharpness otherwise (1B). This disparity is unsurprising, given that these two estimation methods rely on distinct principles -- the closed-form estimator exclusively utilizes regressor values at the decision-making boundary, while the maximum score estimator is sharp in the absence of such regressor values.

 In Illustrative Design 2, maximum score was sharp only in Case 2D whereas the closed-form estimator is sharp only in Case 2A, thus leaving 2B and 2C as cases  where neither of these two estimators is sharp.





\subsubsection{Robustness of the closed-form estimator}
To analyze robustness of the closed-form estimator we use the sequences
of parameters of the data generating process with the property
$\|\alpha(t,n; \alpha_0)-\alpha_0\|=O(1/\sqrt{n}).$ We do so by the same reasoning as in case of the maximum score estimator, given that the core
object of the estimator (\ref{Ich:objective}), the conditional probability 
$P(Y=1\,|\,X=x),$ is represented by a sample mean.
The following theorem characterizes the distribution limit of $\widehat{\mcA}_{CF}$ under those drifting parameter sequences.
\begin{theorem} 
\label{th:drfitCF}
Suppose that Assumption \ref{assume:MS} holds. 
Consider sequence of the data generating processes
$\alpha(n,t;\alpha_0)=\alpha_0+t /\sqrt{n}$ such that for a parameter value $\alpha_0$ in the DGP there are choice probabilities $p(\widetilde{x})=1/2$ for some $\widetilde{x}$ in the support, and $\alpha(n,t;\alpha_0) \in \mcC_k$ for $\mcC_k$ from the set system specified in Theorem \ref{th:MSgeneral2}. Then for
maximizer of (\ref{Ich:objective}) with 
$h_n \rightarrow 0$ and $h_n\sqrt{n} \rightarrow \infty,$
then 
$
\lim_{n \rightarrow\infty}
\PP\left(d_H(\widehat{\mcA}_{CF},\mcA_{CF})=0
\right)
=1$ 
for any $t \neq 0.$ 
\end{theorem}

This theorem demonstrates that the convergence of the maximizer of 
(\ref{Ich:objective}) is not affected by the choice of the drifting sequence as long as it is not entirely contained in the subsets of the parameter space for which $\PP(Y=1\,|\,X=x) \equiv \frac12$ for some $x$ in the support of $X$.

To summarize, the maximizer of (\ref{Ich:objective}), $\widehat{\mcA}_{CF}$, is neither {\it sharp}, nor is  it{\it robust.}

 \subsection{New estimators based on combination ideas} \label{sec:new:estimators} 


Our previous discussion highlighted the sharpness, or lack thereof, exhibited by both maximum score and closed-form estimators. It became evident that merging the underlying concepts of these approaches is necessary to formulate an estimation procedure surpassing either individual method. Essentially, this entails leveraging all regressor values—both those at the decision boundary and those away from it -- in a judicious manner. The primary challenge lies in devising such an approach.

\subsubsection{Definition and analysis of the sharpness of combination estimators}

Our first idea for a new estimator is, in some sense, an obvious one.  It is to directly combine the maximum score and the closed-form estimators through a feasible data-driven ``switching device''. This is how it goes. First, construct estimators for the choice probabilities
$\hat p(\cdot)$ from the sample $\{(y_i,\widetilde{x}_i)\}^n_{i=1}$.  Next, define the minimum deviation of estimated choice probabilities from $1/2$:  $ \widehat{P}=\min_{i=1,\ldots,n}|\hat{p}(\widetilde{x}_i)-\frac{1}{2}|$.

Our proposed ``switching"  estimator is then explicitly defined as
\begin{equation}\label{def:switching}
 \widehat{\mcA}_H = \widehat{\mcA}_{CF} \cdot 1(\widehat{P} < \nu_n)+\widehat{\mcA}_{ms}\cdot 1(\widehat{P} \geq \nu_n)
 \end{equation}
 with $\nu_n \rightarrow 0$ being the tuning
 parameter sequence. Even tough we use the same notation for a tuning parameter here as we used in the definition of the closed-form estimator, these two sequences are not necessarily the same. 
 
 We refer to it as a hybrid estimator. This estimator builds on the idea that $\widehat{P}$ being far enough from zero is  evidence against having choice probabilities being $1/2$, and $\widehat{P}$ being close enough from zero is supportive of the opposite case.  Naturally, in the former case we can rely on the maximum score estimator and, in the latter case, on the closed form estimator.  $\widehat{\cal A}_H$ may both output a set or a singleton in finite samples, depending on the data generating process.

In Illustrative Design 1 with the right choice of the rate of $\nu_n \to 0$ (e.g., $\sqrt{n}\nu_n \to \infty$) this estimator is going to be sharp. When $\alpha_0 \in \{0,-1\}$ (1A), we expect  $\widehat{P}$ in large enough sample to be less than $\nu_n$ and, thus, picking the closed-form estimator which is sharp in Case 1A. When $\alpha_0 \notin \{0,-1\}$ (1B), we expect  $\widehat{P}$ in large enough sample to be greater than $\nu_n$ and, thus, picking the maximum score estimator which is sharp in Case 1B. 

The situation in Illustrative Design 2, however, is very different. Even with the right choice of the rate of $\nu_n \to 0$, this estimator is not sharp. Namely, it will be sharp in Case 2A as the closed-form estimator is then sharp and it will be selected by the ``switching device'' (there will be two sample conditional choice probabilities close to 1/2), and it will also be sharp in Case 2D as the maximum score estimator is then sharp and it will be selected by the ``switching device'' (no sample conditional choice probabilities close to 1/2). In Cases 2B and 2C a
 the ``switching device'' will, once again, select the closed-form estimator (as there will be one sampling conditional choice  probability close to 1/2) but the that estimator is not sharp in these cases.  

 Our results in Illustrative Design 2 also make clear when the hybrid will be sharp in general discrete regressors settings. This is formulated in Theorem  \ref{th:hybrid_gen}. 
\begin{theorem} 
\label{th:hybrid_gen}
Suppose Assumption \ref{assume:MS} holds 
and the tuning parameter is set such that $\nu_n \rightarrow 0$ and $\nu_n^2 n \to \infty$.   

Estimator $\widehat{\mcA}_H$ is sharp only at those values $\alpha_0$ in the DGP at which one of the two original estimators -- maximum score or closed-form -- is sharp. The switching device will then pick that estimator for large enough sample with probability approaching 1. 
\end{theorem}


It is evident that the issues regarding the sharpness of the hybrid estimator, as seen in Illustrative Design 2, arise from the inability of the switching device to distinguish between cases where only one sample conditional choice probability is close to 1/2 and cases where there are multiple such probabilities. In light of this observation, it is prudent to explore more sophisticated switching devices for $K>2$, such as those based on the $(|\mathcal{X}_n|-K+2)$-th order statistic\footnote{This would require $|\mathcal{X}_n|-K+2\geq 1$.} of the $|\mathcal{X}_n|$-dimensional vector of  ${|\hat{p}(\widetilde{x}_i)-1/2|}$ for unique $\widetilde{x}_i$ in finite-sample support $\mathcal{X}_n$ ($|\mathcal{X}_n|$ denotes the cardinality of $\mathcal{X}_n$). This adjustment 
would help in Illustrative Design 2 and would result in the sharpness of the modified hybrid estimator. However, in general discrete regressors scenarios this would not guarantee sharpness as having $K-1$ conditional choice probabilities close to 1/2 does not necessarily guarantee sufficient variation in the respective $\widetilde{x}_i$ to uniquely identify the parameter (the only case when the closed-form estimator is sharp). Consequently, we opt not to pursue this extension formally.

The next estimator we introduce that integrates idea from both maximum score and closed-form estimation relies on modifying the maximum score objective function (\ref{MS:objective}) which uses estimated choice probabilities in the first step. Our proposed modified objective function is 
\begin{equation}
 \label{defineOFC}
 \begin{array}{l}
C_n(\alpha)= \frac1n\sum\limits^n_{i=1} \bigg[ {\bf 1}\left(\hat{p}(x_i)\geq 1/2-\nu_n\right) \cdot {\bf 1}\left(\alpha+x_i\geq 0\right)\\
\hspace{2in}+{\bf 1}\left(\hat {p}(x_i)\leq 1/2+\nu_n\right)\cdot {\bf 1}\left(\alpha+x_i\leq 0\right)\bigg],
\end{array}
\end{equation}
where $\nu_n$ is a tuning parameter akin to that used in the hybrid estimator discussed earlier. We denote the set of maximizers of (\ref{defineOFC}) as $\widehat{\mathcal{A}}_{OFC}$.\footnote{A similar 
2-step rank estimator for heteroskedastic monotone index models was proposed in \citeasnoun{khan2001} for a point identified model.
 An estimator involving a Maximum score objective function with nonparametrically estimated regressors in the first stage was proposed in \citeasnoun{chenetal2014}.}

The objective function (\ref{defineOFC}) blends principles of  the objective functions (\ref{MS:objective}) and (\ref{Ich:objective}). Analogous to the maximum score estimator, it incorporates inequalities of the index function. Notably, by utilizing estimated choice probabilities and formulating inequalities of the index function as weak in both directions, it assigns importance to observations where the index equals zero. 
  A ``slackness" tuning parameter $\nu_n$, as proven later, ensures sharpness of the estimator $\widehat{\cal A}_{OFC}$ in both point and set identified cases.\footnote{A related slack variables idea was introduced in \citeasnoun{komarova2013}.} 

To provide further insight into the functioning of this estimation method, let's examine Illustrative Design 1. 

First, consider  $\alpha_0$ as outlined in Case 1A. In this scenario, the estimated choice probability $\hat{p}(x_i)$ for $x_i=-\alpha_0$ is expected to be distributed around 0.5, with a positive probability (provided the rate of $\nu_n$ is suitably chosen). In the objective function (\ref{defineOFC}), regardless of whether $\hat{p}(-\alpha_0)$ falls below or above 0.5, as long as it lies within the interval $(0.5-\nu_n, 0.5+\nu_n)$, both inequalities $\alpha-\alpha_0 \geq 0$ and $\alpha-\alpha_0 \leq 0$ are equally likely to be satisfied. Consequently, in the limit, this yields the identified set $\{\alpha_0\}$.

When $x_i=1+\alpha_0$, $\hat{p}(1+\alpha_0)$ is expected to deviate from 0.5 by at least $\nu_n$, but this deviation does not influence the shape of the identified set. To elaborate further, when $\alpha_0=0$, $\hat{p}(1)$ exceeds $0.5+\nu_n$ with probability approaching 1, resulting in the inequality $\alpha \geq -1$. When combined with $\alpha=0$, this ultimately converges to 0 in probability. Conversely, when $\alpha_0=-1$, $\hat{p}(0)$ falls below $0.5-\nu_n$ with probability approaching 1, leading to the inequality $\alpha \leq 0$. When combined with $\alpha=1$, this converges to $-1$ in probability.

Now let us consider the partially identified Case 1B in Illustrative Design 1. When $\alpha_0>0$, both $\hat{p}(0)$ and $\hat{p}(1)$ exceed $0.5+\nu_n$ with probability approaching 1, resulting in the limit set  $\mcA_{OFC}$  that includes all the points from parameter set $\mcA$ that satisfy the inequality $\alpha \geq 0$. This differs from $\mcA_0 =(0,+\infty) \cap \mcA$ only in the inclusion of the boundary point 0, thus giving $d_H(\mcA_{OFC}, \mcA_0)=0$. Other sub-cases $\alpha_0 \in (-1,0)$  and $\alpha_0 \in (-\infty,-1)\cap \mcA$ within Case 1B will give analogous results.  

We now provide a general result on the asymptotic behavior of our second proposed combination estimator. 

\begin{theorem}\label{th:OFCsharp}
Suppose Assumption \ref{assume:MS} holds 
and the tuning parameter is set such that $\nu_n \rightarrow 0$ and $\nu_n^2\,n
\rightarrow \infty$. Then for any parameter value $\alpha_0 \in \mcA$ in DGP, 
$d_H(\widehat{\cal A}_{OFC}, {\mcA_0}) \stackrel{p}{\rightarrow}0$. 
\end{theorem}
In particular, Theorem \ref{th:OFCsharp} means that when $\mcA_0$ is a singleton, then $\widehat{\cal A}_{OFC}$ converges in probability to the singleton (as the boundary of $\mcA_0$ is empty).  When $\mcA_0$ is a non-singleton, then even for arbitrarily large samples $\widehat{\cal A}_{OFC}$ may differ from $\mcA_0$ in including some boundary points of $\mcA_0$. This, however, does not affect the Hausdorff distance. 

In summary, Theorem \ref{th:OFCsharp} establishes sharpness of our second combination estimator. 


\subsubsection{Robustness of estimators based
on combination ideas}
Similarly to the analysis of the classic estimators
for semiparametric binary choice model we consider robustness property of the new
estimators constructed in this section by exploring continuity of the 
limit with respect to sequences of parameters of the 
data generating process $\alpha(t,n;\alpha_0)$
with the property $\|\alpha(t,n;\alpha_0)-\alpha_0\|=O(1/\sqrt{n}).$  As observed in the analysis of robustness of the maximum score estimator, its objective function is constructed from the estimated choice probability which in the discrete design is 
a sample average which determined the choice
of the rate of drifting of the parameter sequence 
towards $\alpha_0.$ 

\begin{theorem}
\label{th:Combinationdrift}
Suppose that Assumption \ref{assume:MS} holds. 
Consider sequence of the data generating processes
$\alpha(n,t;\alpha_0)=\alpha_0+t /\sqrt{n}$ such that for a parameter value $\alpha_0$ in the DGP there are choice probabilities $p(\widetilde{x})=1/2$ for some $\widetilde{x}$ in the support, and $\alpha(n,t;\alpha_0) \in \mcC_k$ for $\mcC_k$ from the set system specified in Theorem \ref{th:MSgeneral2}. Suppose that 
the tuning parameter $\nu_n \rightarrow 0$ with
$n^2 \nu_n \rightarrow \infty$ then $
\lim_{n \rightarrow\infty}
\PP\left(d_H(\widehat{\mcA}_{H},\mcA_{CF})=0
\right)
=1$ 
for any  fixed $t \neq 0.$ In contrast whenever $\alpha_0$ is such that $p(\widetilde{x}) \neq 1/2$
for any $\widetilde{x} \in {\mathcal X},$ then $\lim_{n \rightarrow\infty}
\PP\left(d_H(\widehat{\mcA}_{H},\mcA_{ms})=0
\right)
=1$
\end{theorem}

We then provide an analogous result for the
maximizer of (\ref{defineOFC}):

\begin{theorem}
\label{th:OFCdrift}
Suppose that Assumption \ref{assume:MS} holds. 
Consider sequence of the data generating processes
$\alpha(n,t;\alpha_0)=\alpha_0+t /\sqrt{n}$ such for any parameter value $\alpha_0 \in \mcA$ in the DGP  $\widetilde{x}$ in the support, and $\alpha(n,t;\alpha_0) \in \mcC_k$ for $\mcC_k$ from the set system specified in Theorem \ref{th:MSgeneral2}. Suppose that 
the tuning parameter $\nu_n \rightarrow 0$ with
$n^2 \nu_n \rightarrow \infty$ then $
\lim_{n \rightarrow\infty}
\PP\left(d_H(\widehat{\mcA}_{H},\mcA_0)=0
\right)
=1$ 
for any fixed $t \neq 0,$ where $\mcA_0$ is the 
identified set under $\alpha_0.$
\end{theorem}

The main conclusion is that neither estimator has continuity properties
under drifting asymptotics.
Estimator $\widehat{\mcA}_{H}$ converges to the population maximizer of the closed-form estimator for any sequence of parameters converging to the point $\alpha_0$ where $p(\widetilde{x})=\frac12$ for some $\widetilde{x} \in {\mathcal X}.$  Estimator $\widehat{\mcA}_{OFC}$ converges to the identified set corresponding to the parameter in the limit of the parameter sequence $\alpha(n,t;\alpha_0).$ As a result, arbitrarily small change in $\alpha_0$ resulting in equalities
$p(\widetilde{x})=\frac12$ not hold, results in a
discontinuous change in the limit of both estimators.

In other words, neither of the constructed estimators are robust.

\section{Random Set Quantile Estimator}\label{sec:quantile} 

In this section we offer a novel approach to construction of sharp and robust  estimators which can serve as a basis for constructing consistent estimators for identified sets in partially identified discrete choice models. Our approach introduces a new class of estimators grounded in the concept of random set theory, a framework that has been integral to Econometrics since the pioneering work of \citeasnoun{beresteanumolinari2008}. While previous econometric research has primarily concentrated on the concept of selection (or Aumann) expectation within this theory and its associated estimators (see also subsequent studies by \citeasnoun{beresteanumolchanovmolinari2011}) and \citeasnoun{BERESTEANU201217}), our focus diverges due to the distinctive characteristics of our model and its classical estimators within the discrete-only regressors framework. Notably,  such feature as the fluctuating behavior of the maximum score estimator in some scenarios prompts us to adopt a fundamentally different approach from the random set theory.

First, to give some insights on what our estimation approach will deliver, we focus on Illustrative Design 1. For this design Theorem \ref{th:MSgeneral2} implied in Case 1A -- for concreteness, let us take  $\alpha_0=0$, --   $$\widehat{\mcA}_{ms} \stackrel{d}{\to} {\bf A} \equiv t \cdot \underbrace{[0,+\infty) \cap \mathcal{A}}_{\equiv B_1} + (1-t) \cdot \underbrace{[-1,0) \cap \mathcal{A}}_{\equiv B_2},$$
where $t$ is a Bernoulli random variable with parameter $\frac12$. 

If we look at the distribution limit, which is a random set, we can see that $\alpha_0=0$ is \textit{the only point} in the parameter space that happens to be \textit{in the closure of realizations} of random set  ${\bf A}$ in at least $100(1/2+\Delta) \%$ cases for $\Delta>0$. Indeed, $0$ belong in the boundary of both $B_1$ and $B_2$  and no other point simultaneously belongs to the boundary of both sets.  Moreover, the proof of Theorem \ref{th:MSgeneral2}  in the Appendix will make it clear that the same result will hold for $\widehat{\mcA}_{ms}$ in a finite-sample for large enough $n$. 

This naturally brings us to the estimation approach based on the notion of the random set quantile. Namely we define our estimator as the random set $\tau$-th quantile of the closure of the maximum score estimator:
\begin{equation}\label{quantile:estimator}
\widehat{\mcA}_{RSQ,\tau}=q_{\tau} \left(
\overline{\widehat{\mcA}_{ms}}\right). 
\end{equation}
Following the definition of the $\tau$-th quantile of a random set in \citeasnoun{molchanov2006book} (p.176), 
\begin{equation}\label{quantile:estimator2}
q_{\tau} \left(
\overline{\widehat{\mcA}_{ms}}\right) = \left\{p \left(\cdot;\,\overline{\widehat{\mcA}_{ms}}\right) \geq \tau \right\}
\end{equation}
for {\it coverage function } $p\left(u\,;\,\overline{\widehat{\mcA}_{ms}}\right) \equiv \PP\left( u \in
\overline{\widehat{\mcA}_{ms}}\right).$ Note that this notion from \citeasnoun{molchanov2006book} also applies to vector settings. 

An important consideration lies in the selection of suitable quantile indices $\tau$ for our estimation. To foreshadow our formal result, we propose utilizing a $\tau=1/2+\Delta$-th quantile of the closure of $\widehat{\mathcal{A}}_{ms}$, where $\Delta \in (0, 1/2)$.

Now, let's revisit Illustrative Design 1 for the parameter value $\alpha_0=0$ in the DGP. In this scenario, the $\tau$-th quantile of the \textit{closure} $\overline{\widehat{\mathcal{A}}_{ms}}$ with $\tau=1/2+\Delta$, $\Delta>0$, reduces to only $\{0\}$ for sufficiently large $n$, thereby converging in probability to the identified set.\footnote{It's worth noting that for indices below $1/2$, such as $(1/2-\Delta)$-th quantile of $\overline{\widehat{\mathcal{A}}{ms}}$ with $\Delta \in (0,1/2)$, the resulting quantile set is $[-1,+\infty) \cap \mathcal{A}$ for sufficiently large $n$. This set is a superset of the identified set and aligns with the maximization of the population maximum score objective function in this scenario. On the other hand, the $1/2$-th quantile of $\overline{\widehat{A}_{ms}}$ for large enough $n$ can, with additional effort, be demonstrated to be the two-element set ${-1,0}$, failing to asymptotically recover the identified set.}

Given that the concept of a random set quantile may be unfamiliar to econometricians, it's helpful to draw parallels to voting rules for additional clarity. Let's explore this by examining the median of a random set within the framework of a simple discrete model.

Suppose that we can generate infinitely many random samples $\{(x_i,y_i\}_{i=1}^n$ of a fixed size $n$. Each sample casts votes for any number of elements of $\mathcal{A}$ which maximize the objective (\ref{MS:objective}). Essentially, a given sample votes for  its respective maximum score estimate $\widehat{\mcA}_{ms}$. After the completion of set $\widehat{\mcA}_{ms}$ to its closure $\overline{\widehat{A}_{ms}}$,  majority winners are selected -- namely, those elements in $\mathcal{A}$ that are voted for by at least 50\% of the samples. The collection of those majority winners would give us $q_{.5} \left(
\overline{\widehat{\mcA}_{ms}}\right)$. For any arbitrary index $\tau \in (0,1)$, the quantile $q_{\tau}(\overline{\widehat{\mcA}_{ms}})$ comprises those elements from $\mathcal{A}$ that adhere to the 'quota rule' with a threshold of $\tau$. In simpler terms, these elements within $\mathcal{A}$ must garner votes from at least $100\tau \%$ of the samples to be included.



\subsection{Feasible version of the random set quantile estimator} 

The estimator $\widehat{\mcA}_{RSQ,\tau}$ is infeasible due to  the distribution of the random set $\overline{\widehat{\mcA}_{ms}}$ (of maximizers of 
(\ref{MS:objective})) being a population object. Consequently, we have rely on the 
finite sample to approximate the population quantile
of $\overline{\widehat{\mcA}_{ms}}$. 

To illustrate our ideas on how we can proceed with this, let us focus on Illustrative Design 1 and Case 1A -- for concreteness, we take $\alpha_0=0$ as the parameter value in the DGP. Despite the weak limit of the estimator in this case being an essentially a random choice between $[-1,0)$ or $[0,+\infty)$,  in a {\it concrete sample} we only observe just one set -- either $[-1,0)$ or $[0,+\infty)$ (whichever of them maximizes (\ref{MS:objective})) with high probability (other sets can be realized as maximizers with a decreasingly small probability). Consequently, to simulate the uncertainty across samples, we need to employ suitable sampling techniques. Simultaneously, these methods must be constructed in a manner that their impact on cases of $x$ where $p(x) \neq 1/2$ remains inconsequential.
   
To illustrate our proposed approach, we rely on the representation of the maximum score objective function as in (\ref{MS:objective}). We propose a sampling technique that will draw  the conditional choice probability $\hat p_s(x)$, $s =1, \ldots, S$, for each value of discrete regressor in its sample support. Since by the standard Moivre-Laplace theorem for each $x $ in the sample support, 
$$ \sqrt{n}\frac{\hat p(x)-p(x)}{\sqrt{p(x)(1-p(x)}} \stackrel{a}{\longrightarrow} \mathcal{N}\left(0,\,1\right),$$
the for a given sample of size $n$ 
it may seem natural to draw $\hat p_s(x)$ from 
$ \mathcal{N}(\hat p(x), \frac{\hat p(x)(1-\hat p(x))}{n})$. However, if $p(x)=1/2$, then this symmetric way of drawing on both sides of $\hat p(x)$ will give a probabilistic advantage to one of the sets from $[-1,0)$ or $[0,+\infty)$ -- whichever of them was the maximum score estimate in our sample. Indeed, if in Case 1A for parameter value $\alpha_0$ in DGP we have $\hat p(-\alpha_0)>1/2$, then  $\hat p_s(-\alpha_0)$ will be primarily located above $1/2$ as well while we need them $\hat p_s(-\alpha_0)$ to appear in approximately similar proportions on both sides of $1/2$. This leads us to proposing an asymmetric sampling technique. 

Namely, if $\hat p(x)>1/2$ for $x$ in the sample support, then 
$$
\textstyle \hat p_s(x) \sim \mathcal{N}\left( \hat 
p(x), \frac{\hat 
p(x)(1-\hat 
p(x))}{n}\right) \cdot {\bf 1}\left\{\hat p_s(x) > \hat 
p(x) \right\} +\mathcal{N}\left( \hat 
p(x), \tau \frac{\hat 
p(x)(1-\hat 
p(x))}{n}\right) \cdot {\bf 1}\left\{\hat p_s(x) \leq \hat 
p(x) \right\}, $$ 
and if $\hat p(x)<1/2$, then 
$$
\textstyle \hat p_s(x) \sim \mathcal{N}\left( \hat 
p(x), \tau\frac{p(x)\hat 
p(x)(1-\hat 
p(x))}{n}\right) \cdot {\bf 1}\left\{\hat p_s(x) > \hat 
p(x) \right\} +\mathcal{N}\left( \hat 
p(x), \frac{\hat 
p(x)(1-\hat 
p(x))}{n}\right) \cdot {\bf 1}\left\{\hat p_s(x) \leq \hat 
p(x) \right\}. $$ 
The value of $\tau>1$ may be data driven and may depend on  the desired quantile index as well as probabilities of values close to the decision making boundary (defined as  $p(x)=0.5)$.  

Finally, for each $s=1,\ldots,S$ we produce a set 
\begin{equation}  
\label{maxscoretoy_real_sampling}
\widehat{\cal A}_{ms,s} = \arg \max _{\alpha \in {\cal A}} \sum_{x \in \{0,1\}}  (\hat p_s(x)-0.5)\cdot \sgn(\alpha+x) \hat{q}(x). 
\end{equation} 
The feasible version of the quantile estimator 
$\widehat{\mcA}_{RSQ,\tau}$ is approximated from the simulation sample by taking a sample $(0.5+\Delta)$-quantile based on the sample  of $\{\widehat{\cal A}_{ms,s}\}_{s=1}^S$.

\subsection{Sharpness of the random set quantile estimator}
We now formulate our result regarding the asymptotic behavior of the random quantile set estimator. We consider a general discrete regressors setting and, for technical convenience, focus on $\widehat{\mcA}_{RSQ,\tau}$ defined as in 
(\ref{quantile:estimator}). 

To formulate the sharpness result, we first use a refined result on the asymptotic behavior of the maximum score estimator in Theorem \ref{th:MSgeneral2}.

\begin{theorem}
\label{th:MSrefined} 
Suppose that Assumption \ref{assume:MS} holds. 
If for a parameter value $\alpha_0$ in the DGP there are choice probabilities $p(\widetilde{x})=1/2$ for some $\widetilde{x}$ in the support, then 
for any quantile index $\tau> \min_{\ell_1, \ldots, \ell_{[L/2]+1}} p(B_{\ell_1}=1)+ \ldots p(B_{\ell_{[L/2]}}=1)$, 
$$d_H(\widehat{\mcA}_{RSQ,\tau},\mcA_0) \stackrel{p}{\rightarrow} 0, $$
In particular, this is guaranteed if one takes $\tau>1/2$. 
\end{theorem}

The result of  Theorem \ref{th:MSrefined} immediately implies that the estimator is
sharp on the entire parameter space.  

\begin{theorem}
\label{th:QRS_toydesign} 
Suppose that Assumption \ref{assume:MS} holds. Let $\widehat{\mcA}_{RSQ,\tau}=q_{\tau}(\overline{\widehat{A}_{ms}})$ correspond to $\tau$-th quantile of random set $\overline{\widehat{A}_{ms}}$, as defined in \citeasnoun{molchanov2006book} (p. 192). If $\tau=1/2+\Delta$, $\Delta \in (0, 1/2)$, then 
for any $\alpha_0 \in \mcA$
\[d_H\left( \widehat{\mcA}_{RSQ,\tau},\, {\cal A}_0\right) \stackrel{p}{\rightarrow}0.\]
\end{theorem}

We note that there is no difference between various
``parameter regimes" $\alpha_0.$ Regardless of whether
the model is point or partially identified, the $(\frac12+\Delta)$-quantile of the random set of 
closure of the maximum score estimator converges to the 
identified set.

\subsection{Robustness of the random set quantile estimator}\label{rsqe}

In our analysis of robustness of the maximum score
estimator in Section \ref{MS:section} we characterize
the maximizer of (\ref{MS:objective}) as a random set.
Just as in our analysis of sharpness of the 
estimator, in this section for simplicity of exposition we focus on the infeasible quantile estimator
under drifting sequences of the data generating process
$\alpha(t,n;\alpha_0)$ with the property $\|\alpha(
t,n;\alpha_0
)-\alpha_0\|=O(1/\sqrt{n})$ where $\alpha_0$ is the where $p(\widetilde{x})=\frac12$ at least for one $\widetilde{x} \in {\mathcal X}.$ In other words,
the maximum score estimator converges to a random set when the data generating process is indexed by parameter $\alpha_0.$ The following theorem characterizes the structure of the probability along the such parameter sequences.

\begin{theorem} \label{th:QRSEdrift}
Suppose that Assumption 1 holds. Consider sequence of the data generating
processes $\alpha(n, t; \alpha_0) = \alpha_0 + t/\sqrt{n}$ such that for a parameter value $\alpha_0$ in the DGP there exist $\widetilde{x} \in {\mathcal X}$ such that 
choice probabilities $p( \widetilde{x}) = \frac12,$ and $\alpha(n, t; \alpha_0) \in \mcC_k$ for $\mcC_k$ from
the set system specified in Theorem \ref{th:MSgeneral2}. Then 
$
\PP\left(d_H(\widehat{\mcA}_{RSQ,\tau},\mcA(t))=0\right) \rightarrow 1,
$
where deterministic set $\mcA(t)$ has the following properties: 
\begin{itemize}
 \item[(i)] For $t \in \real^{K}$ it can be equal to $\overline{\mcC_k}$ or intersections of $\overline{\mcC_k}$  with any possible subsets
 of $\overline{\mcC_1},\ldots,\overline{\mcC_L};$
 \item[(ii)] for $\|t\| \rightarrow 0,$  
 $\mcA(t)$ approaches the closure of the identified set under $\alpha_0$;
 \item[(iii)] for $\|t\| \rightarrow \infty,$  
 $\mcA(t)$ approaches the closure of the identified set under $\alpha(n,t;\alpha_0)$.
\end{itemize}
\end{theorem}

Thus, estimator $\widehat{\mcA}_{RSQ,\tau}$ is robust: its limit ranges from the closure of the identified set corresponding to the data generating process indexed by the limit of the parameter sequence to the closure of the identified set corresponding to the data generating process indexed by the elements of the parameter sequence. As we established earlier,$\widehat{\mcA}_{RSQ,\tau}$ is also sharp and, unlike the maximum score estimator, it converges to the identified set for all parameter values of the data generating process.\footnote{ 
While this estimator is infeasible,  a consistent estimator of this quantile (see our proposed approach earlier (\ref{th:QRS_toydesign}) would also have this property.}

\setcounter{equation}{0}

\section{Application}\label{application}
The UK General Election in 2019 marked a significant development for the Labour Party as it faced a decline in its constituency victories. With a total of 650 constituencies, Labour secured only 202 seats during this electoral contest, a historic low both in terms of numerical count and proportion since the year 1935. Various media analyses pointed towards the aftermath of the Brexit referendum as a contributing factor to Labour's electoral setbacks. A telling example is a headline from The Guardian that succinctly captured the sentiment: "It was Brexit, not left-wing policies, that lost Labour this election."\footnote{\tiny \url{https://www.theguardian.com/politics/2020/jun/18/key-points-from-review-of-2019-labour-election-defeat}}

We consider the outcome
representing the indicator for a political
party winning a given constituency in 2015
and retaining its seat in the 2019 elections. We focus on the question whether
the ``Leave" vote in the Brexit referendum and its consequences has impacted preferences of the UK voters.

To address our question, we construct 5 factors and measure their impact on the outcome of 2019 election for the Labour. The first is an indicator variable reflecting the ``Leave" vote.
The second is an indicator denoting a winning margin in the 2015 General election of less than 5\%, capturing fiercely contested constituencies before the Brexit referendum.
The third is an ordered variable with values 0, 1, 2. It takes on 0 if the mean income growth from 2015 to 2019 was negative (8.15\% in our data), 1 if positive but below the median growth across all constituencies (41.84\% of constituencies in the data), and 2 if above the median growth (naturally, 50\% of constituencies).
The forth is an indicator that the winning party in a constituency in 2015 was Labour.
We also include an interaction between the
indicators of the Leave vote and Labour's 2015 victory, capturing the potential differential effects of the Brexit referendum on constituencies held by Labour in 2015. 

Table \ref{table:App1}
provides a summary of these variables. It shows that 79.8\% of constituencies 
had the same party win the elections in 2015 and 2019 while elections in 2015 were close within 5\% in 8.7\% of constituencies. 22.9\% of constituencies
voted to ``Leave" and had Labour won in 2015 while overall, Labor won 35.7\% of constituencies.

\begin{table}[!ht]
\caption{Summary statistics.}
\label{table:App1}
\vskip 0.05in 
\par
\begin{center}
\begin{tabular}{lcccc}
\hline\hline
Variable & Mean & St. dev. & Min & Max \\ \hline
Same party wins constituency in GE 2019 & 0.7985 &   0.4015 & 0 & 1 \\ 
Indicator for ``Leave'' vote &     0.6246  &   0.4846    &      0    &      1  \\ 
Indicator if the GE 2015 was within 5\% margin & 0.0877    & 0.2831 &          0  &        1 \\ 
2015-to-2019 mean income growth category & 1.4184 &   0.6380      &    0   &       2 \\
Indicator if Labour won in 2015 & 0.3569 & 0.4795 & 0 & 1 \\

Indicator if Labour won in 2015 $\times$ Indicator for ``Leave'' vote & 0.2292 &     0.4207 &          0   &       1
\\ \hline
\end{tabular} 
\end{center}
\end{table}

We estimate the semiparametric model in 
(\ref{discrete:choice}) under Assumption
\ref{assume:MS}, notably using the 
median restriction in Assumption \ref{assume:MS} (iv).
We set the vector of covariates
$\widetilde{X}=(1,X_1,\ldots,X_5)'$ with the first
element corresponding to the intercept and 5 non-constant covariates we outlined above. We normalize the vector of 
estimated coefficients 
$\widetilde{\alpha}=(\alpha_0,\alpha_1,-1,
\alpha_3,\alpha_4,\alpha_5),$ using
a natural normalization for the coefficient of the indicator for the 2015 elections being close in a given constituency.

In our data, there are 23 unique discrete realizations of $\widetilde{X}$. For 2 elements in the sample support of $\widetilde{X}$, we have $\PP(Y=1|\widetilde{X})=0.5$. For 4 elements, $\PP(Y=1|\widetilde{X})<0.5$, and for the remaining 17 elements, $\PP(Y=1|\widetilde{X})>0.5$. 

We now use the estimators considered in Sections \ref{sec:classic}
and \ref{sec:quantile}.

\paragraph{Maximum Score estimator}. Objective function (\ref{MS:objective}) of the maximum score estimator excludes the points in the support of covariates where $\PP(Y=1| \widetilde{X})=0.5$. For the remaining 21  points in the support of the covariates we construct the following system of inequalities: 
\begin{align*} 
\widetilde{x}'\widetilde{\alpha} \geq 0 & \text{ if } \PP(Y=1|\widetilde{X}=\widetilde{x})>0.5 \text{ (17 inequalities )},\\ 
\widetilde{x}'\widetilde{\alpha} < 0 & \text{ if } \PP(Y=1|\widetilde{X}=\widetilde{x})<0.5 \text{ (4 inequalities )}.
\end{align*}
The set of solutions to this system contains all maximands of the objective function (\ref{MS:objective}). Notably, in our model and data, this system does indeed have a solution. Consequently, the maximum score estimator can be succinctly characterized by the following system of inequalities (with normalization $\widetilde{\alpha}=(\alpha_0,\alpha_1,-1,\alpha_3,\alpha_4,\alpha_5)'$):
$$(\alpha_0,\alpha_1,\alpha_3,\alpha_4,\alpha_5) 
\left(
\begin{array}{ccccccccccccccccc}
     1   &  1  &   1 &    1   &  1   &  1  &   1  &   1   &  1  &   1  &   1   &  1   &  1   &  1   &  1   &  1  &   1\\
     0   &  0   &  0  &   0   &  0  &   0   &  0   &  0  &   1  &   1   &  1  &   1  &   1  &   1  &   1   &  1  &   1\\
     0   &  0  &   1  &   1  &   2  &   2  &   1  &   2  &   0  &   0   &  1   &  1   &  2   &  2  &   0 &    1  &   2\\
     0   &  1   &  0  &   1   &  0  &   1  &   1   &  1    & 0   &  1  &   0  &   1   &  0  &   1  &   0   &  0  &   0\\
     0   &  0  &   0   &  0  &   0   &  0   &  0  &   0  &   0   &  1   &  0  &   1  &   0  &   1  &   0   &  0   &  0\\  
\end{array}
\right) \geq c_1
$$
with 
$c_1=(0,0,0,0,0,0,1,1,0,0,0,0,0,0,1,1,1)$
and 
$$
(\alpha_0,\alpha_1,\alpha_3,\alpha_4,\alpha_5)  \left( 
    \begin{array}{cccc}
    1   &  1 &    1  &   1 \\
     0  &   0  &   1  &   1 \\
     1  &   2  &   0  &   1 \\
     0  &   0  &   1  &   1 \\
     0  &   0   &  1  &   1 \\
     \end{array}
     \right)  < c_2$$
with $c_2=(1,1,1,1)$. 

Estimate $\widehat{\mcA}_{ms}$ obtained from this system of inequalities has a non-empty interior in $\mathbb{R}^5$ and is contained in 
$[0, 1.5]\times [0, +\infty]\times [-0.5, 0.5] \times [0, +\infty) \times (-\infty,0] $. 


We can provide a meaningful interpretation for 
the estimate $\widehat{\mcA}_{ms}$ in light of the fact that $\widetilde{X}=(1,X_1,\ldots,X_5)'$.  
In our model specification, the base group is non-Labour constituencies in 2015 that voted Remain. 
In Table \ref{Tab2} we present raw joint counts of the
outcome of the vote for Labour party in 2015
and the Brexit vote.
    \begin{table}[!ht]
    \begin{center}
            \caption{Joint counts of Labour winning in 2015 and vote in the 2018 referendum across constituencies}\label{Tab2}
            \vskip 0.05in 
        \begin{tabular}{lrr}
        \hline\hline 
           \quad  & Labour won in 2015 & Labour didn't win in  2015\\
           \hline
            Voted ``Leave" &  149 & 257 \\
            Voted ``Remain" & 83 & 161 \\
            \hline
        \end{tabular}
\end{center}
    \end{table} 

Given $X_2$ and $X_3$, the utility indices of the  latent utilities $U^*=\alpha_0+\alpha_1 X_1 -X_2 +\alpha_3 X_3 +\alpha_4 X_4 + \alpha_5 X_5 + \varepsilon$ manifest as follows: \begin{align*}
U^*_{00}&=\alpha_0 -X_2 +\alpha_3 X_3    \text{ (Remain, non-Labour in 2015)} \\
U^*_{01} & =\alpha_0 -X_2 +\alpha_3 X_3 +\alpha_4  \text{  (Remain, Labour in 2015)} \\
U^*_{10} & = \alpha_0+\alpha_1 -X_2 +\alpha_3 X_3    \text{ (Leave, non-Labour in 2015)} \\
U^*_{11} & = \alpha_0+\alpha_1 -X_2 +\alpha_3 X_3 +\alpha_4 + \alpha_5  \text{ (Leave, Labour in 2015)} 
\end{align*}
Following from the directions of the effects (signs of $\alpha_1$, $\alpha_4$, $\alpha_5$) that 
$U^*_{01} \geq U^*_{00}, \quad U^*_{10} \geq U^*_{00}, \quad U^*_{01} \geq U^*_{11}.$
System of inequalities for maximizers of (\ref{MS:objective}) additionally implies that  $\alpha_4+\alpha_5 <0$ and $\alpha_0+\alpha_1+\alpha_4+\alpha_5 \geq 0$ thus giving 
$U^*_{11} < U^*_{10}, \quad U^*_{11} \geq U^*_{00}.$
The inequalities allow for either $\alpha_1\geq \alpha_4$ or $\alpha_1\geq \alpha_4$ thus leaving the relationship between $U^*_{01}$ and $U^*_{10}$ ambiguous. 

To summarize, we conclude that given $X_2$, $X_3$, 
$U^*_{01} \geq U^*_{11} \geq U^*_{00}$
 and $U^*_{10} > U^*_{11} \geq U^*_{00}$
with the inequality between $U^*_{01}$ and $U^*_{10}$ left undetermined. 

At the level of utility indices, one can infer that our findings from the estimate $\widehat{\mcA}_{ms}$ partially align with the perspectives presented in the Guardian article. This partial alignment applies to comparisons between Labour 2015 \& Leave vs. Labour 2015 \& Remain, as well as Labour 2015 \& Leave vs. non-Labour 2015 \& Leave. However, it does not extend to the comparison between Labour 2015 \& Leave and non-Labour 2015 \& Remain, introducing nuances to the interpretation of the underlying relations. 

\paragraph{Closed form estimator.} Next we consider the maximizer of (\ref{Ich:objective}) and 
note that if the tuning parameter $\nu_n$ used to construct the weights is smaller than 0.1, then we can use only 2 elements with $\PP(Y=1|\widetilde{X})\in [0.5-h_n, 0.5+h_n[$ for the estimator. This leads to two equalities (under the same normalization on $\widetilde{\alpha}$ as above):
$$
0=\alpha_0+\alpha_1-1+2\alpha_3+\alpha_4+\alpha_5=0\;\;\mbox{and}\;\;
0=\alpha_0-1
$$
leading to estimate 
$\widehat{\mcA}_{CF} = \{(\alpha_0, \alpha_1, \alpha_3, \alpha_4, \alpha_5): \alpha_0=1,   \; \alpha_1+2\alpha_3 +\alpha_4 +\alpha_5=0\},$
which is a three-dimensional hyperplane in $\real^5$. 

Thus, the closed-form estimator leaves ambiguous the directions and relative (to normalized $\alpha_2=-1$)  effects of $X_1$, $X_3$, $X_4$ and $X_5$ just indicating that their effects combined in a certain way sum up to 0.

\paragraph{Combination estimators.}
We start with the estimator (\ref{def:switching})
based on combination of the estimates $\widehat{\mcA}_{ms}$ and $\widehat{\mcA}_{CF}.$ 
With the tuning parameter $\nu_n$ less than 0.1, we get the combination estimate that coincides with $\widehat{\mcA}_{CF}$ and, thus, has properties
as discussed above. 

Next we consider estimator $\widehat{\mcA}_{OFC}$
maximizing (\ref{defineOFC}). 
While $\widehat{\mcA}_H$ opts between \textit{all} equality constraints provided by maximizers of (\ref{Ich:objective}) and \textit{all} inequality constraints provided by maximizers of (\ref{MS:objective}), $\widehat{\mcA}_{OFC}$ combines \textit{both} equality constraints and inequality information. In our application, for the nuisance parameter $\nu_n \in (0, 0.1) $  obtained estimate is expressed as:
$\widehat{\mcA}_{OFC} = 
\left\{(\alpha_0, \alpha_1, \alpha_3, \alpha_4, \alpha_5): 
\alpha_0=1, \alpha_3=0, \alpha_1+\alpha_4+\alpha_5=0
 \right\}.
$
This estimate reveals that there is no discernible impact ($\alpha_3=0$) of the change in economic well-being on the utility index for the re-election of the incumbent party, accounting for other factors.
 
Building on our earlier findings: 
$U^*_{01} \geq U^*_{11} $ and
$U^*_{10} > U^*_{11}. $
These conclusions persist, but now the additional  constraint $\alpha_1+\alpha_4+\alpha_5=0$ also allows us to assert 
$U^*_{11} = U^*_{00}.$
This finding lends support, at least at the level of utility indices, to the proposition that Labour lost many constituencies due to the Leave vote within those constituencies. Given $X_2$ and $X_3$, the Leave vote in other constituencies resulted in a greater utility index, while the Remain vote across all constituencies had either the same or a greater utility index.

\paragraph{Random set quantile estimation.}  We rely on our  approach to a feasible sample quantile of random set outlines in the text. With this approach, there are four sets each occurring with a probability close to 0.25 (but strictly less than 0.25). 

Each of these sets is a subset of $\widehat{\mcA}_{ms}$: 
\begin{align*}
\widehat{\mcA}_1 &= \widehat{\mcA}_{ms} \cap \{ (\alpha_0, \alpha_1, \alpha_3, \alpha_4, \alpha_5)':  \alpha_0 \geq 1, \; \alpha_0 + \alpha_1 +2 \alpha_3+\alpha_4 + \alpha_5 \geq 1 \} \\
\widehat{\mcA}_2 &= \widehat{\mcA}_{ms} \cap \{ (\alpha_0, \alpha_1, \alpha_3, \alpha_4, \alpha_5):  \alpha_0 \geq 1, \; \alpha_0 + \alpha_1 +2 \alpha_3+\alpha_4 + \alpha_5 < 1 \} \\
\widehat{\mcA}_3 &= \widehat{\mcA}_{ms} \cap \{ (\alpha_0, \alpha_1, \alpha_3, \alpha_4, \alpha_5):  \alpha_0 < 1, \; \alpha_0 + \alpha_1 +2 \alpha_3+\alpha_4 + \alpha_5 \ \geq 1 \} \\
\widehat{\mcA}_4 &= \widehat{\mcA}_{ms} \cap \{ (\alpha_0, \alpha_1, \alpha_3, \alpha_4, \alpha_5):  \alpha_0 < 1, \; \alpha_0 + \alpha_1 +2 \alpha_3+\alpha_4 + \alpha_5 < 1 \} .
\end{align*}
Taking closures of these sets, as well as closures of any other realization of $\widehat{\mcA}_{ms}$ in our sample (although other outcomes occur with negligible probabilities), the feasible $\tau$-th random set quantile for $\tau > 0.5$ must be in the intersection of at least three out of four sets $\overline{\widehat{\mcA}_1}$, \ldots, $\overline{\widehat{\mcA}_4}$. This leads to the set
$\left\{(\alpha_0, \alpha_1, \alpha_3, \alpha_4, \alpha_5)': 
\alpha_0=1, \alpha_3=0, \alpha_1+
\alpha_4+\alpha_5=0
 \right\}$
as our $\tau$-random set quantile estimate. 

Consequently, in this case, a feasible quantile random set estimate aligns with $\widehat{\mcA}_{OFC}$.

\paragraph{Probit/Logit.} Even though we consider model (\ref{discrete:choice}) without parametric specification for the unobserved shock $\epsilon$, one might wonder about the implications of employing parametric estimation methods such as Probit or Logit.

Given that the system of inequalities corresponding to maximizers of (\ref{MS:objective}) has a solution, delineating a set of hyperplanes that perfectly separate two classes of points (those with $\PP(Y=1| \widetilde{X})>1/2$ and those with $\PP(Y=1| \widetilde{X})<1/2$), it can be demonstrated that Probit and Logit estimates must fall within the set $\widehat{\mcA}_{ms}$ after re-normalization to enforce $\alpha_2=-1$. In our model and data, this alignment is evident, as illustrated in the Probit output in Table \ref{table:Probit}. (The conclusions from the  results from the Logit model are analogous; hence, Logit estimation is omitted for brevity.)
\begin{table}[!ht]
\caption{Estimates from the Probit model}
\label{table:Probit}
\begin{center}
\begin{tabular}{lc}
\hline
Variable & Probit    \\ \hline
Indicator for ``Leave" vote  & 0.6997    \\ 
\quad & (0.1516)   \\ 
Indicator for GE 2015  being within 5\% margin & -0.8948     \\ 
\quad & (0.1820)  \\ 
2015-to-2019 mean income growth category & 0.0145      \\ 
\quad & (0.0981)  \\ 
Indicator if Labour won in 2015 & 1.3352     \\ 
\quad & (0.3007)  \\ 
Indicator if Labour won in 2015 $\times$ Indicator for ``Leave" vote & -2.2655     \\ 
\quad & (0.0911)  \\ 
constant & 0.6486      \\ 
\quad & (0.1608)  \\ 
\hline
\end{tabular}%
\end{center}
{\small Robust standard errors in Probit regression are in parentheses.} 
\par
\end{table}
If we take Probit estimates, 
$\widehat{\alpha}_1+\widehat{\alpha}_4+\widehat{\alpha}_5=-0.2307,$
giving a stronger support to the Guardian article statement regarding the impact of Leave vote on the 2019 Labour electoral performance. The test for the null hypothesis 
$H_0: \alpha_1+\alpha_4+\alpha_5=0,$
based on Probit  suggests does not reject the null ($p-value=0.1315$) which is is consistent with estimators $\widehat{\mcA}_{OFC}$ and $\widehat{\mcA}_{RSQ,\tau}$. However, $\widehat{\mcA}_{OFC}$ and random set quantile estimates are more explicit about the relationship $\alpha_1+\alpha_4+\alpha_5=0$ whereas this message appears  convoluted in the output of the Probit model.

\setcounter{equation}{0}

\section{Extensions}
\label{sec:panel data}
In this section we demonstrate that our analysis naturally extends to several
classes of models including the binary choice model with independent noise as well as static and dynamic binary choice panel data models with fixed effects and discrete regressors.

\subsection{Maximum Rank Correlation Estimator}

\label{sec:MRCreview}

Identification argument in \citeasnoun{manski1987} applies in the 
simplified setting where the error term
$\epsilon$ in model (\ref{discrete:choice})
is statistically independent. However,
in case where regressors are continuous,
parameters of the model can be estimated
by exploiting additional information of independence, opening the possibility to construct estimators with 
properties superior to those of the maximum score estimator
of \citeasnoun{manski1987}.

In this section we consider the performance
of this estimator when regressors are discrete. We focus on model (\ref{discrete:choice}) under the following stronger version of Assumption 
\ref{assume:MS} (iii): \vspace{.05in}\\
{\bf Assumption \ref{assume:MS}.} {\it (iii'') $\varepsilon \perp \widetilde{X}$ and distribution density $f_{\epsilon}(\cdot)$
 is above zero and strictly above $L>0$ in some fixed neighborhood of $0.$
} \vspace{.05in}\\
Also, just as in Section \ref{MS:section}, we use parameter normalization $\alpha^k=1$ for one $k=1, \ldots, K$, and denote $\alpha$ as the non-normalized part of $\widetilde{\alpha}$.

From the results in \citeasnoun{bierenshartog}, model (\ref{discrete:choice}) under Assumption \ref{assume:MS} (i), (ii) and (iii'') is not point identified. The identified set $\mcA_0$ is fully
characterized as a collection of 
parameter values $\alpha$ satisfying
\begin{align}
p(\widetilde{x})  > p(\widetilde{x}^*) & \iff \widetilde{x}'\widetilde{\alpha} > {\widetilde{x}}^{*\prime}\widetilde{\alpha}
\label{SIineq} \\
p(\widetilde{x})  = p(\widetilde{x}^*) & \iff \widetilde{x}'\widetilde{\alpha} = \widetilde{x}^{* \prime} \widetilde{\alpha}. 
\label{SIeq}
\end{align}

From this definition of $\mcA_0$ it is clear that depending on the value of the parameter
of the DGP, the identified set can be a singleton or a non-singleton set. Namely, it will be a singleton if there are enough pairs of $\widetilde{x}$  and  $\widetilde{x}^*$, $ \widetilde{x} \neq \widetilde{x}^* $, such that $ p(\widetilde{x})  = p(\widetilde{x}^*)$. to point identify parameter from the respective system of equations $(\widetilde{x}-\widetilde{x}^*)'\widetilde{\alpha}=0$.



\citeasnoun{MRC} introduced the maximum
rank correlation estimator for single index models which include our binary choice models 
 (\ref{discrete:choice}) under Assumption \ref{assume:MS} (iii') of independent random shock $\varepsilon$. This estimator maximizes 
 \begin{equation} 
\label{MRCsample}
MRC_n(\alpha)=\frac{1}{n(n-1)}\usum {\bf 1}  
\left\{ y_i>y_j\right\} \, {\bf 1} \left\{\widetilde{x}_i' \widetilde{\alpha}>\widetilde{x}_i' \widetilde{\alpha}.\right\}
\end{equation}
\citeasnoun{MRC} established the properties of this estimator in single index model under a continuity of a regressor with non-trivial impact (WLOS, this is a regressor with a normalized component of $\widetilde{\alpha}$). In this case  the parameter in the DGP is identified up to a location and the MRC estimator of the parameter subvector excluding intercept is consistent. 

In our binary choice model we can alternatively represent (\ref{MRCsample}) as
\begin{equation}
\label{MRCsample2} MRC_n(\alpha)=\frac{1}{n (n-1)}\sum_{ i\neq j}  \hat{p}\left(\widetilde{x}_i
\right) \left(1-\hat{p}\left( \widetilde{x}_j\right) \right) {\bf 1}\left\{x_i' \widetilde{\alpha}  >x_j' \widetilde{\alpha} \right) \hat{P}\left( \widetilde{X}=\widetilde{x}_i\right) \hat{P}\left(\widetilde{X}=\widetilde{x}_j\right),
\end{equation} 
Let $\widehat{\mcA}_{MRC}$ denote the 
set of maximizers of (\ref{MRCsample}). For the convenience of the technical discussion, we will suppose that the index specification in our model does not contain intercept.


We now analyze the {\it sharpness} of 
$\widehat{\mcA}_{MRC}$ maximizing (\ref{MRCsample}). To do that we construct the population analog of the objective function (\ref{MRCsample}) using the same structure as in representation (\ref{MRCsample2}):
\begin{equation} 
\label{MRCobj_limit2} MRC(\alpha)=\sum_{
\widetilde{x},\, 
\widetilde{x}^* \in \mathcal{X}^2} p(\widetilde{x})(1-p(\widetilde{x}^*)) {\bf 1} \left\{\widetilde{x}'\widetilde{\alpha} > {\widetilde{x}^{*'}}\widetilde{\alpha}\right\} P(\widetilde{X}=\widetilde{x})P(\widetilde{X}=\widetilde{x}^*).
\end{equation}
We denote the maximizer of (\ref{MRCobj_limit2}) by $\mcA_{MRC},$

We note the similarity of objective 
(\ref{MRCobj_limit2}) to  
(\ref{MS:population}) in the way both 
objectives treat the cases of exact 
equality. While the maximum score objective
(\ref{MS:population}) drops the terms
where the choicen probability  $p(\widetilde{x})$ is exactly
$1/2$, objective (\ref{MRCobj_limit2}) drops the terms where
choice probabilities $p(\widetilde{x})$ are the same in 
two different points in the support of 
$\widetilde{X}.$ 

To help clarify our discussion in this section, we introduce a discrete-only illustrative design and use it throughout this section.  
\begin{definition}[Illustrative Design 3]
Take $K=2$ and let  $\widetilde{X}=(X_1,X_2)'$ with the support of $(X_1,X_2)'$ being $\{0,1\}^2$
  We take $\widetilde{\alpha}_0 \equiv (\alpha_0,1)$. Let $\epsilon$ be independent of $\widetilde{X}$.  
\end{definition}

In Illustrative Design 3,  $\mcA_0=\{\alpha_0\}$ whenever the parameter of the data generating process $\alpha_0 \in \{-1,0,  1\}.$ Each of the three cases is based on the equalities $p((1,1)')=p((0,0)')$ or $p((1,0)')=p((0,0)')$ or $p((1,0)')=p((0,1)')$,  respectively. For all other values of parameter $\alpha_0$ in the DGP  the model is partially identified. Namely, (a) $\mcA_0=(-\infty,-1)$ when $\alpha_0<-1$;  (b) $\mcA_0=(-1,0)$ when $-1<\alpha_0<0$;  (c) $\mcA_0=(0,1)$ when $0<\alpha_0<1$; (d) and $\mcA_0=(1,+\infty)$ when $\alpha_0>1$. 

If $\alpha_0 \not\in 
\{-1,0,1\}$ in the DGP,  then ${\mcA}_{MRC}=\mcA_0$. 
When $\alpha_0 \in 
\{-1,0,1\}$, then ${\mcA}_{MRC}$ is a superset of $\mcA_0$ as (\ref{MRCobj_limit2}) ignores the cases $p(\widetilde{x})=p(\widetilde{x}^*)$, $\widetilde{x} \neq \widetilde{x}^*$, which are certainly relevant in the definition and formation of the identified set (as explained above, in our design in these cases the identified sets are singletons).

Namely, $\mcA_{MRC}=(-\infty,0) \cap \mcA$ when $\alpha_0=-1$ (because $p((1,1)')=p((0,0)')$ is ignored by (\ref{MRCobj_limit2})), 
$\mcA_{MRC}=(-1,1) \cap \mcA$ when $\alpha_0=0$ (because $p((1,0)')=p((0,0)')$ is ignored by (\ref{MRCobj_limit2})),  and $\mcA_{MRC}=(0,+\infty)\cap \mcA$ when $\alpha_0=1$ (because $p((1,0)')=p((0,1)')$ is ignored by (\ref{MRCobj_limit2})). Thus, analogously to the maximum score case, the infeasible population MRC maximizer is not sharp when $\alpha_0$ in the DGP results in cases $p(\widetilde{x})=p(\widetilde{x}^*)$, $\widetilde{x} \neq \widetilde{x}^*$, and is sharp otherwise.  

Consequently, for the feasible MRC estimator $\widehat{\mcA}_{MRC}$ in Illustrative Design 3  we expect a fluctuating behavior whenever 
 $\alpha_0 \in \{-1,0,1\}$ in the DGP, which is completely analogous to what we had in the maximum score estimation. Specifically, we can establish that 
\begin{equation}\label{MRC:distribution:limit}
\widehat{\mcA}_{MRC}
\stackrel{d}{\longrightarrow}
\left\{
\begin{array}{ll}
B\cdot (-\infty,-1)+(1-B)\cdot(-1,0),\;\;
& \mbox{if}\;\;\alpha_0=-1,\\
B\cdot (-1,0)+(1-B)\cdot(0,1),\;\;
& \mbox{if}\;\;\alpha_0=0,\\
B\cdot (0,1)+(1-B)\cdot(1,+\infty),\;\;
& \mbox{if}\;\;\alpha_0=1,\\
\end{array}
\right.
\end{equation}
where $B$ is a Bernoulli random variable
with parameter $1/2$.

The similarity of the structure of the distribution limit (\ref{MRC:distribution:limit}) to that 
of the maximum score estimator, allows us 
to establish {\it robustness} of the estimator $\widehat{\mcA}_{MRC}.$ In particular, the selection of sequences
$\alpha(t,n;\alpha_0)=\alpha_0+t/\sqrt{n}$
for $\alpha_0 \in \{-1,0,1\}$ allows us
to establish that 
$$
\sup\limits_{t>0}\lim\limits_{n \rightarrow \infty}\PP \left(
d_H(\widehat{\mcA}_{MRC},\mcA^L)=0
\right)= 1\;\;\mbox{and}\;\;
\sup\limits_{t<0}\lim\limits_{n \rightarrow \infty}\PP \left(
d_H(\widehat{\mcA}_{MRC},\mcA^R)=0
\right) = 1,
$$
where $\mcA^L$ is the identified set whenever parameter of the data generating process takes values above $\alpha_0$
and $\mcA^R$ is the identified set whenever parameter of the data generating process takes values below $\alpha_0$
in some bounded neighborhood of $\alpha_0.$


Taking our discussion from Illustrative Design 3 to general cases, we will be able to conclude that $\widehat{A}_{MRC}$ will not have a probability limit but will have a weak limit representing a mixture of sets whenever there are cases $p(\widetilde{x})=p(\widetilde{x}^*)$, $\widetilde{x} \neq \widetilde{x}^*$, and will be sharp otherwise.

We next take ideas of the closed form estimation (inspired by \citeasnoun{ahnetal2018}) to our case of independent errors. It is based on similar principles as 
the maximizer of (\ref{Ich:objective})
 and relies
on equality restrictions in identification condition (\ref{SIeq}) as they are most important from the perspective of identifying power. Specifically, the estimator is defined as the set of maximizers of 
\begin{equation}\label{Ich:objective:MRC}
I_n(\alpha)=\frac{1}{n(n-1)}
\sum\limits_{i \neq j}w_{ij}((\widetilde{x}_i-\widetilde{x}_j)'\widetilde{\alpha})^2
\end{equation}
with $w_{ij}={\bf 1}\left\{
|\widehat{p}(\widetilde{x}_i)-\widehat{p}(\widetilde{x}_j)|<\nu_n
\right\}$ for the tuning parameter $\nu_n \rightarrow 0.$ If $w_{ij}$ for any pair $\widetilde{x}_i$ and $\widetilde{x}_j$, $\widetilde{x}_i \neq \widetilde{x}_j$, then the estimator is taken to be  the entire parameter space $\mcA$. 

We can establish that the maximizer of (\ref{Ich:objective:MRC})
is sharp only when $\alpha_0$ in the DGP results in cases of point identification (thus, there have to be ``enough'' cases (\ref{SIeq})), as is not sharp otherwise. As for robustness, we can establish that it is robust in the absence of equalities  (\ref{SIeq})) for distinct $\widetilde{x}, \widetilde{x}^*$, and is not robust otherwise.  E.g., in Illustrative Design 3, the CF estimator is  only when $\alpha_0
 \in \{-1,0,1\}$, and is only robust when $\alpha_0
 \notin \{-1,0,1\}$. In general, one can have cases when the closed form estimator is neither sharp nor robust. These results are completely analogous to those in our main model under the median independence.

 Similarly to  Section \ref{sec:new:estimators}, we can construct estimators based on combination ideas. We can combine the MRC and the closed form estimator directly based on whether $p(\widetilde{x})$ is close for two points in the support of the distribution of covariates $\widetilde{X}$, which will serve as a ``switching device''.  This estimator will have same drawbacks as the analogous estimator in Section \ref{sec:new:estimators} -- it will not be universally sharp (it will only be sharp for $\alpha_0$ in the DGP when either the MRC or closed form estimators are sharp) and will not be generally robust either. 
The second approach combines objective functions (\ref{MRCsample2})
 and (\ref{Ich:objective:MRC}) and considers 
 \begin{equation*}
\widehat\mcA_{OFC}= \arg \max _{\theta\in \Theta} 
\usum I[\hat{p}(\widetilde{x}_i)\geq \hat{p}(\widetilde{x}_j)-\nu_n]I[\widetilde{x}'_i \widetilde{\alpha} \geq \widetilde{x}'_j \widetilde{\alpha}]
\end{equation*}
for $\nu_n \to 0$, $n\nu_n^2 \to \infty$.  We can establish that this estimator is sharp. However, they it it not robust for $\alpha_0$ in the DGP that results in cases ${p}(\widetilde{x})={p}(\widetilde{x}^*)$ for $\widetilde{x} \neq \widetilde{x}^*$. 

 Finally, we also can construct the estimator
 based on the idea of the quantile of a 
 random set in \citeasnoun{molchanov2006book}. This estimator exploits the distribution limit
(\ref{MRC:distribution:limit}) of the 
estimator $\widehat{\mcA}_{MRC}$. E.g. in Illustrative design 3, when $\alpha_0 \in \{-1,0,1\}$, the MRC selects the sets above and below the true parameter of the data generating process with equal probabilities (see details in (\ref{MRC:distribution:limit})). As a result, selecting an estimator which for $\Delta>0$ outputs $\tau=1/2+\Delta$-quantile
of the closure of the random set $\widehat{\mcA}_{MRC},$ expressed as
$q_{\tau}\left(\overline{\widehat{\mcA}_{MRC}}\right),$ produces an estimator which
is sharp over the entire parameter space. 
Analogously to Section \ref{sec:quantile}, we can establish that universal sharpness extends to general cases,  and that the estimator is
universally robust too.

\subsection{Discrete choice models for panel data}\label{sec:panel}

\subsubsection{Static panel data model}
Panel setting extends the
cross-sectional model (\ref{discrete:choice}) by assuming the presence of the 
fixed effects with unknown distribution impacting the random utility while also 
allowing multiple observations over time available for each cross-sectional unit.
The version of this model where the error terms follow a logistic distribution with 
time-varying covariates have been analyzed in 
\citeasnoun{anderson70},\footnote{Interestingly, recent work demonstrates how crucial the logistic distribution assumption is for point identification and consistent estimation.
\citeasnoun{chamberlain2010} shows that when the observable covariates have bounded support, the logistic assumption on an unobserved component
is {\em necessary } for point identification.}  which proved that
a conditional maximum likelihood estimator consistently estimates the model
parameters up to scale without additional assumptions about the fixed effects. 

\citeasnoun{manski1987} considered the following semiparametric version of the model considered in 
\citeasnoun{anderson70}: 
\begin{equation}\label{discrete:choice:panel}
 Y_{it}={\bf 1}\left(c_i+{X}_{it}'\widetilde{\alpha}+\epsilon_{it}>0\right)
\end{equation}
where $i=1,2,...n$ are the cross-sectional units and $ t=1,2$ are the time periods. The binary variable $Y_{it}$ and the $K$-dimensional regressor vector $X_{it}$ are each observed and the parameter of 
interest is the $K$ dimensional vector $\widetilde{\alpha}$. The variables not observed in the data are $c_i$, and $\epsilon_{it}$, the former not varying with $t$ and often referred to as 
the ``fixed effect" or the individual specific effect. \citeasnoun{manski1987} imposed no  specific distributions  on unobservables. Under 
 scale normalization for $\widetilde{\alpha},$ unbounded support and continuity 
of distribution of at least one of the covariates, and only additionally maintaining the assumption that error terms
for each cross-sectional unit are known only to be time-stationary with unbounded
support, \citeasnoun{manski1987} established point-identification of coefficients 
$\widetilde{\alpha}.$

Our framework for model (\ref{discrete:choice}) directly extends to model
(\ref{discrete:choice:panel}) in which we consider the setting of all discrete regressors. We maintain Assumption \ref{assume:MS:panel} regarding the 
structure of the data generating process.

\begin{assumption}\label{assume:MS:panel}
\begin{itemize}
 \item[(i)] $\left\{\left((y_i \equiv (y_{i1},y_{i2}),x_i \equiv (x_{i1,x_{i2}}) \right) \right\}^{n}_{i=1}$ is an i.i.d. random sample from the joint distribution $(Y_{i}\equiv(Y_{i1},Y_{i2}),X_i \equiv 
 (X_{i1},X_{i2}))$ induced by (\ref{discrete:choice:panel}) for some $\widetilde{\alpha}_0 \in \mcA
  \subset \real^K.$
  \item[(ii)] Parameter space $\mcA$ is a compact subset of $\real^{K}$ such that for
  dimension $k \in \{1,2,\ldots,K\},$ $\widetilde{\alpha}^k \equiv 1$ for all $\widetilde{\alpha} \in \mcA.$
  \item[(iii)] Distribution of  ${X}_i$ is discrete with support $\mathcal{X}$.  The support of the random vector ${X}_{i2}-{X}_{i1}$ does not lie in any proper linear subspace of $\real^K$. 
  \item[(iv)] $F_{\epsilon_{i1}|{X}_i,c_i}(\cdot|\cdot)=F_{\epsilon_{i2}|{X}_i,c_i}(\cdot|\cdot)$ for all values in the support of $c_i$ and ${X}_i$.
\item[(v)] The support of $\epsilon_{it}\,\big|\,{X}_i,c_i$ is $\real$ with the cdf strictly increasing
at each support point.
\end{itemize}
\end{assumption}

Assumption \ref{assume:MS:panel} (iii) deviates from the continuity assumption of \citeasnoun{manski1987} and, ultimately, leads to a loss of point identification of
parameter $\widetilde{\alpha}.$ We illustrate below that, closely following the case of the cross-sectional
model (\ref{discrete:choice}) the identified set
can either be a singleton or a non-singleton convex
set.

The identified set for parameter $\widetilde{\alpha}_0$ is characterized as the set of solutions in $\mathcal{A}$ to  
\begin{align} 
P(Y_{i2}=1|{X}_i={x}_i)  \lessgtr P(Y_{i1}=1|{X}_i={x}_i) & \quad \Leftrightarrow \quad ({x}_{i2}-{x}_{i1})'\widetilde{\alpha} \lessgtr 0, \label{panel_idcond1}\\
P(Y_{i2}=1|{X}_i={x}_i)  = P(Y_{i1}=1|{X}_i={x}_i) & \quad \Leftrightarrow \quad ({x}_{i2}-{x}_{i1})'\widetilde{\alpha} = 0. \label{panel_idcond2}
\end{align}

\begin{definition}[Illustrative Design 4]
For expositional convenience, we
consider a simplified version of model (\ref{discrete:choice:panel}) in which 
${x}_{it} \in \mathcal{X}_0 \equiv \{(0,1)', (0,1)', (1,0)', (1,1)'\}$, $t=1,2$, $\widetilde{\alpha}=(1,\alpha).$
and  $(\varepsilon_{i1}, \varepsilon_{i2})$ is from $(c_i, {X}_i)$. 
\end{definition}

Suppose  in Illustrative Design 4  $\alpha_0=1$ in the DGP. When $x_{i1}=(0,1)'$ and  $x_{i2}=(1,0)'$ (or the other way around), we are in situation (\ref{panel_idcond2})
which leads us to the equalities identifying $\alpha$ (we would have the same equalities if we additionally conditioned on $c_i$) as they lead to 
$({x}_{i2}-{x}_{i1})'(1,\alpha)'=1-\alpha=0,$
thus meaning that the identified set is $\mathcal{A}_0=\{1\}$ (it is also consistent with inequalities (\ref{panel_idcond1}). If e.g. $\alpha_0=1/2$ in the DGP, then the only conditions defining the identified set are inequalities in the form of (\ref{panel_idcond1}) and we get $\mcA_0=(0,1)$. 


Similarly to our prior analysis we can evaluate the performance of the 
classic estimators for model (\ref{discrete:choice:panel}) under Assumption \ref{assume:MS:panel}.  We start with the classic maximum score estimator and express it for model  (\ref{discrete:choice:panel}) as the maximizer of the objective function of the 
{\it conditional maximum score} proposed by \citeasnoun{manski1987}: 
\begin{equation}\label{MS:objective:panel}
MS_n(\widetilde{\alpha})=    \avg (y_{i2}-y_{i1}){\sign}\left((x_{i2}-x_{i1})'\widetilde{\alpha} \right)|.
\end{equation}
The corresponding population objective function takes the form
\begin{equation}\label{MS:population:panel} \textstyle
MS(\widetilde{\alpha})=\sum\limits_{({x}_{i1},{x}_{i2}) \in \mathcal{X} } {\sign}(({x}_{i2}-{x}_{i1})'\widetilde{\alpha})\left(P\left(Y_{i2}=1
|{X}_i={x}_i\right) - P\left(Y_{i1}=1 | {X}_i={x}_i\right) \right) 
\end{equation}
We can see that the sum (\ref{MS:population:panel}) completely ignores (zeros out) cases when $x_{i2} \neq x_{i1}$ and  $P\left(Y_{i2}=1
|{X}_i={x}_i\right) = P\left(Y_{i1}=1 | {X}_i={x}_i\right)$, which, as seen from (\ref{panel_idcond1})-(\ref{panel_idcond2}), are essential in shaping the identified set.  
This already indicates an issue similar to that of the population maximum score objective function, which in its turn ignored or zeroed out cases of conditional choice probabilities equal to $1/2$. Thus, we conclude that the maximizer of  (\ref{MS:population:panel}) will in general be a superset of the identified set $\mcA_0$. 

In Illustrative Design 4, let us once again take $\alpha_0=1$ in the DGP. We note that  (\ref{MS:population:panel}) will have non-zero entries in the following points in the support of regressors: (i) ${x}_{i\tau}=(1,1)'$, ${x}_{i,\tau'}=(0,0)';$ 
(ii) ${x}_{i\tau}=(1,1)'$, ${x}_{i,\tau'}=(1,0)';$
(iii) ${x}_{i\tau}=(1,1)'$, ${x}_{i,\tau'}=(0,1)';$
(iv) ${x}_{i\tau}=(0,0)$, ${x}_{i,\tau'}=(1,0);$
(v) ${x}_{i\tau}=(0,0)$, ${x}_{i,\tau'}=(0,1),$  for $(\tau,\tau')=(1,2)$ or $(2,1).$
This objective function is maximized for $\widetilde{\alpha}=(1,\alpha)$ which satisfies inequalities: 
$\alpha>0,$ and $1+\alpha>0.$ 
As a result, the set of maximizers of (\ref{MS:population:panel}) (over $\mcA$) is  $(0,+\infty) \cap \mcA$ which is a superset of the identified set $\mcA_0=\{1\}$. In Appendix \ref{staticpanelproofs} we show that the maximizer of the sample objective function (\ref{MS:objective:panel}) for the simple design considered here, just like in 
case of the cross-sectional semiparametric discrete choice model, converges
to a random set when model (\ref{discrete:choice:panel})  under
Assumption \ref{assume:MS:panel} has cases $P\left(Y_{i2}=1
|{X}_i={x}_i\right) = P\left(Y_{i1}=1 | {X}_i={x}_i\right)$ for $x_{i2} \neq x_{i1}$. This means that 
the maximum score estimator maximizing (\ref{MS:objective:panel}) is not universally  sharp.

%


The two-step closed form estimator, similar to that constructed in \citeasnoun{ichimura1994}, relies exclusively on 
conditions (\ref{panel_idcond2}) for $x_{i2} \neq x_{i1}$,  
unlike the \citeasnoun{manski1987} maximum score estimator. To write its formal objective function, it is convenient to denote $\Delta p_{12}(x_i)=\EE\left[ Y_{i2}-Y_{i1} \,|\,X_i=x_i\right]$, and $\widehat{\Delta p_{12}}(x_i)$ denote its sample analogue. The objective function for the estimator is constructed
by selecting observations with small 
values of $\widehat{\Delta p_{12}}(x_i):$
\begin{equation}\label{Ich:objective:panel}
I_n(\alpha)=-\avg w_i\left((x_{i2}-x_{i1})' \widetilde{\alpha})^2 \right),
\end{equation}
where $w_i={\bf 1}\{|\widehat{\Delta p_{12}}(x_i)| <\nu_n\}$. Sequence $\nu_n \rightarrow 0$ is the tuning
parameter for this estimator. The only cases informative in (\ref{Ich:objective:panel}) for the parameter value $\widetilde{\alpha}$ will be those with $|\widehat{\Delta p_{12}}(x_i)| <\nu_n$ and $x_{i2}-x_{i1} \neq 0$. The default value of the 
estimator is the whole parameter space $\mathcal{a}$ if $|\widehat{\Delta p_{12}}(x_i)| <\nu_n$ either does not hold or only holds when $x_{i1}=x_{i2}$. The reliance of the maximizer
of (\ref{Ich:objective:panel}) on the near zero value of $|\widehat{\Delta p_{12}}(x_i)|$ for some $x_i$ in $\mathcal{X}$ demonstrates that for the semiparametric binary
choice model in the panel data settings we observe
the same behavior of the closed form estimator
as we did for the closed form estimator in the 
cross-sectional model in Section \ref{ichimura}. Namely, this estimator is not sharp universally on the parameter space and is  not generally robust.

We can proceed to construct 
new estimators based on combination ideas and analogous to those constructed for the 
cross-sectional setting in Section \ref{sec:new:estimators}. We can also  construct the 
random set quantile estimator similar to that in Section \ref{sec:quantile}.

The behavior of the maximum score estimator and the closed form estimator is
completely analogous to that in the cross sectional settings. Namely, neither the conditional maximum score nor 
the closed form estimator estimators is sharp, though 
the conditional maximum score estimator is robust. The estimator that combines the conditional maximum score and the closed form estimators directly using a switching device would be neither universally sharp nor robust. The second combination estimator would modify the maximum score objective function  (\ref{MS:objective:panel}) and consider 
\begin{multline*}\sum\limits_{({x}_{i1},{x}_{i2}) } \mathbf{1}(({x}_{i2}-{x}_{i1})'\widetilde{\alpha} \geq 0) \cdot \mathbf{1}\left(\widehat{P}\left(Y_{i2}=1
|{X}_i={x}_i\right) - \widehat{P}\left(Y_{i1}=1 | {X}_i={x}_i\right) \geq -\nu_n\right) 
 \\ 
+ \mathbf{1}(({x}_{i2}-{x}_{i1})'\widetilde{\alpha} \leq 0) \cdot \mathbf{1}\left(\widehat{P}\left(Y_{i2}=1
|{X}_i={x}_i\right) - \widehat{P}\left(Y_{i1}=1 | {X}_i={x}_i\right) \leq \nu_n\right),  \quad
\end{multline*}
where is the summation  is over $({x}_{i1},{x}_{i2})$ in the sample support, and $\nu_n \to 0$. Analogously to the cross-section case in Section \ref{sec:new:estimators} this combination estimator is universally sharp but not universally robust.  





Finally, the random set quantile estimator is both 
sharp and robust following from the convergence of the 
maximizer of the conditional maximum score objective
to a random set which is a mixture of deterministic sets each of whom contains  the identified set $\mcA_0$ on its boundary. 

\subsubsection{Dynamic panel data model}

An important extension of model (\ref{discrete:choice:panel}) adds the dependence
of the random utility on the lagged realizations of the 
discrete outcome. A simple form of that model 
has only dependence on one lag and takes the form
\begin{equation}\label{discrete:choice:panel:dynamic}
Y_{it}={\bf 1}\left\{c_i+  X_{it}'\widetilde{\alpha}+\gamma Y_{i,t-1} +\epsilon_{it}\geq 0 \right\},
\end{equation}
with $t=1,\ldots,T$ and $i=1,\ldots,n.$ Note that 
model (\ref{discrete:choice:panel:dynamic}) differs
from model (\ref{discrete:choice:panel}) by one additional term and, respectively, one additional parameter $\gamma$ which needs to be recovered
from the data. Just as in model (\ref{discrete:choice:panel}), the unobserved components are represented by  (a) a time invariant  fixed effect $c_i$ which  captures the systematic correlation of the unobservables over time; (b)    an idiosyncratic error term $\epsilon_{it}$ which is randomly sampled both over time and the cross-sectional units. The parameter $\gamma$ is of special interest as it measures the effect of state dependence in the model.

There is a rich literature in econometric theory which studies (\ref{discrete:choice:panel:dynamic}) and develops estimators for its parameters
under different conditions on the distribution of
covariates $X_{it},$ fixed effect $c_i$ and the random shock $\epsilon_{it}.$  
In this section we  focus on the setting considered in \citeasnoun{honorekyriazidou}, which proposed an estimator based on a conditional maximum score objective function\footnote{To our knowledge, this was the first paper to consider semiparametric identification and estimation of a dynamic binary choice panel data model.
While they focused exclusively on point identification, other recent work also considers partial identification as we do here. See for example \citeasnoun{kpt2022} and references therein,  and subsequently \citeasnoun{crz2023}, and \citeasnoun{gaowang2023}.
These other approaches do discuss sharpness but not robustness as we do here. 
There is also a very recent literature under {\em parametric} and serially independent assumptions on $\epsilon_{it}$, notably an i.i.d {\em logit} assumption - see \citeasnoun{kitazawa2022}, \citeasnoun{dobronyietal2023}, \citeasnoun{honoreweidner2023}, and references therein. These approaches attain moment conditions that rely  on the functional form of the logistic distribution.} The identification argument there relies on 
the presence of at least 3 time periods (with 4 periods of outcome observations) and the overlapping support of regressors such that one can find observations where those values are close in periods 2 and 3. Then one can verify the sign condition
similar to (\ref{panel_idcond1})-(\ref{panel_idcond2}), conditional on those close observations across periods.

In this section we consider the setting where assumption that regressors have continuous distribution used in \citeasnoun{honorekyriazidou} does not hold and 
instead use Assumption \ref{assume:MS:panel}
with the following modification.

\noindent {\bf Assumption \ref{assume:MS:panel}.} {\it
(i') $\left\{\left((y_i \equiv (y_{i0},y_{i1},y_{i2},y_{i3}),x_i \equiv (x_{i1,x_{i2}},x_{i3}) \right) \right\}^{n}_{i=1}$ is an i.i.d. random sample from the joint distribution $(Y_{i}\equiv(Y_{i0},Y_{i1},Y_{i2},Y_{i3}),X_i \equiv 
 (X_{i1},X_{i2},X_{i3}))$ induced by (\ref{discrete:choice:panel:dynamic}) for some $(\widetilde{\alpha}_0,\gamma_0) \in \mcA
  \subset \real^{K+1}$
  and some distribution of initial values $Y_{i0}.$ \vspace{.05in}\\
(iii') Distribution of regressor vector  ${X}_i$ is discrete with the support $\mathcal{X} \subset \real^{3K}$.  The support of  ${X}_{i2}-{X}_{i1}$ does not lie in any proper linear subspace of $\real^K$
and $\PP(X_{i2}=X_{i3})>0.$
  }

The  objective function in
\citeasnoun{honorekyriazidou} is 
\begin{equation}\label{MS:objective:panel:dynamic}
MS_n(\widetilde{\alpha},\gamma)=\avg
{\bf 1}\left\{x_{i2}=x_{i3}\right\}(y_{i2}-y_{i1})\sign
\left((x_{i2}-x_{i1})'\widetilde{\alpha}+\gamma\,(y_{i3}-y_{i0})\right).
\end{equation}
The corresponding population objective function can be written as
\begin{equation}\label{MS:objective:panel:dynamic:population}
\begin{array}{l}
MS(\widetilde{\alpha},\gamma)=\sum\limits_{{x_{i}
\in \mathcal{X}} \atop {y_{i3},y_{i0} \in \{0,1\}}} \mathbf{1} (x_{i2}=x_{i3}) \sign
\left((x_{i2}-x_{i1})'\widetilde{\alpha}+\gamma\,(y_{i3}-y_{i0})\right) \times \\
\hspace{0.1in}  \big(
P\left(Y_{i2}=1 \big|  X_i=x_i, Y_{i0}=y_{i0}, Y_{i3}=y_{i3}\right)
 -P\left(Y_{i1}=1 \big|  X_i=x_i, Y_{i0}=y_{i0}, Y_{i3}=y_{i3}\right) \big)
\end{array}
\end{equation}

We expositional convenience we will consider an simplified design given in Definition \ref{def:design5}. 
\begin{definition}[Illustrative Design 5] \label{def:design5}. Let $X_{it}$ be $K=2$-dimensional with support $\{0,1\}^2$. 
Let $\varepsilon_i= (\varepsilon_{i1}, \varepsilon_{i2}, \varepsilon_{i3})$ be independent of $(c_i, x_i)$. Let $\varepsilon_{it}$ be independent across time and have  strictly increasing c.d.f. Also, suppose the distribution of the initial outcome $Y_{i0}$ is independent of $(X_i, c_i, \varepsilon_i)$.  Consider normalization $\widetilde{\alpha}=(\alpha,1)'$. 
\end{definition}

In Illustrative Design 5  let  $\alpha_0=-1$ and $\gamma_0=-1$ in the DGP. We show that in this case the identified set is a singleton.  

To establish this, consider the following events defined by values $d_0,\,d_3 \in \{0,1\}$: 
\begin{align*} A(d_{0}, d_3) & = \{Y_{i0}=d_0,  Y_{i1}=0, Y_{i2}=1, Y_{i3}=d_3 \}, \\ 
B(d_{0}, d_3) &= \{Y_{i0}=d_0,  Y_{i1}=1, Y_{i2}=0, Y_{i3}=d_3 \}. 
\end{align*}
Then 
\begin{multline*} P(A(0, 1) | c_i, X_i=x_i, X_{i2}=X_{i3})  \\ 
= (1-F_{\varepsilon}(x_{i1}'\widetilde{\alpha} +c_i)) F_{\varepsilon}(x_{i2}'\widetilde{\alpha} + c_i)  F_{\varepsilon}(x_{i3}'\widetilde{\alpha} -1+c_i) P(Y_{i0}=0)
\end{multline*}
\begin{multline*} P(B(0, 1) | c_i, X_i=x_i, X_{i2}=X_{i3})  \\ 
= F_{\varepsilon}(x_{i1}'\widetilde{\alpha} +c_i)) (1-F_{\varepsilon}(x_{i2}'\widetilde{\alpha} -1+ c_i))  F_{\varepsilon}(x_{i3}'\widetilde{\alpha} +c_i) P(Y_{i0}=0).
\end{multline*}
If we consider values $x_{i1}= (0,0)'$ and $x_{i2}=x_{i3}=(0,1)'$, respectively, then 
$x_{i1}'\widetilde{\alpha}=0,$ $x_{i2}'\widetilde{\alpha}=1,$ and  $x_{i3}'\widetilde{\alpha}=1.$ As a result, 
$$\begin{array}{l}
\PP(A(0, 1) | c_i, X_{i1}=(0,0), X_{i2}=X_{i3}=(0,1)) \\
\hspace{2.5in}= \PP(B(0, 1) | c_i, X_{i1}=(0,0), X_{i2}=X_{i3}=(0,1)).
\end{array}
$$

This implies immediately the  identifiability restriction $(x_{i2}-x_{i1})'\widetilde{\alpha} + \gamma\,(y_{i3}-y_{i0})=0$  which takes the form 
$ 1+\gamma=0.$
Analogously, 
$$
\begin{array}{l}
\PP(A(0, 1) | c_i, X_{i1}=(1,1), X_{i2}=X_{i3}=(0,1)) \\
\hspace{2.5in}= \PP(B(0, 1) | c_i, X_{i1}=(1,1), X_{i2}=X_{i3}=(0,1)).
\end{array}
$$
This implies  the  identifiability restriction 
$ -\alpha+\gamma=0.$
The combination of equations
$1+\gamma=1$ and $-\alpha+\gamma=0$ allows us to  conclude that the identified set is a singleton $\{((-1,1,-1)'\}$.

We now construct the set of maximizers of the 
{\it population} objective function (\ref{MS:objective:panel:dynamic:population} 
under the normalization $\widetilde{\alpha}=(\alpha,1)'$. The elements in the sum in 
(\ref{MS:objective:panel:dynamic:population})
will be zero for $x_i$ such that  $x_{i2}=x_{i3}$ if  
$$\PP(A(d_0, d_3) | c_i, X_i=x_i) = \PP(B(d_0, d_3) | c_i, X_i=x_i).
$$
This is because for those values of covariates
$$
\begin{array}{l}
\PP(Y_{i2}-Y_{i1}=1 | c_i, X_i=x_i, Y_{i0}=d_0, Y_{i3}=d_3)\\
\hspace{2in}= \PP(Y_{i2}-Y_{i1}=-1 | c_i, X_i=x_i,  Y_{i0}=d_0, Y_{i3}=d_3),
\end{array}
$$
thus resulting 
in $$\EE\left[Y_{i2}-Y_{i1} \big|  X_i=x_i,  Y_{i0}=d_0, Y_{i3}=d_3\right]=0.$$

This is analogous to the behavior of the maximum score
objective function both in the cross sectional case (\ref{discrete:choice})
and in the case of static panel data model (\ref{discrete:choice:panel}) where population objective
function drops the terms that are most important  in shaping the identified set and will give point identification in some cases (such as in case $\alpha_0=1$, $\gamma_0=-1$ in Illustrative Design 5). 

To further simplify, we assume that fixed effect $c_i$ are binary. Then we have the following system of inequalities defining the  maxmimzer of the population objective function corresponding to :  $
\alpha+\gamma  <-1$,   $\alpha+\gamma  <1,$ 
$ \alpha-\gamma <1,$  $ -\alpha+\gamma<1.$ 
Removing redundant inequalities leaves three relevant inequalities
$\alpha+\gamma  <-1,$ 
 $\alpha-\gamma <1,$   $-\alpha+\gamma<1.$ 

\begin{figure}[tbp]
\centering
\caption{Maximizer of the population objective function (\ref{MS:objective:panel:dynamic:population}) in the dynamic panel data illustrative design}

\begin{tikzpicture}[scale=0.5]



\draw [very thick ] (0,-1) -- (-1,0);
\draw [very thick ] (-1,0) -- (-6,-5);
\draw [very thick ] (0,-1) -- (-6,-7);


\begin{scope}[on background layer]
\fill[blue!20] (-6,-5)--(-1,0)--(0,-1);
\end{scope}

\begin{scope}[on background layer]
\fill[blue!20] (-6,-5)--(0,-1)--(-6,-7);
\end{scope}




\node at (-1,-1) [circle,fill,inner sep=1.5pt, ]{};

\draw[very thick,-> ] (-6,0) -- (2,0) node[right] {$\alpha$};
\draw[very thick,-> ] (0,-7) -- (0,2) node[above] {$\gamma$};




\end{tikzpicture}
\label{fig_maximizer_dynamicpanel}
\end{figure}
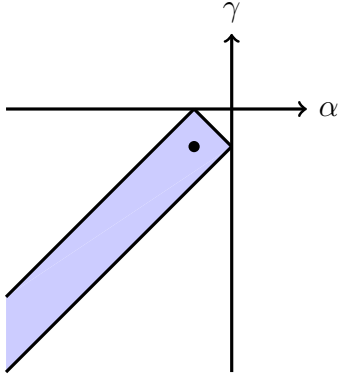

The solution set to these inequalities can be described as 
\begin{equation*} \left\{\lambda \cdot (a,g)' \big| \, a=t, g=-t -1, \; \; t \in [-1,0], \quad \lambda \geq 1\right\}.
\end{equation*}
It includes the point $(\alpha_0=-1,\gamma_0=-1)$ as an interior point (take $t=-1/2$, $\lambda=2$).  Thus, the maximizer of the \textit{population} Honore-Kyriazidou objective function is a (unbounded) superset of the identified set. This maximizer is displayed in Figure \ref{fig_maximizer_dynamicpanel}. 

The maximizer of the \textit{sample} objective function (\ref{MS:objective:panel:dynamic}),  analogously to the cross sectional case, will fluctuate among a finite number of disjoint sets with probabilities bounded away from zero and one. Each of such sets will have the true parameters $\alpha_0=-1$ and $\gamma_0=-1$ of the DGP at the boundary and the union of all such sets will give the described above maximizer of the population objective function (again, analogous to our illustrative example in the cross sectional case).

We can also construct the
closed form estimator by identifying points 
in the support of covariates which satisfy $x_{i2}=x_{i3}$ and for some fixed values of the outcomes 
in the initial and the last period gives equal conditional  probabilities of choice in periods 1 and 2: $P\left(Y_{i2}=1 \big|  X_i=x_i, Y_{i0}=y_{i0}, Y_{i3}=y_{i3}\right)
 =P\left(Y_{i1}=1 \big|  X_i=x_i, Y_{i0}=y_{i0}, Y_{i3}=y_{i3}\right)$. With discrete covariates, both of those probabilities
can be estimated as corresponding sample averages.
The parameters
can be estimated by minimizing with respect to $(\widetilde{\alpha},\gamma)$ the quadratic objective that consists of  terms 
$
\left((x_{i2}-x_{i1})'\widetilde{\alpha}+\gamma\,(y_{i3}-y_{i0})\right)^2
$
multiplied by weights that are non-zero only if $x_{i2}=x_{i3}$ and 
$\widehat{P}\left(Y_{it}=1 \big|  X_i=x_i, Y_{i0}=y_{i0}, Y_{i3}=y_{i3}\right), t=1,2$, are close to each other. As
before, the estimator would output the entire parameter
space if no points with properties described above are found.

Similar to previous models, we can construct new estimators based on combination of the output and
combination of objectives of the conditional
maximum score estimator and the closed-form two-step
estimator.

We can once again show that the random set quantile of the feasible estimator for quantile indices strictly above $1/2$ will give the true parameter value $\alpha_0=-1$ and $\gamma_0=-1$ for large enough sample sizes. 

All our estimation techniques extend from Illustrative Design 5 
 to a general case of discrete regressors.   Similar to our previous observation in Section 
\ref{sec:panel} these observations allow us to conclude
that the maximizer of the conditional maximum score
objective function (\ref{MS:objective:panel:dynamic})
will produce an estimator which is not sharp on the 
entire parameter space but robust. In contrast, the 
closed form estimator and the combination estimator that directly conditional maximum
score and closed form estimator  are neither sharp nor robust.
The estimator that combines the objective functions of the conditional maximum
score and closed form estimator is  sharp everywhere
on the parameter space but not robust.
Finally, the random set quantile estimator will be 
both sharp and robust on the entire parameter space.

\section{Conclusions}
\label{sec:conclusion}
In this paper we study semiparametric discrete choice models when covariates are discrete, violating the assumption of the continuity of distribution of regressors required to establish point identification. The parameters of the model are generally only partially identified. However, depending on the true value of the parameters of the data generating process the identified set can be a singleton or a non-singleton set. We focus on the question if this behavior of the identified set is accurately captured by a given estimator. We propose two criteria for evaluation of a given estimator. Sharpness of an estimator for a given true parameter of the data generating is the property where it converges in probability to the identified set corresponding to the true parameter value. Robustness of an estimator is the property of continuity of the distribution limit of an estimator with respect to local changes in the parameter of the data generating process. We explore classic estimators for discrete choice models including the maximum score estimator and the closed-form two-step estimator. We find that the maximum score estimator is not sharp everywhere on the parameter space, however, it is robust everywhere. In contrast, the closed form estimator is sharp only in parts of the parameter space where the model is point identified and is not robust at those points.

To construct estimators with better properties,  we consider a direct combination of the two classic estimators using a ``switching'' device as well a combination of their objective functions. The latter combination idea allows us to construct estimators which are sharp, and this is not true for the latter combination  idea. Both combination estimators fail to be robust.

We then propose a novel class of estimators based on the concept of a quantile of a random set. It uses the random set output by the maximum score estimator and produces an estimator which is both sharp and robust on the entire parameter space. We illustrate the performance of our novel estimator and compare it with classic estimators by analyzing the impact of the Brexit referendum vote on outcomes of the 2019 UK General Elections. Our new estimator both performs better and provides more meaningful results than the alternatives. We also show that our framework extends to other important settings including the maximum rank correlation estimator for the discrete choice model under independence as well as static and dynamic panel data models with discrete outcomes.

{\normalsize 
\bibliographystyle{econometrica}
\bibliography{overallkkn}
}

\appendix

\section{Appendix}

\setcounter{equation}{0}

\subsection{Proofs in Sections \ref{sec:classic} and \ref{sec:quantile}}\label{appendix:proofs}

\textbf{Properties of the infeasible maximum score estimator in Section \ref{MS:sharpness}} \\ Let us show that for all values of parameter
$\widetilde{\alpha}_0$ of the DGP, the infeasible estimator $\widehat{\cal A}_{ms, INF}$   converges in probability to the
maximizer ${\cal A}_{ms}$ of the population objective function
$MS(\alpha)$. Indeed, take any $\varepsilon>0$. Then $\widehat{\cal A}_{ms, INF}$   and ${\cal A}_{ms})$ are different 
only when some population support points are not in the sample support.  This means that
$$P(d_H(\widehat{\cal A}_{ms, INF}, {\cal A}_{ms}) ) \leq \sum_{\widetilde{x} \in \mathcal{X}} (1-P(\widetilde{X}=\widetilde{x}))^n \to 0$$
as $n \to \infty$ because $\mathcal{X}$ consists of a finite number of points. \vspace{.2in}\\
\textbf{\bf Identified set for Illustrative Design 1 }\\ 
Consider Illustrative Design 1.  The limiting maximum score objective function is \[\sum_{x \in \{0,1\}}  (p(x)-0.5)\cdot \sgn(\alpha+x) P(X=x),
\]
Let $\alpha_0=0$ in the DGP, in which case $p(0)=0.5$ and $p(1)>0.5$. Then $\mcA_0$ is obtained as the solution in $\mcA$ to the following relations: 
$$\alpha=0, \quad \alpha+1>0,$$
thus giving us $\mcA_0=\{0\}$. In contrast, the case $x=1$ is the only one that effectively enters the limiting objective function. Hence, ${\cal A}_{ms}$ is then given as the set of $\alpha \in \mcA$ that solve $\alpha+1\geq 0$. 

Case $\alpha_0=-1$ in the DGP is analogous: $p(0)<0.5$ and $p(1)=0.5$ with $x=0$ being the only one that effectively enters the limiting objective function. Hence, ${\cal A}_{ms}$ is then given as the set of $\alpha \in \mcA$ that solve $\alpha< 0$. On the other hand, $\mcA_0$ is the solution in $\mcA$ to the following two relations: 
$$\alpha+1=0, \quad \alpha<0,$$
thus giving $\mcA_0=\{-1\}$. 

Now take $\alpha_0 \notin \{0,1\}$ in the DGP. Then $p(0) \neq 0.5$ and $p(1)\neq 0.5$. Focusing on case $\alpha_0>0$ in the DGP, we obtain $p(0) > 0.5$ and $p(1) > 0.5$, thus ${\cal A}_{ms}$ is then given as the set of $\alpha \in \mcA$ that solve $\alpha+1\geq 0$, $\alpha\geq 0$, ultimately resulting in $[0,=\infty) \cap \mcA$. The identified set $\mcA_0$ is the solution to 
$$\alpha+1>0, \quad \alpha>0$$ 
thus giving $\mcA_0=(0,+\infty)$. 

Other cases of $\alpha_0 \notin \{0,1\}$ in the DGP can be considered analogously. 
$\blacksquare$ \vspace{.2in}\\
\textbf{Distribution limit of the maximum score estimator in Illustrative Design 1.}\\
Start by taking $\alpha_0=0$ in the DGP. We obtain $[0,+\infty) \cap \mathcal{A}$ as $\widehat{\cal A}_{ms}$ when $\hat{p}(1)>0.5$ and $\hat{p}(0)>0.5$. Set $\left[-1,0\right)\cap \mathcal{A}$ is obtained as $\widehat{\cal A}_{ms}$ when $\hat{p}(1)>0.5$ and $\hat{p}_0<0.5$. Both of these situations happen with probabilities approaching 0.5 as $n \rightarrow \infty$. 

Indeed, $\sqrt{n}(\hat{p}(0)-0.5, \hat{p}(1)-p(1))' \stackrel{d}{\to} \mathcal{N}((0,0)'. \Sigma)$ for a  p.d. $\Sigma$. Since $p(1)>0.5$, then $P(\hat{p}(1)>0.5) \to 1$ as $n \to \infty$. From the symmetry of the distribution of $\hat{p}(0)-0.5$ conditional on $\hat{p}(1)>0.5$. it is easy to conclude that 
$$P \left(\hat{p}(0) >0.5, \hat{p}(1) >0.5 \right)=P \left(\hat{p}(0)<0.5, \hat{p}(1)>0.5 \right).$$ 
The Gaussianity of the limit distribution also implies that $P \left(\hat{p}(0) =0.5, \hat{p}(1) >0.5 \right) \to 0$. All these facts then summarily give 
$$P \left(\hat{p}(0) >0.5,\hat{p}(1) >0.5 \right) \to 0.5, \quad  P \left(\hat{p}(0) <0.5, \hat{p}(1) >0.5 \right) \to 0.5$$
as $n \to \infty$. 


Other sets possible as realization of $\widehat{\cal A}_{ms}$ are $[-1,+\infty)\cap \mathcal{A}$ (when $\hat{p}(1) >0.5,  \hat{p}(0) =0.5$), or $[0,+\infty)\cap \mathcal{A}$ (when $\hat{p}(1) =0.5,  \hat{p}(0) >0.5$), or $(-\infty,0)\cap \mathcal{A}$ (when $\hat{p}(1) =0.5,  \hat{p}(0) <0.5$), or $\mathcal{A}$ (when $\hat{p}(1) =0.5,  \hat{p}(0) =0.5$), or $(-\infty,-1)\cap \mathcal{A}$ (when $\hat{p}(1) <0.5,  \hat{p}(0) \leq 0.5$). When $\hat{p}(1) <0.5,  \hat{p}(0) > 0.5$, then either $[0,+\infty)\cap \mathcal{A}$ or $(-\infty,-1)\cap \mathcal{A}$ can be the minimand. All these situations, however, occur
with probability approaching 0 as $n \to \infty$. 

For any $0<\varepsilon<1$, $P\left(d_H(\widehat{\cal A}_{ms}, {\cal A}_{ms}) > \varepsilon \right) \to 0.5$  
as $n \rightarrow \infty.$

Case $\alpha_0=-1$ in the DGP is analogous to $\alpha_0=0$: now 
$$P \left(\hat{p}(0) <0.5,\hat{p}(1) >0.5 \right) \to 0.5, \quad  P \left(\hat{p}(0) <0.5, \hat{p}(1) <0.5 \right) \to 0.5,$$
giving $[-1,0) \cap \mathcal{A}$ and $(-\infty,-1) \cap \mathcal{A}$ each with probability $1/2$ in the limit. 

Now consider $\alpha_0 \notin \{0,-1\}$. For convenience, take $\alpha_0>0$. Then 
$$P \left(\hat{p}(0) >0.5,\hat{p}(1) >0.5 \right) \to 1$$
as $n \to \infty$ thus giving $\widehat{\mcA}_{ms}$ as $[0,+\infty) \cap \mcA$ with probability approaching 1. Other cases within the $\alpha_0 \notin \{0,-1\}$ setting are considered analogously. $\blacksquare$ \vspace{.2in}\\
\textbf{Proof of Theorem \ref{th:MSgeneral}}. First, consider $\widetilde{\alpha}_0$ in the DGP such that  $p(\widetilde{x})\neq 1/2$, for all $\widetilde{x} \in \mathcal{X}$. Then
$P(\cap_{\widetilde{x} \in \mathcal{X}} \sign(\hat{p}(\widetilde{x})-0.5)= \sign({p}(\widetilde{x})-0.5)) \to 1$  
as $n \to \infty$, thus implying that w.p.a.1 $\widehat{\mcA}_{ms}$ collects all the $\widetilde{\alpha} \in \mcA$ that solve the following system of inequalities: 
\begin{align}
\label{ineqaux1}
\sign(\hat{p}(\widetilde{x})-0.5)=1 & \Rightarrow \widetilde{x}'\widetilde{\alpha} \geq 0 \notag \\
\sign(\hat{p}(\widetilde{x})-0.5)=-1 & \Rightarrow \widetilde{x}'\widetilde{\alpha} < 0.
\end{align}
Note that $\mcA_0$ itself collects 
all the $\widetilde{\alpha} \in \mcA$ that solve the system of inequalities where all the inequalities are strict: 
\begin{align}
\label{ineqaux2}
\sign({p}(\widetilde{x})-0.5)=1 & \Rightarrow \widetilde{x}'\widetilde{\alpha} > 0 \notag \\
\sign({p}(\widetilde{x})-0.5)=-1 & \Rightarrow \widetilde{x}'\widetilde{\alpha} < 0.
\end{align}
Since the model is well specified, then $\mcA_0$ is non-empty. The fact that it solves a system of strict inequalities implies that it has interior in $\real^{k-1}$ (after normalization).  The non-emptiness of $\mcA_0$ implies that having non-strict inequalities in (\ref{ineqaux1}) does not lead to the reduction of the dimension of the solution set5 (before it is intersected in $\mcA$) since assuming otherwise would have lead us to conclude that $\mcA_0$ is empty. Thus, the only difference between the solution set to system (\ref{ineqaux2}) and the solution set to system (\ref{ineqaux1}) is  that the former may include some points at the boundary of the convex polyhedron solving the latter. In any case, we ultimately obtain that the after being intersected with $\mcA$, the two sets may only differ at the boundary, hence, ultimately implying that the Hausdorff distance between these two sets is 0. To summarize, this means that  
$d_H\left( \widehat{\mcA}_{ms}, \mcA_0\right) =o_p(1)$.  Thus, the maximum score estimator in this case is asymptotically sharp. 

For $\widetilde{\alpha}_0$ in the DGP that results in $p(\widetilde{x})= 1/2$ for some $\widetilde{x} \in \mathcal{X}$ the statement will be implied by Theorem \ref{th:MSgeneral2} which gives a much more refined result for this case. $\blacksquare$ \vspace{.2in}\\
\textbf{Proof of Theorem \ref{th:MSgeneral2}. } There are many facts to be established here, so we can take it one step at a time.  

The sample support w.p.a.1 coincides with population support of $\widetilde{X}$. Without a loss of generality, let the first $M$, $M\geq 1$, points in $\cal{X}$ be those at the decision making boundary  $p(\widetilde{x}) = 1/2$ while the rest are not. Let us denote the collection of the first $M$ support points as $\mathcal{X}_{db}$. Of course, as explained earlier, the maximizer $\mathcal{A}_{ms}$ of the population maximum score objective function is purely defined by $\widetilde{x} \notin \mathcal{X}_{db}$ through inequalities 
\begin{align}
p(\widetilde{x})>1/2 & \quad \Rightarrow \quad \widetilde{x}'\widetilde{\alpha}\geq 0,     \label{ineq1} \\
  p(\widetilde{x})<1/2 & \quad \Rightarrow \quad \widetilde{x}'\widetilde{\alpha} < 0,
     \label{ineq2}
\end{align}    
for all $\widetilde{x} \notin \mathcal{X}_{db}$. Set $\mathcal{A}_{ms}$ is a superset of $\mcA_0$. 

We  first want to show an intermediate result that for all $\widetilde{x} \notin \mathcal{X}_{db}$, the  inequalities (\ref{ineq1}) and (\ref{ineq2}) defined as above also all hold w.p.a.1. This conclusion stems from the following two facts.  The first fact is  
$$P\left( \cap_{\widetilde{x} \in \cap \mathcal{X}\backslash \mathcal{X}_{db}} \sign(p(\widetilde{x})-1/2) \cdot \sign(\hat p(\widetilde{x})-1/2) >0\right) \stackrel{p}{\to} 1$$
meaning that for all $\widetilde{x} \notin \mathcal{X}_{db}$ simultaneously all $\hat p(\widetilde{x})$ are on the same side of 1/2 as their population  analogues w.p.a.1. The second fact is that \textit{any} combination of inequalities in the form $\widetilde{x}'\widetilde{\alpha}\geq 0$ or $\widetilde{x}'\widetilde{\alpha}< 0$ for all $\widetilde{x} \in \mathcal{X}_{db}$ that results in an non-empty set has a non-empty overlap with $\mcA_{ms}$ and, is thus, consistent with inequalities delivered by $\widetilde{x} \notin \mathcal{X}_{db}$, implying that these support points have a positive input in the maximization of the maximum score objective function w.p.a. 1. This establishes our first intermediate result which in particular implies that w.p.a.1 $\widehat{\mcA}_{ms}$ is contained in ${\mcA}_{ms}$.

Our second intermediate observation relies on the fact that 
\begin{equation}\label{CLT}\sqrt{n}(\hat p(\widetilde{x}_1)-p(\widetilde{x}_1), \ldots, p(\widetilde{x}_M)-p(\widetilde{x}_M))' \stackrel{d}{\to} \mathcal{N}(0, \Sigma)\end{equation}
for some p.d. $\Sigma$, which implies 
that 
$$\sum_{(s_1, \ldots, s_M \in \{+,-\}^M}
P\left( \cap_{\widetilde{x} \in \mathcal{X}_{db}} s_m \sign(p(\widetilde{x})-1/2)  >0\right) \stackrel{p}{\to} 1.$$

This means that any combination of inequalities $\hat{p}(\widetilde{x})-1/2>0$ or  $\hat{p}(\widetilde{x})-1/2<0$ inequalities for $\widetilde{x} \in \mathcal{X}_{db}$ happens with  positive probability asymptototically (a combination  has to include an inequality for each $\widetilde{x} \in \mathcal{X}_{db}$). 
Each this case results in a maximum score estimate $\mcC_{\ell}$ which is a subset of $\mcA_{ms}$ (given  our first intermediate result above and, thus, maintaining that for all $\widetilde{x} \notin \mathcal{X}_{db}$, the  inequalities (\ref{ineq1}) and (\ref{ineq2}) all hold w.p.a.1). Different combinations of signs $(s_1,\ldots, s_M)$ result either in disjoint or identical maximum score estimates $\mcC_{\ell}$. Indeed,  for different collections of signs $(s_1,\ldots, s_M)$ and $(s^*_1,\ldots, s^*_M)$ the sample maximum score objective function may happen to be optimized either at the same collection of inequalities from 
\begin{equation}
\label{coll_ineqs} 
\widetilde{x}'\widetilde{\alpha}\geq 0 \text{ or } \widetilde{x}'\widetilde{\alpha}< 0, \quad \widetilde{x} \in \mathcal{X}_{db}
\end{equation}
intersected with $\mcA_{ms}$,  or different 
collections of inequalities intersected with $\mcA_{ms}$ (by different we mean that 
for at least one $m=1, \ldots, M$, one collection has $\widetilde{x}_m'\widetilde{\alpha}\geq 0$ whereas the other collection has $\widetilde{x}_m'\widetilde{\alpha} < 0$). By construction, different collections result in two disjoint estimates $\mcC_{\ell}$ and $\mcC_{\ell^*}$. 

Now let us show that in the setting of this theorem there are at least two disjoint estimates $\mcC_{\ell}$ and $\mcC_{\ell^*}$. Indeed, since $M \geq 1$, then there are at least two different collections of inequalities (\ref{coll_ineqs}) that have non-empty solutions. For concreteness, suppose that we have a collection of inequalities 
\begin{equation} 
\label{ineq3}\widetilde{x}'\widetilde{\alpha}\geq 0 , \quad \widetilde{x} \in \mathcal{X}_{db}
\end{equation}
that has a non-empty solution and, thus, a non-empty intersection with $\mathcal{A}_{ms}$. This intersection  would be a maximum score estimate for the collection of signs $s_1=\ldots=s_M=+$ (that is, when all $\hat{p}(\widetilde{x})-1/2>0$ for all $\widetilde{x} \in \mathcal{X}_{db}$). 

\begin{itemize}
\item[(a)] If the solution to (\ref{ineq3}) has an interior in $\real^{K-1}$, then it is a convex polyhedron and we can choose $\widetilde{x} \in \mathcal{X}_{db}$ whose inequality $\widetilde{x}'\widetilde{\alpha} \geq 0$ forms the polyhedron's edge with this edge not being at the boundary of $\mcA_{ms}$. Suppose such $\widetilde{x}$ happened to be $\widetilde{x}_1$. Then for the collection of signs $s_1=-, s_2=\ldots=s_M=+$ the system of inequalities 
\begin{equation*} 
\widetilde{x}_1'\widetilde{\alpha} < 0 , \quad \widetilde{x}_m '\widetilde{\alpha} \geq  0, \quad m=2, \ldots, M,
\end{equation*}
will have a non-empty solution too, which, of course will once again intersect with $\mcA_{ms}$, This intersection will be a maximum score estimate in a sample where 
 $\hat{p}(\widetilde{x}_1)-1/2<0$ but $\hat{p}(\widetilde{x}_m)-1/2>0$, $m=2, \ldots, M$. 
 
\item[(b)] Now suppose that the solution to (\ref{ineq3}) does not have an interior even though it is non-empty. This means that there is  $\widetilde{x}_{m}$ and  collection of $\gamma_{\ell} \geq 0$, $\ell=1, \ldots, M$, $\ell \neq m$, such that $$\widetilde{x}_m =-\sum_{\ell \neq m} \gamma_{\ell} \widetilde{x}_{\ell}$$ 
(and $\gamma_{\ell}>0$ for at least one $\ell \neq m$). Reversing the inequality for $\widetilde{x}_m$ to 
$$\widetilde{x}_m' \widetilde{\alpha}<0$$ which corresponds to the change in sign $s_m$ to $-$ and, thus, to the change to a realization $\hat{p}(\widetilde{x}_m)-1/2<0$ then will result
in a non-empty solution and, thus, when intersected with $\mcA_{ms}$ yet another  maximum score  estimate corresponding to a set of realizations of $\hat{p}(\widetilde{x}_m)-1/2<0$ and $\hat{p}(\widetilde{x}_{\ell})-1/2>0$ for $\ell \neq m$. If this non-empty solution has an interior then we proceed as in (a). It this non-empty solution does not an interior, then we conduct another round of revision of inequalities as in (b), etc. until we get to the point of a solution with a non-empty interior.   

\end{itemize} 

Thus, at this stage we have shown that in the setting of this theorem there are at least two disjoint maximum score estimates $\mcC_{\ell}$ and $\mcC_{\ell^*}$ occurring with a probability bounded away from zero asymptotically, which allows us to conclude that the weak limit is  indeed random. 

There are several other things we need to establish. 

One  of them is establishing that every maximum score estimate $\mcC_{\ell}$ obtained as an intersection of  a solution to a combination of inequalities (\ref{ineq3}) (a combination  has to include an inequality for each $\widetilde{x} \in \mathcal{X}_{db}$) contains  $\mcA_0$ at its boundary.  Let us fix a specific $\mcC_{\ell}$. Some of the inequalities among (\ref{ineq3}) that are shaping $\mcC_{\ell}$ must be relevant (that is, removing them changes the set) and some may be irrelevant (removing them does not change a set). Let $\mathcal{X}_{db}(\mcC_{\ell})$ denote the subset of $\mathcal{X}_{db}$ that contains $\widetilde{x}$ that give relevant inequalities for $\mcC_{\ell}$. The boundary of $\mcC_{\ell}$ consists of the following two components: (a) the boundary of $\mcA_{ms}$ intersected with the closure $\overline{\mcC}_{\ell}$; (b) the solution to $\widetilde{x}'\widetilde{\alpha}=0$ that should for all $\widetilde{x} \in \mathcal{X}_{db}(\mcC_{\ell})$. The second component obviously contains $\mcA_0$. 

The second remaining thing is establishing  that the intersection of $[L/2]+1$ of sets $\overline{\mcC}_{\ell}$ gives us $\overline{\mathcal{A}}_0$. 

First, consider the case when some of the points of ${\mathcal{A}}_0$ are in the interior of $\mcA_{ms}$. Then for each $\widetilde{x}_m \in \mathcal{X}_{db}$ the hyperplane $\widetilde{x}_m' \widetilde{\alpha}=0$ splits $\mcA_{ms}$ on is own into two subsets with each subset having an interior. This implies that given the partitioning created by all the other hyperplanes $\widetilde{x}' \widetilde{\alpha}=0$, $\widetilde{x} \neq \widetilde{x}_m$, adding the last hyperplane  results in one of the following two situations. 
\begin{itemize}
\item[(i)] The hyperplane passes through interiors of sets in the already existing partitioning thus determining the final partitioning (that is, the final collection of sets $\{\mcC_{\ell}\}$) and ensuring that half of the sets in the final partitioning have $\widetilde{x}_m'\widetilde{\alpha} \geq 0$ whereas the other half have $\widetilde{x}_m'\widetilde{\alpha} <0$.  
\item[(ii)]\footnote{Note that situation (ii) does not happen if there is a constant term among covariates.} The hyperplane goes only through the boundary of subsets in the already existing partitioning. This means that $\widetilde{x}_m$ coincides with another $\widetilde{x}_h \in \mathcal{X}_{db}$ up to a scalar. Then $\widetilde{x}_m$ either does not modify partitioning if $\widetilde{x}_m =c \widetilde{x}_h$ for some $c>0$, $c\neq1$, $\widetilde{x}_h \in \mathcal{X}_{db}$. Otherwise (that is, if $\widetilde{x}_m =c \widetilde{x}_h$ for $\widetilde{x}_h \in \mathcal{X}_{db}$ only happens for $c<0$), it creates further partitioning by carving out $\widetilde{x}_m'\widetilde{\alpha}=0$ in the prior partitioning.  This ensures that at least half of the sets in the final partitioning have $\widetilde{x}_m'\widetilde{\alpha} \geq 0$ and at least half of the sets in the final partitioning have $\widetilde{x}_m'\widetilde{\alpha} \leq 0$.  
\end{itemize}
To summarize, for each $\widetilde{x}_m \in \mathcal{X}_{db}$  there is guaranteed to be $[L/2]$ sets among $\{\mcC_{\ell}\}$ that satisfy inequality $\widetilde{x}_m'\widetilde{\alpha} \geq 0$ and there is guaranteed to be $[L/2]$ sets among $\{\mcC_{\ell}\}$ that satisfy inequality $\widetilde{x}_m'\widetilde{\alpha} \leq 0$. This ensures that the intersection of any $[L/2]+1$ sets $\{\overline{\mcC}_{\ell}\}$ consists only of the set of $\widetilde{\alpha}$ such that 
$$\widetilde{x}'\widetilde{\alpha} = 0 \quad \text{ for all } \widetilde{x} \in \mathcal{X}_{db}$$
intersected with the closure of $\overline{\mcA}_{ms}$. This gives the closure $\overline{\mcA}_{0}$ of the identified set. 

Now, consider the case when all of the points of ${\mathcal{A}}_0$ are at the boundary of $\mcA_{ms}$ (note that $\mcA_{ms}$ will  include some of its boundary points but does not necessarily contain all of its boundary -- generally it will be neither closed nor open).  Then the above arguments apply analogously to this situation with the only modification of everything being considered as projected on the relevant part of the boundary of $\mcA_{ms}$ (since $\mcA_{ms}$ is polyhedron, the relevant part of the boundary belongs to a hyperplane). $\blacksquare$ \vspace{.2in}\\
\textbf{Proof of Theorem  \ref{th:MSrobustness}.}
Let $\alpha(n,t;\alpha_0)=\alpha_0+t/\sqrt{n} \in 
\mcC_k$ with $\alpha_0$ being the point in $\mcA$
such that $p(\bar{x}_m)=\frac12$ for $M$ points $\bar{x}_m \in {\mathcal X}$ on the decision making
boundary, as in the proof of Theorem \ref{th:MSgeneral2} whenever $\alpha_0$ is the 
true parameter of the data generating process. 
Let $p^{(n,t)}(\bar{x}_m)$ denote the probability $P(Y=1\,| \widetilde{X}=\bar{x}_m)$ for the data generating process
corresponding to parameter $\alpha(n,t;\alpha_0).$ Given that the conditional density of $\epsilon$ is bounded away
from zero, then
$$
p^{(n,t)}(\bar{x}_m)=\frac12+\frac{1}{\sqrt{n}}
f_{\epsilon|\widetilde{X}}(0\,|\,\bar{x}_m)\bar{x}_m't+o(n_{-1/2}).
$$
Construct the $M \times K$ matrix
$$
\Pi(\alpha_0)=\left(f_{\epsilon|\widetilde{X}}(0|\bar{x}_1)\bar{x}_1,
\ldots,f_{\epsilon|\widetilde{X}}(0\,|\,\bar{x}_M)\bar{x}_M
\right)'.
$$
This means that we can express convergence in distribution
(\ref{CLT}) under the data generating process
indexed by $\alpha(n,t;\alpha_0)$ as
\begin{equation}\label{CLT:drift}\sqrt{n}(\hat p(\widetilde{x}_1)-\frac12), \ldots, p(\widetilde{x}_M)-\frac12)' \stackrel{d}{\to} \mathcal{N}\left(\Pi(\alpha_0)\,t, \Sigma\right).\end{equation}

This means that whenever $t \rightarrow 0,$ then 
the mean of distribution limit approaches 0 and
the distribution coincides with (\ref{CLT}). This means that the distribution of the maximizer
of (\ref{MS:objective}) coinsides with that 
in Theorem \ref{th:MSgeneral2}.

Since $\alpha(n,t;\alpha_0) \in \mcC_k,$ then
a collection of inequalities
$$
\bar{x}'\widetilde{\alpha}(n,t;\alpha_0) \geq 0\;\;
\mbox{or}\;\;\bar{x}'\widetilde{\alpha}(n,t;\alpha_0) < 0,\;\;\bar{x} \in {\mathcal X}_{db}
$$
hold with $\widetilde{\alpha}(n,t;\alpha_0)$
containing $\alpha(n,t;\alpha_0)$ and a normalized
$k$-th component. Since $p(\bar{x})=\frac12$
under $\alpha_0$ for each $\bar{x} \in {\mathcal X}_{db},$
which means that, respectively,
$$
\bar{x}'t \geq 0\;\;
\mbox{or}\;\;\bar{x}'t < 0,\;\;\bar{x} \in {\mathcal X}_{db}
$$
for a given chosen $t.$ Thus, as $n \rightarrow \infty,$ each $\hat{p}(\bar{x}) \lessgtr \frac12$ for each $\bar{x} \in 
{\mathcal X}_{db}.$ However, given that 
set $\mcC_k$ is defined by the set of inequalities
$$
\bar{x}'\widetilde{\alpha} \geq 0\;\;
\mbox{or}\;\;\bar{x}'\widetilde{\alpha} < 0,\;\;\bar{x} \in {\mathcal X}_{db},
$$
then with probability approaching 1 the maximizer
of (\ref{MS:objective}) approaches the set $\mcC_k.$
In other words,
$$
\PP(B_k(t)=1) \rightarrow 1,\;\;\mbox{as}\;\;n \rightarrow \infty,
$$
where $B_k(t)$ is the dummy variable equal to 1
if $\mcC_k$ is the maximizer of (\ref{MS:objective}) in a given sample of size $n.$
$\blacksquare$ \vspace{.2in}\\
\textbf{Proof of Theorem \ref{th:closedformPOPgeneral}.} 
For a given $\alpha_0 \in \mcA$ in the DGP let $M$ denote the number of  points in $\mathcal{X}$ such that $P(\widetilde{x})=1/2$. WLOG, suppose these are the first $M$ points $\widetilde{x}_1$, \ldots, $\widetilde{x}_M$ in $\mathcal{X}$ (of course, it is possible for $M$ to be 0 in case of no points at the decision boundary).  

First, consider $\alpha_0$ in the DGP such that the model is point identified. This means that under such $\alpha_0$ we have $M\geq K-1$  and the respective system 
$$\widetilde{x}_{\ell}'\widetilde{\alpha}=0, \quad \ell=1, \ldots, M,$$
has a unique solution.  This unique solution has to be $\alpha_0$  (otherwise we would get a contradiction with in the definition of the identified set). Hence, $\mcA_{CF}=\mcA_0$.   

Second, consider $\alpha_0$ in the DGP such that the model is not point identified. This means that under such $\alpha_0$ the system 
$$\widetilde{x}_{\ell}'\widetilde{\alpha}=0, \quad \ell=1, \ldots, L,$$
has multiple solutions. The solution system is a non-singleton  convex set.  Since the identified set in addition to the system of equations above is formed by inequalities 
$$\widetilde{x}'\widetilde{\alpha}>0 \quad \text{if } p(\widetilde{x})>1/2,$$
$$\widetilde{x}'\widetilde{\alpha}<0 \quad \text{if } p(\widetilde{x})<1/2,$$
then in general the non-singleton set defined by just equations is a superset of the identified set. Hence, generally in this case $\mcA_{CF} \supseteq \mcA_0$. $\blacksquare$ \vspace{.2in}\\
\textbf{Proof of Theorem \ref{th:closedformtoy1}}. 
This result relies on the limit
$$(\hat{p}(\widetilde{x}_1)-{p}(\widetilde{x}_1), \ldots, \hat{p}(\widetilde{x}_{|\mathcal{X}|})-{p}(\widetilde{x}_{|\mathcal{X}|}))' \stackrel{d}{\rightarrow} \mathcal{N}(0,\Sigma)$$ 
for some p.d. $\Sigma$ (here we use some ordering of $|\mathcal{X}|$ components in   $\mathcal{X}$. Suppose the first $M$ points in $\mathcal{X}$  are such that $p(\widetilde{x})=1/2$ ($M$ can be any between 0 and $|\mathcal{X}|$). 

Then it is easy to establish that for each $\ell=1, \ldots, M$, 
$$P(|\hat{p}(\widetilde{x}_{\ell})-1/2|< h_n) = P(\sqrt{n}|\hat{p}(\widetilde{x}_{\ell})-1/2|< \sqrt{n}h_n) \to 1$$
as $n\,h_n^2 \to \infty$. At the same time,  for each $\ell=M+1, \ldots, |\mathcal{X}|$, 
$$P(|\hat{p}(\widetilde{x}_{\ell})-1/2| \geq  h_n)  \geq P(\sqrt{n}|{p}(\widetilde{x}_{\ell})-1/2| -\sqrt{n}h_n \geq  \sqrt{n}|\hat{p}(\widetilde{x}_{\ell})-{p}(\widetilde{x}_{\ell})|) \to 1$$
because $|{p}(\widetilde{x}_{\ell})-1/2|>0$ and $h_n \to 0$ imply that $\sqrt{n}|{p}(\widetilde{x}_{\ell})-1/2| -\sqrt{n}h_n \to +\infty$ as $n \to \infty$. 

Taking into account that $P(C_n \cap D_n) \to 1$ if $P(C_n) \to 1$ and $P(D_n) \to 1$ as $n \to \infty$, we can now conclude that 
$$P\left( \max_{\ell =1, \ldots, M} |\hat{p}(\widetilde{x}_{\ell})-1/2|< h_n, \min_{\ell =M+1, \ldots, |\mathcal{X}|} |\hat{p}(\widetilde{x}_{\ell})-1/2|\geq h_n\right) \to 1 \quad \text{as } n \to \infty.$$

This implies that w.p.a. 1 $\widehat{\mcA}_{CF}$ solves the same system of equations as $\mcA_{CF}$. This gives $P(\widehat{\mcA}_{CF} \neq \mcA_{CF}) \to 0$ which immediately implies that $d_H(\widehat{\mcA}_{CF}, \mcA_{CF}) \stackrel{p}{\to} 0$. $\blacksquare$ \vspace{.2in}\\
\textbf{Proof of Theorem \ref{th:drfitCF}.} 
As we established in (\ref{CLT:drift}) for 
the sequence of parameters of the 
data generating process $\alpha(n,t;\alpha_0)$: $$\sqrt{n}(\hat p(\widetilde{x}_1)-\frac12, \ldots, p(\widetilde{x}_M)-\frac12)' \stackrel{d}{\to} \mathcal{N}\left(\Pi(\alpha_0)\,t, \Sigma\right).$$
Then for $M$ points in $\widetilde{x}_{\ell} \in {\mathcal X}$ where
$p(\widetilde{x}_{\ell})=\frac12$ under the parameter of the data generating process $\alpha_0:$
$$
\PP(|\hat{p}(\widetilde{x}_{\ell})-1/2|< h_n) \geq  P(\sqrt{n}|\hat{p}(\widetilde{x}_{\ell})-p(\widetilde{x}_{\ell})|< \sqrt{n}h_n+(\Pi(\alpha_0)\,t)_{\ell}) \to 1,
$$
since $n^2h_n \rightarrow \infty.$
Similarly, for all points $\widetilde{x} \in {\mathcal X}$
where $p(\widetilde{x}) \neq \frac12:$
$$
\PP(|\hat{p}(\widetilde{x})-1/2| \geq  h_n) \geq  P(\sqrt{n}|\hat{p}(\widetilde{x})-1/2|+ \sqrt{n}h_n \geq \sqrt{n}(\hat{p}(\bar{x})-p(\bar{x}))) \to 1,
$$
since drifting does not affect the distribution
limit of the conditional probability of the outcome
for the points not on the decision boundary.
$\blacksquare$ \vspace{.2in}\\
\textbf{Proof of Theorem \ref{th:hybrid_gen}.} Let $\alpha_0$ in the DGP be such that the maximum score estimator is sharp. This means that $p(\widetilde{x})\neq 1/2$ for any $\widetilde{x} \in \mathcal{X}$, which in its turn implies that 
$$P(|\hat{p}(\widetilde{x}) - 1/2| \geq \nu_n) \geq P(\sqrt{n}|{p}(\widetilde{x})-1/2| -\sqrt{n}\nu_n \geq  \sqrt{n}|\hat{p}(\widetilde{x})-{p}(\widetilde{x})|) \to 1$$
because $|{p}(\widetilde{x})-1/2|>0$ and $\nu_n \to 0$ imply that $\sqrt{n}|{p}(\widetilde{x})-1/2| -\sqrt{n}\nu_n \to +\infty$ as $n \to \infty$. Taking into account that $P(C_n \cap D_n) \to 1$ if $P(C_n) \to 1$ and $P(D_n) \to 1$ as $n \to \infty$, we can now easily conclude that 
$P\left( \min_{ \widetilde{x} \in \mathcal{X}} |\hat{p}(\widetilde{x})-1/2|\geq \nu_n\right) \to 1$ as $n \to \infty.$
Then w.p.a.1 $\widehat{\mcA}_{H} = \widehat{\mcA}_{ms}$ and, thus, given the sharpness of the maximum score estimator, 
$$d_H(\widehat{\mcA}_{H},\mcA_0) = d_H(\widehat{\mcA}_{ms}, \mcA_0) + o_p(1) =o_p(1).$$

Let $\alpha_0$ in the DGP be such that the closed form estimator is sharp (necessarily, this means that $\mcA_0=\{\alpha_0\}$ and $d_H(\widehat{\mcA}_{CF}, \mcA_0) \stackrel{p}{\to} 0$). It implies then there are  enough $\widetilde{x} \in \mathcal{X}$ such that $p(\widetilde{x})=1/2$ to identify ${\alpha}_0$ from the system 
$$\widetilde{x}'\widetilde{\alpha} =0 \quad \text{ for } \widetilde{x} \in \mathcal{X} \text{ s.t. } p(\widetilde{x})=1/2.$$ 
In particular, this means that there is at least one probability of choice at the decision boundary and that whenever $ p(\widetilde{x})=1/2$, 
$$P(|\hat{p}(\widetilde{x})-1/2|<\nu_n)
=P(\sqrt{n}|\hat{p}(\widetilde{x})-1/2|< \sqrt{n}\nu_n) \to 1$$
as $n\nu_n^2 \to \infty$. This immediately gives  
$P\left( \min_{ \widetilde{x} \in \mathcal{X}} |\hat{p}(\widetilde{x})-1/2|< \nu_n\right) \to 1$ as $n \to \infty$, 
which in its turn gives
$d_H(\widehat{\mcA}_H, \mcA_0) = d_H(\widehat{\mcA}_{CF}, \mcA_0)+o_p(1) = o_p(1)$. 

Now suppose that neither the maximum score nor the closed form estimator is sharp. This means that there are  $\widetilde{x}\in \mathcal{X}$ such that $p(\widetilde{x})=1/2$ (hence, the maximum score is not sharp) but it is not enough of them to identify the parameter in the DGP (hence, the closed form estimator is sharp). By the argument above we can establish 
that 
$P\left( \min_{ \widetilde{x} \in \mathcal{X}} |\hat{p}(\widetilde{x})-1/2|\geq \nu_n\right) \to 1$ as $n \to \infty$. Hence, 
$d_H(\widehat{\mcA}_H, \mcA_0) = d_H(\widehat{\mcA}_{CF}, \mcA_0)+o_p(1).$
Since $d_H(\widehat{\mcA}_{CF}, \mcA_0) \neq o_p(1)$ dues to the closed form estimator not being sharp, we conclude $d_H(\widehat{\mcA}_H, \mcA_0) \neq o_p(1)$. $\blacksquare$ \vspace{.2in}\\
\textbf{Proof of Theorem \ref{th:OFCsharp}.} 
For a given $\alpha_0$ in the DGP, let $M$ denote the number of support points such that $p(\widetilde{x})=1/2$,  $0 \leq M \leq |\mathcal{X}|$. Without a loss of generality, these are the first $M$ points in $\mathcal{X}$. Using the same techniques as in the proof of Theorem  \ref{th:hybrid_gen}, we can establish that 
(a) $P(\min_{\ell=1, \dots, M} |\hat{p}(\widetilde{x}_{\ell})-1/2|\leq \nu_n) \to 1$; (b) for every $\widetilde{x} \in \mathcal{X}$ such that $p(\widetilde{x})>1/2$, it holds that $P(\hat{p}(\widetilde{x})-1/2>0) \to 1$; (c)  for every $\widetilde{x} \in \mathcal{X}$ such that $p(\widetilde{x})<1/2$, it holds that $P(\hat{p}(\widetilde{x})-1/2<0) \to 1$, as $n \to \infty$. Then with probability approaching 1, $\widehat{\mcA}_{OFC}$ solves the following system in $\mcA$: 
\begin{align*}
\widetilde{x}_{\ell}' \widetilde{\alpha} &=0, \quad \ell=1, \ldots, M, \\
\widetilde{x}_{\ell}' \widetilde{\alpha} &>0, \quad \text{if} \; \; {p}(\widetilde{x})>1/2, \\
\widetilde{x}_{\ell}' \widetilde{\alpha} &<0, \quad \text{if} \; \; {p}(\widetilde{x})<1/2,
\end{align*}
which is exactly the definition of $\mcA_0$. Thus, $d_H(\widehat{\mcA}_{OFC}, \mcA_0) = o_p(1)$. $\blacksquare$

Thus, in the construction of $\hat \alpha_{CF}$  observations with $x_i=0$ are the only ones taken into account with probability approaching 1. This means that $\hat \alpha_{CF,n}$ will be zero with probability approaching 1 ultimately implying that  $d_H(\hat \alpha_{CF,n}, {\cal A}_{0, \alpha_n}) \stackrel{p}{\nrightarrow} 0$. 
$\blacksquare$


\vskip 0.1in 

\textit{Proof of Theorem \ref{th:Combinationdrift}. } By the convergence result (\ref{CLT:drift}) for
each point $\widetilde{x}_{\ell},$ $\ell=1,\ldots,M$
where $p(\widetilde{x})=\frac12$ under $\alpha_0,$
then under $\alpha(n,t;\alpha_0):$
$$
\PP(|\hat{p}(\widetilde{x}_{\ell})-\frac12|<\nu_n)
\geq 
\PP(\sqrt{n}|\hat{p}(\widetilde{x}_{\ell})-{p}(\widetilde{x}_{\ell})|<\nu_n \sqrt{n}-(\Pi(\alpha_0)\,t)_{\ell}) \rightarrow 1.
$$
This means that along this parameter sequence 
$\widehat{\mcA}_H$ with probability approaching 1
solve the system of equalities
$
\widetilde{x}_{\ell}'\widetilde{\alpha}=0
$ for points $p(\widetilde{x}_{\ell})=\frac12.$
In other words, the limit of the estimator coincides
with $\mcA_{CF}$ regardless of the value $t.$

If $p(\widetilde{x})\neq \frac12$ under $\alpha_0$
for any $\widetilde{x},$ then the sign of $\hat{p}(\widetilde{x})-\frac12$ with probability approaching 1 coincides with the sign of ${p}(\widetilde{x})-\frac12$ (where $p(\cdot)$ is the 
conditional probability of $Y=1$ under $\alpha_0$) regardless of parameter $t$
in the sequence $\alpha(n,t;\alpha_0).$
This is the set $\mcC_k \equiv \mcA_{ms}.$
$\blacksquare$ \vspace{.2in}\\
\textbf{Proof of Theorem \ref{th:OFCdrift}.} 
By the argument following the proof of Theorem \ref{th:Combinationdrift} we note that 
under the sequence $\alpha(n,t;\alpha_0),$
the sign of $\hat{p}(\widetilde{x})-\frac12$
with probability approaching 1 coincides
with the sign of ${p}(\widetilde{x})-\frac12$,
where $p(\widetilde{x})$ is the conditional probability of $Y=1$ under the data generating
process with parameter $\alpha_0.$
In other words, the drifting sequence $\alpha(n,t;\alpha_0)$ does not impact the limit
and its properties are as characterized in the 
proof of Theorem \ref{th:OFCsharp}, i.e., it is 
the identified set $\mcA_0.$
$\blacksquare$ \vspace{.2in}\\
\textbf{Proof of Theorem \ref{th:MSrefined}. } 
Theorem \ref{th:MSgeneral2} has established  that the intersection of $[L/2]+1$  sets $\overline{\mcC}_{\ell}$ gives us $\overline{\mathcal{A}}_0$. Considering all possible collection of $[L/2]$  sets from $\{\overline{\mcC}_{\ell}\}_{\ell=1}^L$ we can find a collection that delivers the minimum sum $p(B_{\ell_1}=1)+ \ldots p(B_{\ell_{[L/2]}}=1)$ of probabilities of realizations of the $[L/2]$ sets in this collection in the distribution limit. We denote 
$\tau^*$ this minimum sum. Then  for any quantile index $\tau>\tau^*$ the $\tau$-quantile of the closure of the random set in the distribution limit has to include only points obtained from at least one  intersection  of $[L/2]+1$ sets among $\{\overline{\mcC}_{\ell}\}_{\ell=1}^L$. But, from what we have shown, this has to coincide with $\overline{\mcA}_0$. 

It is also straightforward to see that by definition it is always true that $\tau^* \leq 1/2$. Hence, one can always consider indices $\tau>1/2$ to recover $\overline{\mcA}_0$ as a $\tau$-quantile of a random set.  $\blacksquare$ \vspace{.2in}\\
\textbf{Proof of Theorem \ref{th:QRS_toydesign}.}
 The case when there are points $\widetilde{x} \in \mathcal{X}$ such that $p(\widetilde{x})=1/2$ follows immediately from Theorem \ref{th:MSrefined}. 

 The case when there are no points $\widetilde{x} \in \mathcal{X}$ such that $p(\widetilde{x})=1/2$ follows from the first part of Theorem \ref{th:MSgeneral}. $\blacksquare$ \vspace{.2in}\\
\textbf{Proof of Theorem \ref{th:QRSEdrift}.}
Following the proof of Theorem  \ref{th:MSrobustness}, under 
 $\alpha(n,t;\alpha_0)=\alpha_0+t/\sqrt{n} \in 
\mcC_k$ with $\alpha_0$ being the point in $\mcA$
such that $p(\bar{x}_m)=\frac12$ for $M$ points $\bar{x}_m \in {\mathcal X}$ on the decision making
boundary,  we establish 
$$\sqrt{n}(\hat p(\widetilde{x}_1)-\frac12), \ldots, p(\widetilde{x}_M)-\frac12)' \stackrel{d}{\to} \mathcal{N}\left(\Pi(\alpha_0)\,t, \Sigma\right).$$
This means that the limit of the maximum score estimator along the drifting parameter sequence
takes the value on set $\mcC_j$ with probability
$p(B_j(t)=1)$ equal to the probability that 
$j$-th element of normal random vector
$\xi \sim \mathcal{N}\left(\Pi(\alpha_0)\,t, \Sigma\right)$ exceeds all other elements.

By construction of $\widehat{\mcA}_{RSQ,\tau},$
we consider sets $\overline{\mcC}_j$ corresponding
to the closure of the outcomes of the maximum score
estimator. For a given $t,$ the limit of $\widehat{\mcA}_{RSQ,\tau}$ then takes the value
at the intersection of the collection of sets $\overline{\mcC}_{\ell_1},\ldots,\overline{\mcC}_{\ell_p}$ with the highest value of the sum of probabilities $p(B_{\ell_1}(t)=1)+\ldots+p(B_{\ell_p}(t)=1).$
Provided that $t'\widetilde{x} \lesseqgtr0$ for
$\widetilde{x} \in {\mathcal X}$ where, respectively
$p(\widetilde{x})\lesseqgtr \frac12$ for the data
generating process indexed by $t,$ then set $\overline{\mcC}_k$ (the closure of the set of inequalities $t'\widetilde{x} \lesseqgtr0$) is included in collections of sets in the limit of 
$\widehat{\mcA}_{RSQ,\tau}.$

When $\|t\| \rightarrow 0$ then $p(B_1(0))=\ldots=p(B_M(0)=1).$ This means that 
collection $\overline{\mcC}_1,\ldots,\overline{\mcC}_M$
has the highest sum of probabilities. Its intersection is the closure of the identified set
$\overline{\mcA}_0$ under $\alpha_0.$

When When $\|t\| \rightarrow +\infty,$ then 
at each point $\widetilde{x}_1,\ldots,\widetilde{x}_M,$
$\widehat{p}(\widetilde{x}_j)$ approaches respectively, $0,$ $\frac12$ or $1.$
$\widehat{p}(\widetilde{x}_j) \stackrel{p}{\longrightarrow} 0$ if $t'\widetilde{x}_j<0,$ $\widehat{p}(\widetilde{x}_j) \stackrel{p}{\longrightarrow} \frac12$ if $t'\widetilde{x}_j=0,$
and 
$\widehat{p}(\widetilde{x}_j) \stackrel{p}{\longrightarrow} 1$ if $t'\widetilde{x}_j>0.$
At the same time, set $\overline{\mcC}_k$ is the 
closure of the intersection of the same inequalities
$t'\widetilde{x}_j \lesseqgtr 0$ for $t \in \mcC_k.$
Therefore, the limit of $\widehat{\mcA}_{RSQ,\tau}$
is $\overline{\mcC}_k,$ which is the closure identified
set under $\alpha(n,t;\alpha_0) \in \mcC_k$. $\blacksquare$

\vskip 0.1in

\section{Uniform convergence of the objective function (\ref{MS:objective}).}\label{appendix:uniform}

In this Appendix we give a more advanced coverage of the properties of the maximum score objective function in our toy design. Our analysis is aimed at explaining the 
main structural differences
between the settings of the 
classic maximum score of \cite{manski1975} and the corresponding empirical
process-based arguments in \cite{kimpollard} and our setting. This will provide another explanation of why we obtain a fluctuating behavior of the maximum score estimator in Illustrative Design 1 for $\alpha_{0} \in \{0,-1\}$ appying the result in Theorem \ref{th:MSgeneral2}.

Consider the empirical objective function 
$$MS_n(\alpha)= \avg \psi(y_i,x_i,\alpha), $$
where $\psi(y_i,x_i,\alpha)=y_i{\bf 1}\left\{\alpha+x_i\geq 0\right\}+(1-y_i){\bf 1}\left\{\alpha+x_i < 0\right\}.$
The corresponding population counterpart is 
$$
MS(\alpha)=E[\psi(Y,X,\alpha)].
$$
with $E[\psi(Y,X,\alpha)]=\sum\limits_{x \in \{0,1\}}(P(Y=1\,|\,X=x)\mbox{sign}(\alpha+x)q(x)).$
We use notation $q(x)=P(X=x).$

Suppose that $\alpha_0$ is the true parameter value and consider the class of functions
$\mcF_{\delta}$ consisting of functions 
$$
f(y,x)=\psi(y,x,\alpha)-E[\psi(y,x,\alpha)]-\psi(y,x,\alpha_0)+E\left[\psi(y,x,\alpha)\right]
$$
indexed by $|\alpha-\alpha_0| \leq \delta$ for a given bound $\delta.$ By construction the corresponding class is a VC class of functions.
Note that:
$$
\psi(y,x,\alpha)-\psi(y,x,\alpha_0)=
(1-2y)\,I[\mbox{sign}(x+\alpha) \neq  \mbox{sign}(x+\alpha_0)]\mbox{sign}(\alpha-\alpha_0)
$$
If we set $\alpha_0=0,$ this expression further simplifies to:
$$
\psi(y,x,\alpha)-\psi(y,x,\alpha_0)=
(1-2y)\,I[x+\alpha<0]
=\frac12-y+(y-\frac12)\mbox{sign}(x+\alpha).
$$
Note that the maximum variance of this
object does not depend on the 
size of the neighborhood $\delta \leq 1,$
meaning that 
$$
\sup\limits_{f \in \mcF_{\delta}}
\frac{1}{n}\sum\limits^n_{i=1}f(y_i,x_i)=
O_p(\sqrt{\frac{\log\,n}{n}})
$$
via a standard Hoeffding's bound. This
means that the corresponding
empirical objective function is 
not stochastically equicontinuous.

Consider a modified version of the objective function
$$
MS_n(\alpha)=
(\widehat{p}(0)-\frac12)(1-\widehat{q}(1))\;\mbox{sign}(\alpha)+
(\widehat{p}(1)-\frac12)\widehat{q}(1)\;\mbox{sign}(1+\alpha).
$$
Its population counterpart under
$\alpha_0=0$ is
$$
MS(\alpha)=(p(1)-\frac12)\;
q(1)\;\mbox{sign}(1+\alpha).
$$
In both these specifications, we use
a modified definition of the $\mbox{sign}$ function:  $\mbox{sign}(x)=2{\bf 1}\{x \geq 0\}-1.$
The standard approach in the analysis
of extremum estimators is to study
the behavior of the centered process
$
MS_n(\alpha)-MS(\alpha)-
\left(MS_n(\alpha_0)-MS(\alpha_0) \right).
$
The rationale for this is that if the centered process is ``small" in
expectation, then the 
supremum of the empirical objective
function can linked with the supremum
of the population objective function.
If the population objective function
is ``locally quadratic," then the rate
of convergence of the estimator
balances the expectation of the centered process in the neighborhood
$\delta$ of the maximizer of the 
population objective function
and the $\delta^2$ arising from 
the second order expansion of the 
population objective function. 
In case of standard maximum score
estimator, it can be shown that 
$$
{\mathbb E}\left[ \sup\limits_{|\alpha-\alpha_0| \leq \delta}\left|
MS_n(\alpha)-MS(\alpha)-
\left(MS_n(\alpha_0)-MS(\alpha_0) \right)
\right|\right]=O(\sqrt{\delta/n}). 
$$
Then, if $MS(\alpha)-MS(\alpha_0) \geq
-H\,\delta^2$ is some neighborhood
of $\alpha_0,$ then the tradeoff implies
$H\,\delta^2=O(\sqrt{\delta/n})$
which would set $\delta=O(1/n^{1/3}).$
This is the convergence rate of the 
classic maximum score estimator
under point identification.

In our case the distribution of process $MS_n(\alpha)-MS(\alpha)$ can then be analyzed
directly as a function of a vector
of random variables $(\widehat{p}(1),\widehat{p}(0),\widehat{q}(1)).$
We note that 
$$
\sqrt{n}\left(
\begin{array}{c}
\widehat{p}(1)-p(1)\\
\widehat{p}(0)-\frac12\\
\widehat{q}(1)-q(1)\\
\end{array}
\right) \stackrel{d}\longrightarrow
{\mathcal N}\left(
0,\,\Sigma,
\right)
$$
where $\Sigma$ is a function of $p(1),\,p(0)$ and $q$.

Using the standard delta-method
we obtain
$$
\sqrt{n}(MS_n(\alpha)-MS(\alpha))
\rightsquigarrow
\mbox{sign}(\alpha)\,Z^0+
\mbox{sign}(1+\alpha)\,Z^1,
$$
where $(Z^0,\,Z^1)'$ is the mean zero Gaussian random vector.
We note that $MS_n(\alpha)-MS(\alpha)$
is doubly robust with respect to 
$q(1)$ and, thus, this asymptotic representation is valid up to terms
of order $O_p(1/\sqrt{n}).$

Thus, we can write:
$$
\sqrt{n}MS_n(\alpha)=\sqrt{n}MS(\alpha)+
\mbox{sign}(\alpha)\,Z^0+
\mbox{sign}(1+\alpha)\,Z^1
+o_p(1).
$$
Naturally,
$\mbox{Argmax}_{\alpha}MS_n(\alpha)$
is one (or several) of the intervals: $(-\infty,-1),$ $[-1,0)$
and $[0,+\infty)$ as well as their 
finite endpoints (and, with vanishing probability it can also be pairwise unions sets of points 0 or 1.).
Then:
$$
\widetilde{\mathcal A}_{ms}=
\left\{
\begin{array}{l}
(-\infty,-1),\; \mbox{if}\, Z^1<-\sqrt{n}C^*, Z^1+Z^0<-2\sqrt{n}C^*,\\
\mbox{[}-1,0),\;\mbox{if} \,Z^1>-2\sqrt{n}C^*,\,
Z^0<0,\\
\mbox{[}0,\,+\infty),\;\mbox{if}\,Z^1+Z_0>-2\sqrt{n}C^*,\,Z^0>0,\\
\end{array}
\right.
$$
where $C^*=(p(1)-\frac12)q(1)+o_p(1/\sqrt{n})$
Therefore,
$$
P\left(d_H \left(\widetilde{\mathcal A}_{ms},\,(-\infty,0) \right) =0\right) \rightarrow 0,
$$
and
$$
P\left(d_H \left(\widetilde{\mathcal A}_{ms},\,[0,1) \right)=0 \right)\rightarrow \frac12,\;\;
P\left(d_H \left(\widetilde{\mathcal A}_{ms},\,[1,+\infty) \right)=0 \right)\rightarrow \frac12,
$$
once again confirming the fluctuating behavior of $\widetilde{\mathcal A}_{ms}$ in case $\alpha_0=0$ (and analogous in case $\alpha_0=-1$).

For the centered process for $\alpha \in [\alpha_0-\delta,\,\alpha_0+\delta]$ for some small $\delta>0$
$$
\sqrt{n}\left(MS_n(\alpha)-MS(\alpha)-(MS_n(\alpha_0)-MS(\alpha_0))\right)
\rightsquigarrow
-(1+\sign(\delta))Z^0=-2\,Z^0.
$$
In other words, compared to the behavior of the 
maximum score objective function with continuous
regressors, the expected value of the recentered process for the case
with all discrete regressors does not depend on  the size of the neighborhood around the true parameter $\alpha_0$.

\paragraph{Behavior under sequences $\alpha(n,t;\alpha_0)$:}

As already described earlier, parameter drifting in the focal case $\alpha_0=0$ (and analogous case $\alpha_0=-1$) can 
bridge two regimes -- one regime with the fluctuating behavior between two sets $[-1,0)$ and $[0,+\infty)$ with equal probabilities and the second one just giving $[0,+\infty)$ (coincides with $\mathcal{A}_{0, \alpha_n}$) with probability 1.

For the drifting sequence $\alpha(n,t;\alpha_0)=
\alpha_0+t/\sqrt{n}$ we set $h=f_{\epsilon|\widetilde{X}}(0|\widetilde{x}_m)t'\widetilde{x}_m,$
where $p(\widetilde{x}_m)=\frac12.$ Then
$$
\sqrt{n}(MS^{(n)}_n(\alpha)-MS^{(n)}(\alpha))
\rightsquigarrow
\mbox{sign}(\alpha)\,Z^0+
\mbox{sign}(1+\alpha)\,Z^1,
$$
Then 
$$
\sqrt{n}MS^{(n)}_n(\alpha)=
\left\{
\begin{array}{l}
-Z^0-Z^1-h(1-q(1))-\sqrt{n}C^*,\;
\mbox{if}\,\alpha<-1,\\
-Z^0+Z^1-h(1-q(1))+\sqrt{n}C^*,\;
\mbox{if}\,-1 \leq \alpha<0,\\
Z^0+Z^1+h(1-q(1))+\sqrt{n}C^*,\;
\mbox{if}\,\alpha>0.\\
\end{array}
\right.
$$
Then 
$$
\argmax_{\alpha \in \real}\sqrt{n}MS^{(n)}_n(\alpha)=\left\{
\begin{array}{l}
(-\infty,-1),\;\mbox{if}\;Z^1<-\sqrt{n}C^*-h(1-q(1)),\;
Z^1+Z^0<-\sqrt{n}C^*-h(1-q(1)),\\
\mbox{[}-1,0),\;\mbox{if}\;Z^0<-h(1-q(1)),\;Z^1>-\sqrt{n}C^*,\\
\mbox{[}0,+\infty),\;\mbox{if}\;Z^0>-h(1-q(1)),\;Z^1+Z^0>-\sqrt{n}C^*-h(1-q(1)).
\end{array}
\right.
$$
This means that the
limit random set takes the form 
$$
{\bf A}_t
=\left\{
\begin{array}{l}
\mbox{[}-1,0),\;\mbox{with probability}\;
\Phi(-\tau),\\
\mbox{[}0,+\infty),\;\mbox{with probability}\;
1-\Phi(-\tau),\\
\end{array}
\right.
$$
with $\tau=2h/\sqrt{1-q(1)}.$
As $\tau$ varies from $0$ 
to $+\infty,$ the distribution of the random set varies from equal randomization between $[-1,0)$
and $[0,+\infty)$ to selecting
a fixed set $[0,+\infty).$ Thus, this
choice of the drifting sequence indeed bridges the two cases.

\paragraph{Closed-form estimator for the simple design}
Now limit the parameter space to a compact set $\mcA.$ Consider the baseline setting where $\alpha_0=0.$
Then we can express the closed-form estimator via 
estimated probability $\widehat{p}(0)$ as:
$$
\widetilde{\mcA}_{CF}=\left\{
\begin{array}{l}
\{0\},\;\mbox{if}\;\left|\widehat{p}(0)-\frac12\right|<\delta_n,\\
\mcA,\;\mbox{otherwise.}
\end{array}
\right.
$$
This estimator outputs the point estimate $0$ if the estimated probability 
$P(Y=1|X=0)$ is close to $\frac12$ and outputs the entire parameter space 
otherwise. In this case we can calibrate the threshold sequence $h_n=\frac{t\,c_n}{\sqrt{n}},$
where $c_n \rightarrow \infty$ is a slowly diverging sequence such that $c_n^2/n \rightarrow 0.$ 

Provided that in the baseline design $\alpha_0 = 0,$ then 
$$
\sqrt{n}\left(\widehat{p}(0)-\frac12\right) \stackrel{d}{\longrightarrow}(1-q(1)) Z^0,
$$
and by the dominated convergence theorem $P\left(
\left| \sqrt{n}(\widehat{p}(0)-\frac12) \right| \leq tc_n
\right) \rightarrow 1.$ This means that $\widetilde{\mcA}_{CF,n} \rightsquigarrow \{0\}.$

In contrast, if $\alpha_0 > 0,$ then $p(0)>\frac12.$ As a result, $P\left(
\left| \sqrt{n}(\widehat{p}(0)-\frac12) \right| \leq tc_n
\right) \rightarrow 0$ and $\widetilde{\mcA}_{CF} \rightsquigarrow \mcA.$

The limit of the closed-form estimator exhibits discontinuity in $\alpha_0$ and converges
to a singleton $\{0\}$ when $\alpha_0=0$ and, otherwise, converges to the parameter space $\mcA.$

\section{Static Panel Data Model Discussion}\label{staticpanelproofs}
We focus on slightly modified Illustrative Design 4 where 
$$
Z_{it}=X^1_{it}+\alpha\,X^2_{it}-c_i
-\epsilon_{it}
$$
with $i=1,\ldots,n,$ $t=0,1$, and the observed outcome variable is 
$Y_{it}={\bf 1}\{Z_{it} \geq 0\}.$ In this design, the support of $X_{it}$ is taken to be $\{0,1\}^2$. The modification is applied to the formulation of the unobservable part as $-c_i
-\varepsilon_{it}$.


The maximum score objective function
for this model takes the form
$$
\frac{1}{n}
\sum^n_{i=1}(y_{i1}-y_{i0}){\sign}
\left(x^{1}_{i1}-x^{1}_{i0}+\alpha\,(
x^{2}_{i1}-x^2_{i0}
)\right).
$$

Denote by $p^{+}(t^1,t^2)=P(Y_{i1}-Y_{i0}=1\,|\,X^1_{i1}-X^1_{i0}=t^1,
X^2_{i1}-X^2_{i0}=t^2)$ with 
$t^1,\,t^2 \in \{-1,0,1\}$
and $p^{-}(t^1,t^2)=P(Y_{i1}-Y_{i0}=-1\,|\,X^1_{i1}-X^1_{i0}=t^1,
X^2_{i1}-X^2_{i0}=t^2).$
In addition, $q(t^1,t^2)=P(X^1_{i1}-X^1_{i0}=t^1,
X^2_{i1}-X^2_{i0}=t^2).$
We now replace these probabilities
with their sample counterparts and modify the objective function 
accordingly. 
This leads to
$$
\begin{array}{ll}
&
(\widehat{p}^+(1,1)-\widehat{p}^{-}(1,1))\widehat{q}(1,1)\sign(1+\alpha) 
+(\widehat{p}^+(-1,-1)
-\widehat{p}^-(-1,-1))\widehat{q}(-1,-1))\sign(-1-\alpha)\\
&+
((\widehat{p}^+(1,-1)-\widehat{p}^{-}(1,-1))\widehat{q}(1,-1) \sign(1-\alpha) 
+(\widehat{p}^+(-1,1)
-\widehat{p}^-(-1,1))\widehat{q}(-1,1))\sign(\alpha-1)\\
&+
(\widehat{p}^+(0,1)-\widehat{p}^{-}(0,1))\widehat{q}(0,1)\sign(\alpha) 
+(\widehat{p}^+(0,-1)
-\widehat{p}^-(0,-1))\widehat{q}(0,-1))\sign(-\alpha)\\
&+
(\widehat{p}^+(1,0)-\widehat{p}^{-}(1,0))\widehat{q}(1,0)
-(\widehat{p}^+(-1,0)
-\widehat{p}^-(-1,0))\widehat{q}(-1,0))
\end{array}
$$
The last two terms do not impact the 
location of the maximum and we omit it from further analysis. We denote
the first six terms of this
expression $Q_n(\alpha).$

Denote $U_{i1}=\epsilon_{i1}+c_i$ and
$U_{i0}=\epsilon_0+c_i$. Then 
$$
\begin{array}{l}
p^+(1,1)=P\left(U_{i1}  \leq 1+\alpha,\,U_{i0}>0 \right),\quad p^-(1,1)=P\left(U_{i1}  > 1+\alpha,\,U_{i0} \leq 0 \right),\\
p^+(-1,-1)=P\left(U_{i1}  \leq 0,\,U_{i0}>1+\alpha \right),\quad 
p^-(-1,-1)=P\left(U_{i1}  > 0,\,U_{i0} \leq 1+\alpha \right),\\ 
p^+(1,-1)=P\left(U_{i1}  \leq 1,\,U_{i0}>\alpha \right), \quad 
p^-(1,-1)=P\left(U_{i1} > 1,\,U_{i0} \leq \alpha \right), \\
p^+(-1,1)=P\left(U_{i1}  \leq \alpha,\,U_{i0}>1 \right), \quad 
p^-(-1,1)=P\left(U_{i1} > -\alpha,\,U_{i0} \leq 1 \right),
\end{array}
$$

$$
\begin{array}{l}
p^+(0,1)=P\left(U_{i1}  \leq 1+ \alpha,\,U_{i0}>1 \right)+
P\left(U_{i1}  \leq  \alpha,\,U_{i0}>0 \right)
,\\
p^-(0,1)=P\left(U_{i1}  > 1+\alpha,\,U_{i0} \leq 1 \right)+P\left(U_{i1}  > \alpha,\,U_{i0} \leq 0 \right),\\
p^+(0,-1)=P\left(U_{i1}  \leq 1,\,U_{i0}>1+\alpha \right)
+
P\left(U_{i1}  \leq 0,\,U_{i0}>\alpha \right)
,\\
p^-(0,-1)=P\left(U_{i1}  > 1,\,U_{i0} \leq 1+\alpha \right)
+
P\left(U_{i1}  > 0,\,U_{i0} \leq \alpha \right). 
\end{array}
$$

Consider the setting where $\epsilon_{it}$
is i.i.d. Then, 
under $\alpha_0=0$ we have  
$$
\begin{array}{l}
p^+(1,1)=p^-(-1,-1)=p^+(1,-1)=p^-(-1,1)=
E\left[ F_{\epsilon}(1-c_i)(1-F_{\epsilon}(-c_i))\right], \\
p^-(1,1)=p^+(-1,-1)=p^-(1,-1)=p^+(-1,1)=
E\left[ F_{\epsilon}(-c_i)(1-F_{\epsilon}(1-c_i))
\right], \\ 
p^+(0,1)=p^-(0,1), \quad p^+(0,-1)=p^-(0,-1). 
\end{array}
$$

As a result, with $\alpha_0=0$ the \textit{population} objective
function takes the form 
$$
\begin{array}{ll}
Q(\alpha)=&(({p}^+(1,1)-{p}^{-}(1,1))({q}(1,1)\sign(1+\alpha)-q(-1,-1)\sign(-1-\alpha))\,\\
&+(({p}^+(1,1)-{p}^{-}(1,1))({q}(1,-1)\sign(1-\alpha) -q(-1,1)\sign(\alpha-1)). 
\end{array}$$
Wecan show  that ${p}^+(1,1)-
{p}^{-}(1,1)=E[F_{\epsilon}(1-c_i)-F_{\epsilon}(-c_i)]$ and we will suppose it is strictly positive  (e.g., it holds  if the distribution of $\varepsilon$  has convex support]).  Then function $Q(\alpha)$ is maximized on the 
set $\mcA_{ms}=(-1,1).$

By the standard CLT
$$
\sqrt{n}\left(
\begin{array}{l}
\widehat{p}^{\pm}(t^1,t^2)-{p}^{\pm}(t^1,t^2)\\
\widehat{q}(t^1,t^2)-{q}(t^1,t^2)
\end{array}
\right)
\stackrel{d}{\longrightarrow}
{\mathcal N}\left(0,\,\Sigma\right),\;
t^1,\,t^2 \in \{-1,0,1\}.
$$
In this case,
$$
\sqrt{n}(Q_n(\alpha)-Q(\alpha))
\rightsquigarrow \sign(1+\alpha)\,Z^0
+\sign(\alpha)\,Z^1+\sign(1-\alpha)Z^2,
$$
where $Z^{0,1,2}$ are independent Gaussian random variables. 
Then the maximizer $\widehat{\mcA}_{ms}$ of the empirical objective function
 is such that
$$
P\left(d_H(\widehat{\mcA}_{ms},
(-1,0))=0\right) \rightarrow \frac12,\,
P\left(d_H(\widehat{\mcA}_{ms},
(0,1))=0\right) \rightarrow \frac12.
$$
This can also be concluded from the fact that in the sample objective function the terms $\widehat{p}^+(0,1)-\widehat{p}^{-}(0,1)$ and $\widehat{p}^+(0,-1)-\widehat{p}^{-}(0,-1)$ will fluctuate between strictly positive and strictly negative values with approximately equal probabilities leading to the fluctuating behavior of the estimator. 

At the same time, we can show that whenever $\alpha_0 \in (0,1),$ then 
$$
P\left(d_H(\widehat{\mcA}_{ms},
\mcA_{ms})=0\right) \rightarrow 1.
$$
Indeed, in this case
$p^{+}(0,1)-p^{-}(0,1)=-(p^{+}(0,-1)
-p^-(0,-1))=E\left[
F_{\epsilon}(1+\alpha_0-c_i)-F_{\epsilon}(1-c_i)\right]>0.$ The population objective
function is 
$$
\begin{array}{ll}
Q(\alpha)=&(F_{\epsilon}(1+\alpha_0-c_i)-F_{\epsilon}(1-c_i))({q}(1,1)\sign(1+\alpha)-q(-1,-1)\sign(-1-\alpha))\,\\
&+(F_{\epsilon}(1+\alpha_0-c_i)-F_{\epsilon}(1-c_i))({q}(1,-1)\sign(1-\alpha) -q(-1,1)\sign(\alpha-1)) \, \\
&+(F_{\epsilon}(1+\alpha_0-c_i)-F_{\epsilon}(1-c_i))({q}(0,1)\sign(\alpha) -q(0,-1)\sign(-\alpha)) 
\end{array}$$
is maximized at $\mcA_{ms}=(0,1)$. At the same time, now in the sample objective function we will always have positive estimates $\widehat{p}^+(1,1)-\widehat{p}^{-}(1,1)$, $\widehat{p}^+(1,-1)-\widehat{p}^{-}(1,-1)$, $\widehat{p}^+(0,1)-\widehat{p}^{-}(0,1)$ and negative estimates $\widehat{p}^+(-1,-1)-\widehat{p}^-(-1,-1)$, $\widehat{p}^+(-1,1)-\widehat{p}^-(-1,1)$, $\widehat{p}^+(-1,0)-\widehat{p}^-(-1,0)$. This means that 
with probability approaching 1 in 
the sample objective function maximizer will be $(0,1)$.

\section{Simulation experiment for feasible quantile random set estimator.}
\label{simulation}
\begin{figure}\centering
{\label{a}\includegraphics[width=.45\linewidth]{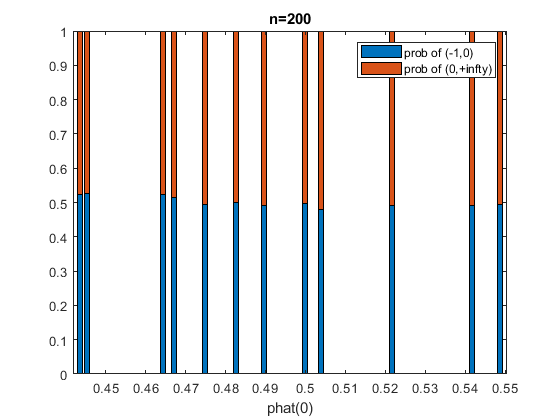}}\hfill
{\label{b}\includegraphics[width=.45\linewidth]{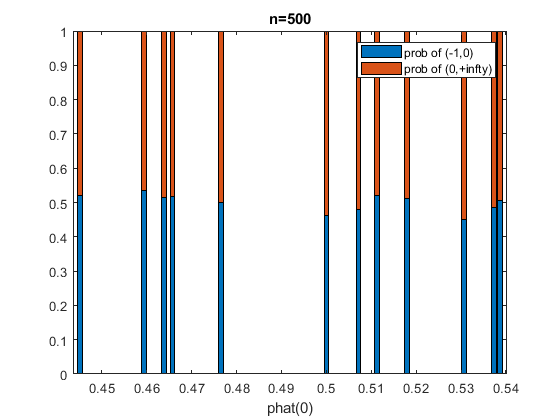}}\par 
{\label{c}\includegraphics[width=.45\linewidth]{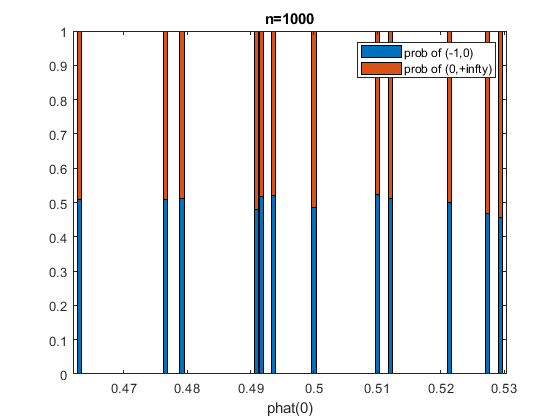}}
\hfill
{\label{d}\includegraphics[width=.45\linewidth]{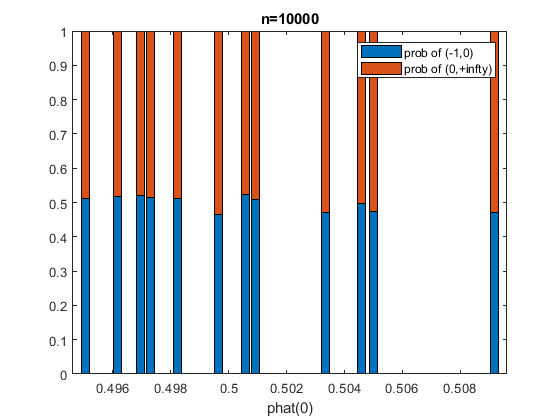}}
\caption{Illustration of the probabilities when  $(-1,0)$ or $(0,+\infty)$ delivers the maximum value to the maximum score objective function. These are sample probabilities obtained under sampling $\hat p_s(x)$, $x \in \{0,1\}$. $s=1, \ldots, 2000$. }
\label{fig:MolchanovFeasible1}
\end{figure}

Consider $\alpha_0=0$ in Illustrative Design 1. Then $p(0)=1/2$ and 
$p(1)>\frac{1}{2}$. Let $q_0=0.7$ and $\varepsilon \sim \mathcal{N}(0, \sigma^2)$ with $\sigma=0.5$. We choose sample sizes $n=200$. $n=500$, $n=1000$ and $n=10000$.  Figure \ref{fig:MolchanovFeasible1} is representative of the results we get for this case. The illustrations in Figure \ref{fig:MolchanovFeasible1} are for 12 different realizations of $\hat p(0)$ in the sample. The blue vertical bars represent probabilities with which $(-1,0)$ (or its closure or a half-closure) is $\widehat{\cal A}^{\ast}_{ms,s}$, and the red vertical bars represent probabilities with which $(0,+\infty)$ (or its closure) is $\widehat{\cal A}^{\ast}_{ms,s}$. Even though  for graphical quality we chose to limit our illustration to 12 different realizations of $\hat p(0)$, the patterns we see in those graphs are representative of what would happen for other realizations of $\hat p(0)$. In particular, we conclude that if we take $\Delta=0.05$, then the $(0.5+\Delta)$th sample quantile of the closure of the maximum score estimand is $\mathcal{A}_0=\{\alpha_0\}$ with an extremely high probability. Namely, for 1,000 different random sample of size $n=200$ (and, hence, potentially for 1,000 different realizations of $\hat p(0)$) only in 2.2\% cases the sample $0.55$th quantile obtained in the described above way is a strict superset of $\{0\}$. For 1,000 different random sample of size $n=500$ it is 1.7\%. For 1000 different random sample of size $n=1000$ it is 2.2\%. For 1,000 different random sample of size $n=10000$ it is 2\%. If we slightly increases the quantile index and take $0.5+\Delta=0.6$, then all these  percentages for all mentioned $n$ are 0. 

In the final comment we note that infrequently  $\hat p_s(x)$ may be drawn outside of their natural $[0,1]$ range,. In this case, one may want to consider 
\begin{equation*}  
\widetilde{\cal A}^{\ast}_{ms,s} = \arg \max _{\alpha \in {\cal A}} \sum_{x \in \{0,1\}}  (\min\{\max\{0,\hat p_s(x)\},1\}-0.5)\cdot \sgn(\alpha+x) \widehat{P}(X=x), \quad s=1, \ldots, S. 
\end{equation*} 
instead of (\ref{maxscoretoy_real_sampling}) but this does not change our conclusions in any way. 
\pagebreak

\end{document}